\documentclass[10pt]{article}
\usepackage{cite}
\usepackage{ulem}
\usepackage{color}
\usepackage{epsfig}
\usepackage{graphicx}
\usepackage{subfigure}
\usepackage{ulem}
\usepackage{amsfonts}
\usepackage{epsfig,bm}
\usepackage{graphicx}
\usepackage{comment}
\newcommand{\be}{\begin{equation}}
\newcommand{\ee}{\end{equation}}
\newcommand{\ba}{\begin{eqnarray}}
\newcommand{\ea}{\end{eqnarray}}
\newcommand{\nn}{\nonumber\\}
\begin{document}
\begin{titlepage}
\title{Thermodynamic Geometric Stability of Quarkonia states}
\author{}
\date{Stefano Bellucci $^{a}$ \thanks{\noindent bellucci@lnf.infn.it},
Vinod Chandra$^{b}$ \thanks{joshi.vinod@gmail.com} \ and
Bhupendra Nath Tiwari $^{a}$ \thanks{\noindent bntiwari.iitk@gmail.com}\\
\vspace{0.5cm}
$^{a}$INFN-Laboratori Nazionali di Frascati\\
Via E. Fermi 40, 00044 Frascati, Italy.\\ \vspace{0.5cm}
$^b$Department of Theoretical Physics, \\ Tata Institute of Fundamental Research,\\
Homi Bhabha Road Mumbai-400005, India.}

\maketitle \abstract{We compute exact thermodynamic geometric
properties of the non-abelian quarkonium bound states from the
consideration of one-loop strong coupling. From the general
statistical principle, the intrinsic geometric nature of strongly
coupled QCD is analyzed for the Columbic, rising and Regge
rotating regimes. Without any approximation, we have obtained the
non-linear mass effect for the Bloch-Nordsieck rotating strongly
coupled quarkonia. For a range of physical parameters, we show in
each cases that there exists a well-defined, non-degenerate,
curved, intrinsic Riemannian manifold. As the gluons become softer
and softer, we find in the limit of the Bloch-Nordsieck
resummation that the strong coupling obtained from the Sudhakov
form factor possesses exact local and global thermodynamic
properties of the underlying mesons, kaons and $D_s$ particles.}
%$D$-brane charge= electric charge
\vspace{1.5cm}

\vspace{2mm} {\bf Keywords}: Thermodynamic Geometry, Quarkonia,
Massive Quarks, QCD Form Factor. %Bessel Functions,

{\bf PACS}: 02.40.-k; 14.40.Pq; 12.40.Nn; 14.70.Dj

\end{titlepage}

\section{Introduction}

Quarkonia are the central objects to explore the non-perturbative
nature of non-abelian gauge theories \cite{KhozeRingwald}. The
fate of the heavy quarkonia in hot QCD medium describes the nature
of confined-deconfinement phase transition
\cite{MatsuiSatz,Karsch,nora} in QCD and thereby the nature of the
inter-quark forces\cite{vin1,vin2,vin3}. An understanding of such
issues is essential towards the formation of hot and dense states
of quark gluon plasma matter in the heavy ion collision
\cite{Star} and in the early universe \cite{Aguiar}. Thus, our
prime focus is to address the stability issues of vacuum
constituted by various quarkonia states. This is in direct concern
with the fact that the quark matter formation is an important
state of art in the particle physics and its behavior is an
intense topic of experimental research in the hot QCD media
\cite{Phenix}. This motivates us to investigate the nature of
stability when the quarkonia possess a non-zero mass and spin.
Following the notions of the quarkonium vacuum, we notice that an
extensive study is required for a proper understanding of the
thermodynamic stability and global statistical correlation in the
non-abelian gauge theories. To the best of our knowledge, this is
perhaps the first attempt in this direction.

%On the other hand,
It is by now well established that the thermodynamic intrinsic
geometric examinations are important in the configurations
involving black holes in string theory
\cite{9601029v2,9504147v2,0409148v2, 9707203v1,0507014v1,
0502157v4, 0505122v2} and $M$-theory \cite{0209114,0401129,
0408106,0408122}, possessing a set of rich stability structures
\cite{0606084v1,SST,bnt, BNTBull, BNTBull08}. There exists a
wealth of literature where thermodynamic geometry plays an
important role towards the understanding the physics of
black-holes, black branes, black rings and various higher
dimensional objects. Attentions were paid on the equilibrium
perspective of black holes, and thereby one explicates the nature
of the pair correlations and associated stability of the
solutions. Recently, the thermodynamic geometry has emerged as one
of the most powerful tool in the understanding covariant
statistical correlations and the limit of the underlying
microscopic-macroscopic duality issues for the 1/2- BPS black
holes in string theory \cite{BNTBull08}. There have been several
general notions explored in the condensed matter physics
\cite{RuppeinerRMP, RuppeinerA20,RuppeinerPRL,RuppeinerA27,
RuppeinerA41, RuppeinerPRD78,anyon,tsal}, motivating us to
consider the quarkonia with a set of equilibrium parameters and
analyze the underlying parametric pair correlation functions and
global correlations. We find thus that the intrinsic geometric
investigation possesses an exact mathematical nature of the
massive fluctuating quarkonia. In this concern, we provide a brief
account of the concerned strong interactions and thermodynamic
geometry in the next section.

Given a definite covariant intrinsic geometric description of the
quarkonia, we examine (i) the conditions of thermodynamic
stability of the configuration, (ii) its parametric global
correlation function in terms of a chosen set of fluctuating
parameters. In this analysis, one can enlist the complete set of
non-trivial parametric correlation functions of the quarkonia. The
similar considerations remain valid over the black hole
configurations in general relativity \cite{0601119v1,
0512035v1,0304015v1, 0510139v3}, attractor black holes
\cite{9508072v3, 9602111v3,new1,new2,0805.1310} and Legendre
transformed chemical configurations \cite{Weinhold1, Weinhold2}.
Interestingly, we have explored the issue of thermodynamic
geometry to the hot QCD \cite{BNTSBVC} and offered the detailed
picture of such duality relations between the thermodynamic
geometry and quark-number susceptibilities. The issue of the quark
matter stability is quite novel in its own right and calls for
further investigations of the thermodynamic geometric structure of
the phases of non-abelian gauge theories and quark matter
production \cite{Baluni}.

The thermodynamic geometric consideration is examined for the
strong coupling of massless quarks which reveal limited effects of
the QCD calculations, which have been known for the gluons since
decades \cite{JacobKroll, LiSterman}. The intrinsic geometric
analysis of the linear Regge rotating quarkonium configuration
\cite{Polyakov} offers the thermodynamic stability of quark matter
with a precise account of non-zero Regge mass. This consideration
supports the formation of the topological Heisenberg spin glass
\cite{Dzyaloshinskii} and vector boson pair production at the
hadron colliders \cite{CampbellEllis}. Further, the thermodynamic
geometric properties take an intrinsic account of the
proton-proton and proton-antiproton scattering, hadronic total
cross sections through the soft gluon summation in impact
parameter space \cite{GrauPancherieSrivastawa}, and production of
the gluons \cite{LiptovFadin}. Thus, the differential geometry
plays an important role in the study of the physics of quarkonia.

In this paper, we examine the thermodynamic stability properties
of the quarkonia system described by two and three parameters. We
find that our analysis uniformly applies to all possible QCD
particles with varying masses from the range of the charmonium
states to the bottomonium states. This follows from the fact that
the effective mass of the theory is kept very arbitrary; therefore
the finite temperature behavior of the quarkonia is captured
through the present analysis. From the thermodynamic geometric
viewpoint of the hot QCD \cite{BNTSBVC}, it seems quite
interesting that we can reveal the near equilibrium behavior of
quarkonia over an ensemble of hot QCD states. To do so, we
introduce the thermodynamic metric tensor through the variations
of the strong coupling, and focus our attention on the
thermodynamic properties when gluons become softer and softer
\cite{GrauPancherieSrivastawa}. Our consideration reveals the
thermodynamic geometric nature of very soft gluons and thereby
describes the stability properties of non-abelian quarkonium bound
states with and without the mass.

As explained in the section $2$, the problem thus posed starts
from the very definition of Sudhakov form factor \cite{Sudakov},
and thus the intrinsic geometric understanding of the QCD coupling
in the limit of the vanishing transverse momentum energies. Notice
that our analysis considers all types of QCD states, as we do not
truncate the theory but work with the full Bessel function
underlying in the Sudhakov form factor. We initiate our analysis
where the computation requires only the first few excited states
in which one may think that only the charmonium states play the
dominant role in the quark matter formation. The analysis is
continued for the mass-less Regge rotating quarkonia, and finally
all possible mass states are considered through the strong
coupling. Our study is two folded, (i) the investigations on the
zero temperature properties of quarkonia, and (ii) extensions to
the finite temperature, as well as very high temperature
properties of the quarkonium vacuum fluctuations in the
non-abelian gauge theories. This shows the universality of our
consideration of the thermodynamic geometry, that applies for all
possible mass excitations of the quarkonium vacuum, and thus the
stability of quark matter formations.

The effect of finite mass is explored for the Regge mass and
arbitrary mass quarkonia with the one-loop Bloch-Nordsieck
resummation \cite{BlochNorsieck}. Following the Polyakov arguments
\cite{Polyakov}, our analysis shows that the rising potential
vacuum can be thought of as a collection of decaying and weakly
interacting particles. However, the Regge rotating configuration
offers almost the same stability structure and is also nearly
weakly interacting. Subsequently, we show that the global
thermodynamic correlation of  the Regge rotating states remains
almost the same as that of the massless states, up to the sign of
the rotation term. With the large quark mass, the strong coupling
acquires a large twist and thus thermodynamically the
configuration becomes highly unstable. For experimental reasons,
the graphical characterizations are offered for large QCD cut-off,
where the quarkonium vacuum attains the Regge rotating regime.
Such an investigation clarifies the thermodynamic stability
properties of the quark matter formation. With this motivation, we
shall focus our attention on the (i) quark masses and their
dependence on the decay constants, (ii) soft gluons and meson
decays, (iii) very soft gluons and (iv) strong coupling hadronic
form factor.

% We consider the following effects into the present consideration.
% \begin{itemize}
% \item mass dependence of decay constants
% \item soft gluons and resummations
% \item meson decays
% \item very soft gluons
% \item hadronic from factor.
% \end{itemize}

%Based on the nature of the quarkonium states,
The paper is organized as follows. In section 2, we present a
brief account of the strong interactions present in the quarkonium
formation and its thermodynamic geometry. In section 3, we
consider the non-rotating massless quarks and examine their
thermodynamic geometric stabilities. Subsequently, we extend the
analysis for the non-zero angular momentum. In section 4, we
explore generic cases in the framework of the Sudhakov form factor
in the limit of Bloch-Nordsieck resummation for the massive
quarkonia, with and without rotations. In section 5, we discuss
some concluding issues towards the thermodynamic geometric picture
of quark matter formation and the stability of underlying QCD
particles.

\section{Phases of the Strong Interaction}
In this section, we shall set-up the formulation of problem and
outline the notion of the intrinsic thermodynamic geometry. The
non-perturbative gluonic effects gives the relation of effective
QCD theories with the index of the potential, momentum scale, mass
and angular momentum, if any, and thus the consideration for the
analysis. Most efficiently, our intrinsic geometric model is
designed to provide the critical values for the parameters of the
gluonic effects. It is worth mentioning the importance of the
present method that the remaining values of the parameters would
be an stable mode for the considered type of quarkonia.

\subsection{Quarkonium Bound States}
In the physics of quarks, the non-perturbative gluonic effects are
addressed via the considerations of Bloch and Nordsieck
\cite{BlochNorsieck}. Some quantities of known interest are
intrinsic transverse momentum of Drell-Yan pairs
\cite{TangermanMulders}, impact parameter distribution of partons
and rise of total cross section \cite{Pancheri}, form factors and
decay constants \cite{PagelsStokar} of the particles. A gluon is
said to be soft, if the transverse momentum $k_{\bot}$ is kept
small. In this limit, the infrared limit of the charmonium is
always unobservable. This follows from the fact that a momentum
integration is required to perform, if one wants to keep a
vanishingly small transverse momentum, i.e., $k_{\bot}\rightarrow
0$ for the gluons. In the case of abelian theory, the results
follow directly from the properties of the underlying Poisson
distribution and their numerical counterparts.

In the case of non-abelian theory, the consideration of Sudhakov
form factor turns out to be an efficient tool for the soft gluons,
which is described as below. Let us recall that the QCD coupling
$\alpha_s(k_\bot)$ never lies near the limit $k_\bot\rightarrow
0$, and thus the QCD effects are limited for the bound state thus
formed after the  resummation. This initiates the requirement for
the integration over the $k_\bot$ and thus the necessity of
Sudhakov form factor. Let us illustrate the computations for
decays processes associated with the pions, viz., $ \pi^0
\rightarrow \gamma \gamma$ and $ \pi \rightarrow \mu \nu $.

During the process, when a virtual photon hits the quarks, soft
gluons are emitted with $k_{\bot} \rightarrow 0$ and a non-zero
total momentum $k^2 \ne 0$, satisfying the standard transverse
momentum Poisson distribution $d^2 P(k_{\bot})$. The associated
Poisson summed vertex and the transverse limiting momenta are
respectively given by \ba \label{eq1} \Gamma_{\pi \rightarrow
\gamma \gamma} &\sim& \frac{d^2 P(k_{\bot})}{d k_\bot^2},\nn
\Pi(0)&=& \frac{d^2 P(k_{\bot})}{d k_{\bot}^2}|_{k_{\bot}= 0} \nn \\
&=&\int d^2 \overrightarrow{b} \exp(-h(b)),\nn \ea
 where the exponent defined as \ba h(b)= \int d^3n_k
(1-\exp(-i\overrightarrow{b}.\overrightarrow{k_{\bot}})) \ea is
known the Sudhakov form factor. Recall that one is required to
take the vanishing transverse momentum limit for a decay, which is
done systematically with the help of the Sudhakov form factor, see
for a review \cite{Collins}. We now illustrate the case of
massless quarks. For the quark mass $m_q=0$, in turns out that the
Sudhakov form factor reduces to the following integral
 \ba h(b)= \int \frac{dk_l}{2k}\frac{dk_{\bot}}{k_{\bot}^2}
(1-\exp(-i\overrightarrow{k_{\bot}}.\overrightarrow{b})), \ea
where $k_{\bot} \in (0, m_{P/2})$ is due to a physical reason that
on average each soft gluon can take as much as half of the initial
center of mass energy. Herewith, the consideration of Sudhakov
form factor comes for a model of limiting transverse momentum QCD
coupling $\alpha_s(k_\bot)|_{k_\bot \rightarrow 0}$. Inspired by
the behavior of relativistic wave functions near the origin for a
QCD potential \cite{Durand, Kuhn} for the Rechardson potential for
the quark bound states, one arrives at the transverse momentum
dependent strong QCD coupling
 \ba \alpha_s(k_\bot^2)= \frac{12 \pi}{(33-2N_f)}
\frac{p}{\ln(1+p(\frac{k_\bot}{\Lambda_{QCD}})^{2p})} \ea in the
limit of one loop gluon exchange potential. Towards the
determination of the index $p$, an interesting argument follows
from the consideration of Polyakov \cite{Polyakov} which we shall
explore further from the perspective of thermodynamic geometry in
the subsequent discussion of the present paper. Before doing so,
let us consider the joint effects of the (i) confinement and (ii)
rotation, and make a platform to describe the properties of the
thermodynamic geometry. In this case, the configuration is
described by Regge trajectories with the leading order effective
potential \ba \label{Regge} V(r,J)= \frac{J(J+1)}{r^2}+Cr^{2p-1}.
\ea Such an effective potential is inspired from the limiting QCD
strong coupling \ba \label{eq2} \alpha_s(Q^2)=
b^{-1}\frac{p}{\ln(1+p(\frac{Q^2} {\Lambda_{QCD}^2})^p)}\ea with
the matching of the pre-factor as $b:= (33-2N_f)/12 \pi$. In the
momentum space, the net effective potential offers the right
quarkonium bound states, after taking account of the one-loop
exchange terms. In the sense of one-loop quantum effects, we
propose that the "quantum" nature of quark matters follows
directly from the consideration of the thermodynamic intrinsic
geometric potential of Richardson types. The index $p$ defines the
nature of the potential \cite{Polyakov}, i.e. whether the
quarkonium lies in the Coulomb phase $(p=0)$ or in the rising
phase $(p>1/2)$. Notice further that the Regge behavior may also
be determined from interpolating values of the index $p$, which is
the matter of the subsequent sections. For the future purpose, it
is worth mentioning that the minimization of the effective
potential containing the confining and rotation effects yields the
value $p=5/6$ for the index, which corresponds to the linear Regge
regime. We shall exploit these facts, while dealing with the
thermodynamic geometry of rotating (massive) quarkonia.

\subsection{Thermodynamic Geometry}

In this subsection, we offer a brief account of the essential
features of thermodynamic geometries from the perspective of the
application to the strongly coupled QCD. We shall focus our
attention on the thermodynamic geometric nature of quarkonium
bound states with a finite number of parameters carried by the
effective field theories.

Let us consider the framework of the intrinsic Riemannian geometry
whose covariant metric tensor is defined as the Hessian matrix of
the QCD coupling, with respect to a finite number of arbitrary
parameters carried by the soft gluons and quarks. This
consideration yields the space spanned by the $n$ parameters of
the strong QCD coupling $\alpha_s$, which in the present
consideration, exhibits a n-dimensional intrinsic Riemannian
manifold $ M_n $. Following the notion of \cite{bnt,BNTBull,
BNTBull08, RuppeinerRMP,RuppeinerA20,RuppeinerPRL,RuppeinerA27,
RuppeinerA41,RuppeinerPRD78} for the thermodynamic geometry, the
components of the covariant metric tensor are defined as

\be \label{eq3} g_{ij}:=\frac{\partial^2
\alpha_s(\vec{x})}{\partial x^j
\partial x^i}, \ee

where the vector $\vec{x} \in M_n $. In the quarkonium effective
configuration, there are only a few in physical parameters which
make the analysis fairly simple. As mentioned in the foregoing
section, the variables of the interest in the present study of
quarkonium are the momentum scale parameter, $Q^2:=q$, the mass
$M$ and the angular momentum $J$, if any. Thus, let us first
illustrate the consideration of thermodynamic geometry for the
non-rotating configurations, viz., $J=0$. In this case, we can
take the QCD coupling $\alpha_s(q,p)$ as a function of ${q,p}$ and
thereby may explore the local and global thermodynamic properties
towards the stability of the quarkonia on $qp$-surface. For a
given QCD coupling $\alpha_s(q,p):= A$, an equilibrium quarkonium
configuration is achieved at the points, where the first
derivatives

\ba \frac{\partial A(q,p)}{\partial q}&=&0,\ \ \frac{\partial
A(q,p)}{\partial p}=0 \ea vanish identically, implying the
existence of the equilibrium data $\{q_0,p_0\}$ on an intrinsic
Riemannian surface $(M_2(R),g)$. In order to verify whether the
quarkonium configuration is  minimally coupled at $\{q_0,p_0\}$,
we may follow the set-up of thermodynamic geometry and define the
components of the thermodynamic metric tensor as

\ba \label{eq4} g_{qq}&=& \frac{\partial^2 A}{\partial q^2},\ \
g_{qp}= \frac{\partial^2 A}{{\partial q}{\partial p}},\ \ g_{pp}=
\frac{\partial^2 A}{\partial p^2}.\nn \ea

In the present case of the $(M_2(R),g)$, it follows that the
determinant of the metric tensor is

\ba \label{eqn5} \Vert g(q,p) \Vert &= &A_{qq}A_{pp}- A_{qp}^2.
\ea Explicitly, we can calculate $\Gamma_{ijk}$, $R_{ijkl}$,
$R_{ij}$ and $ R $ for the above two dimensional thermodynamic
geometry $(M_2,g)$ of the non-rotating quarkonia and may easily
see that the scalar curvature is given by

\ba \label{eqn6} R(q, p)&=& -\frac{1}{2 \Vert g(q,p) \Vert^{2}}
(A_{p p}A_{qqq}A_{q p p}- A_{qq}A_{qpp}^2-A_{pp} A_{qqp}^2
\nonumber \\ &&+ A_{qp}S_{qqp}A_{qpp} +A_{qq}A_{qqp} A_{ppp}
+A_{qp}A_{qqq}A_{ppp}).\ea

As a global intrinsic geometric invariant, the scalar curvature
accompanies information of the correlation volume of underlying
quark fluctuations. The scalar curvature further explicates the
nature of long range global correlations and phase transitions, if
any, deriving from a given phase. In this sense, we anticipate
that the set of particles corresponding to the specific decay, are
statistically interacting, if the underlying quarkonium
configuration has a non-zero thermodynamic scalar curvature.
Incrementally, we notice  that the configurations under present
consideration are allowed to be effectively attractive or
repulsive, and weakly interacting in general. For the two
dimensional thermodynamic geometry \cite{bnt} defined as the
intrinsic Riemannian surface $(M_2(R),g)$, the relation of the
thermodynamic scalar curvature to the thermodynamic curvature
tensor is given as \be \label{eq7}
 R(q,p)=\frac{2}{\Vert g \Vert}R_{qpqp}.\ee

The intrinsic geometric analysis thus provides a set of physical
indications encoded in the geometric quantities, e.g., the scalar
curvature and possible geometrically non-trivial invariants. For a
given quarkonium configuration, the underlying analysis would
involve an ensemble or subensemble equilibrium configuration
forming the statistical basis for the Gaussian distribution of the
particle, which is ensured for any distribution in the late time
limit. With this brief introduction to the thermodynamic geometry,
we shall now proceed to systematically analyze the underlying
stability structures for the parametric fluctuations of the
massless non-rotating quarkonia.

\section{Massless Quarkonia}
In the present section, we shall examine the nature of the QCD
coupling $\alpha_s(q,p)$ from the thermodynamic stability
properties. Firstly, we analyze the massless quarkonium for
non-rotating configurations and then take take an account of the
rotation in the next subsection. In the subsequent analysis, we
shall adopt the notation $L=\Lambda_{QCD}^2$.

\subsection{Non-rotating Quarkonia}
As stated earlier, the thermodynamic metric in the quarkonium
parameter space is given by the Hessian matrix of the strong
coupling with respect to the intensive variables, which in this
case are the two distinct parameters $\{q,p\}$ carried by the
quark matters. Considering Eqn.(\ref{eq2}), one obtains the
following expression:

\ba A(q,p) := \frac{p}{b \ln(1+p(q/L)^p)}. \ea

for the strong QCD coupling. To compute the thermodynamic metric
tensor in the parameter space, we employ the Eqn.(\ref{eq4}),
which leads to the following expression for the components of the
metric tensor

\ba g_{qq} &=& \frac{p^3}{bq^2} \frac{n^Q_{11}}{r^Q_{11}}, \ \
g_{qp} = \frac{p^2}{b q}\frac{n^Q_{12}}{r^Q_{12}}, \ \
g_{pp}=\frac{1}{b} \frac{n^Q_{22}}{r^Q_{22}}. \ea

In this framework, we observe that the geometric nature of
parametric pair correlations offers the notion of fluctuating
quarkonia. In order to simplify the subsequent notations, let us
define the logarithmic factor as
 \ba l(p):=\ln(1+p (q/L)^p). \ea

In this case, we find that the factors in the numerator of the
local pair correlation functions are expressed as follows

\ba n^Q_{11}&:=& 2 p^2 (q/L)^{2 p}+  l(p)((q/L)^p- p(q/L)^p
+(q/L)^{2 p} p), \nn n^Q_{12}&:=& 2 (q/L)^{2 p}p+ l(p)(2
(q/L)^{2p} p^2 \ln(q/L)-3 (q/L)^p \nn && -2 (q/L)^{2 p} p -(q/L)^p
p \ln(q/L)), \nn n^Q_{22}&:=& 2 (q/L)^{2 p}( p + 2p^2\ln(q/L) +
p^3 \ln(q/L)^2)\nn && - l(p)( (q/L)^p +(q/L)^{2 p} p +4 (q/L)^p p
\ln(q/L)\nn && +2 (q/L)^{2 p} p^2 \ln(q/L)+(q/L)^p p^2\ln(q/L)^2).
\ea

We further notice an interesting conclusion for the denominator of
the local pair correlation functions, and we find for the massless
non-rotating configuration that the denominators $\{r^Q_{ii}| \
i=1,2,3\}$ of all the local pair correlation functions take
uniform value $ l(p)^3 \exp{(2 l(p))}$.

Thus, the fluctuating quarkonium configuration may be easily
analyzed in terms of the parameters of the underlying effective
theory. Moreover, it is evident that the principle components of
the metric tensor, signifying self pair correlations remain
positive definite functions. In a given QCD phase, this happens
when the parameters $\{ q, p \}$ are confines in the domain
\ba\{\mathcal D:= (q,p) \in \ M_2| \ n^Q_{11}>0, \ n^Q_{22}>0
\}.\ea Over this domain of $\{ q, p \}$, it is worth mentioning
that the massless non-rotating quarkonia is well-behaved and
locally stable. The global stability is offered by computing the
determinant of the metric tensor, and requiring it to be positive
definite. Following the Eqn.(\ref{eqn5}), we find further for the
generic value of the parameters that the Gaussian fluctuations
form a stable set of correlations over $\{ q, p \}$, if the
determinant of the metric tensor \ba \label{deteqQ} \Vert g \Vert
= \frac{p^3}{b^2 q^2 l(p)^5 \exp{(3 l(p))}} n^Q_g \ea remains a
positive function on the intrinsic $qp$-surface $(M_2(R),g)$.
Explicitly, we obtain that the numerator of the determinant of the
metric tensor can be expressed as \ba n^Q_g(q,p):&=& 2 (q/L)^{3
p}(p+3 p^2+2 p^2 ln(q/L)+2  p^3 ln(q/L)+ p^3 ln(q/L)^2)\nn &&-
l(p)(2 (q/L)^{2 p}+ p(q/L)^{3 p}+2 p^2 (q/L)^{3 p} ln(q/L) +4p^2
(q/L)^{3 p}\nn && +4 p (q/L)^{2 p} ln(q/L)+ p^2 (q/L)^{2 p}
ln(q/L)^2 +2p^2 (q/L)^{2 p} ln(q/L)\nn && +7 p (q/L)^{2 p}).\ea

The behavior of the determinant of the metric tensor shows that
such massless non-rotating quarkonium becomes unstable for
opposite values of $q$ and $p$ for positive $n^Q_g(q,p)$. For
generic $q$ and $p$, the nature of the determinant of the metric
tensor is depicted in the Eqn.(\ref{deteqQ}), showing that the
quarkonia become unstable in the vanishing limit of $q$ or $p$. It
is worth mentioning for a common sigh of the index parameter $p$
and $b$, that the non-rotating massless quarkonia is stable in the
regions of the $(M_2(R),g)$, where the function $n^Q_g(q,p)$ picks
up the positive sign.

\begin{figure}
\hspace*{0.5cm}
\includegraphics[width=8.0cm,angle=-90]{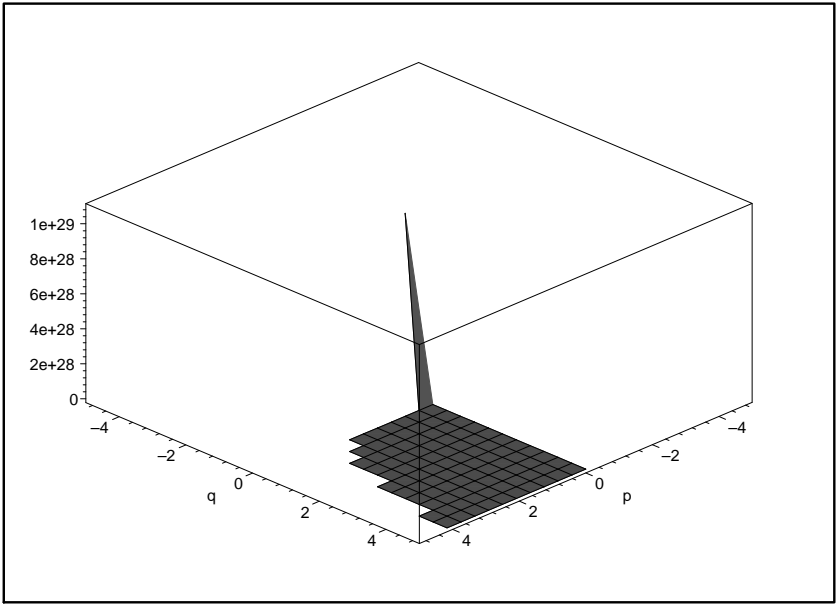}
\caption{The determinant of the metric tensor plotted as a
function of the scale and index parameter $q, p$, describing the
fluctuations in massless non-rotating quarkonia.} \label{det3dPQ}
\vspace*{0.5cm}
\end{figure}

\begin{figure}
\hspace*{0.5cm}
\includegraphics[width=8.0cm,angle=180]{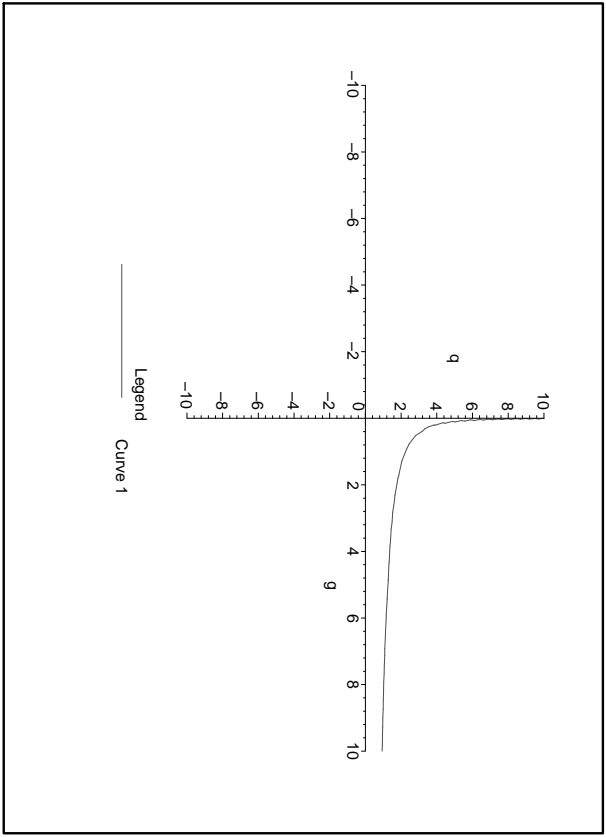}
\caption{The determinant of the metric tensor plotted as a
function of $q$, describing the fluctuations in massless
non-rotating quarkonia near the Coulomb regime.} \label{det2dPQc}
\vspace*{0.5cm}
\end{figure}

\begin{figure}
\hspace*{0.5cm}
\includegraphics[width=8.0cm,angle=180]{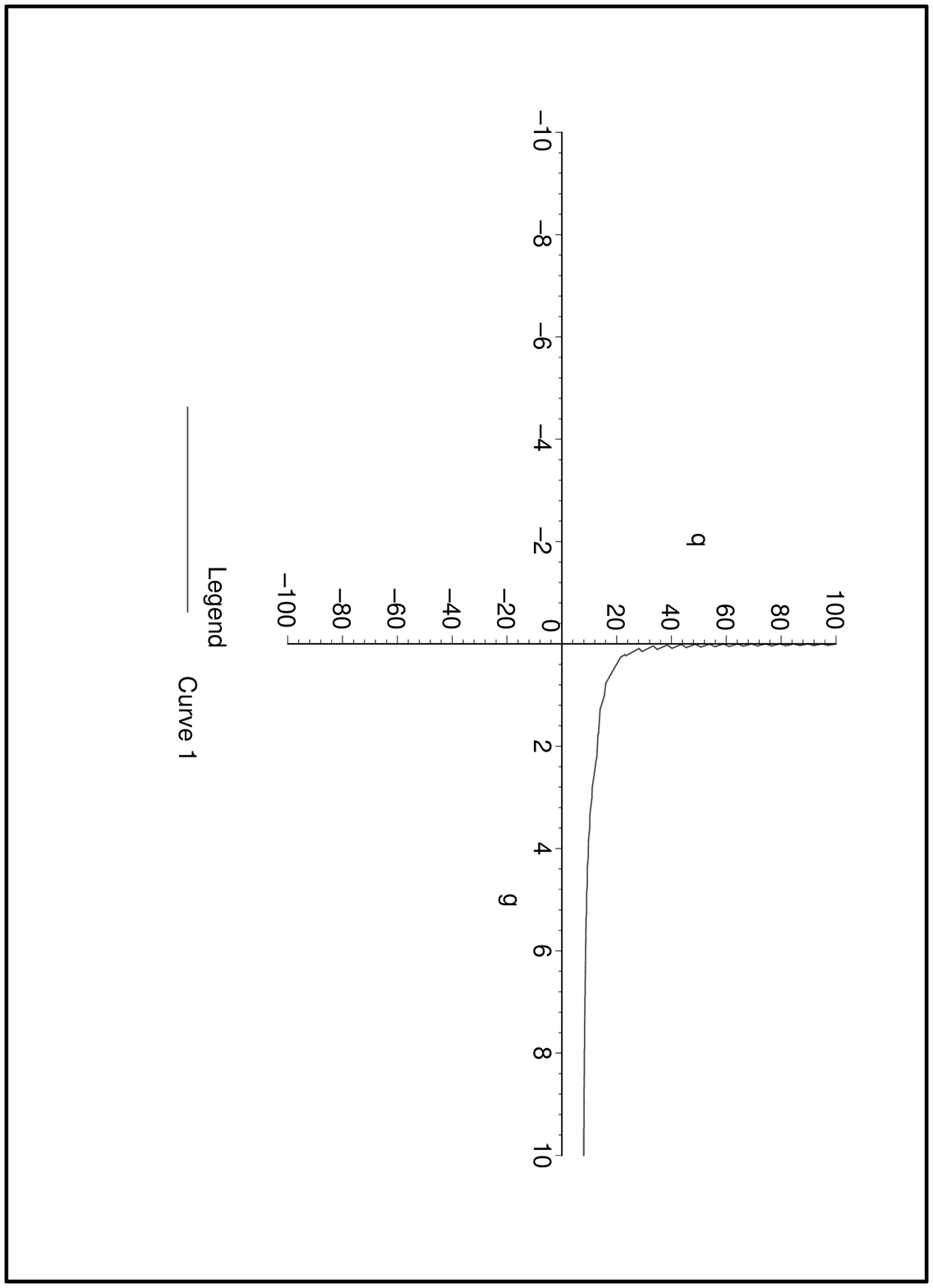}
\caption{The determinant of the metric tensor plotted as a
function of  $q$, describing the fluctuations in massless
non-rotating quarkonia in the raising regime.} \label{det2dPQr}
\vspace*{0.5cm}
\end{figure}

Let us explore the functional behavior of the associated scalar
curvature, in order to examine the mixing effects of the
parameters forming the intrinsic surface. In this case, there is
only one non-trivial component of the Riemann Christoffel tensor
$R_{qpqp}$. As per the definition of the Eqn.(\ref{eqn6}), our
computation shows that the scalar curvature reduce to the
following specific form \ba R(q,p) = \frac{b l(p)}{2p^2 (n^Q_g)^2}
(n^{(0)Q}_R+n^{(1)Q}_Rl(p) +n^{(2)Q}_Rl(p)^2+n^{(3)Q}_Rl(p)^3).
\ea The factors of the numerator of the scalar curvature take the
following expressions

\ba n^{(0)Q}_R&=& \ln(q/L)^4 -16 (q/L)^{2 p} p^4 \ln(q/L)-16
(q/L)^{2 p} p^7 \ln(q/L)^3\nn && -56 (q/L)^{2 p} p^6 \ln(q/L)^2
-48 (q/L)^{2 p} p^6 \ln(q/L)-4 (q/L)^{2 p} p^7 \ln(q/L)^4\nn &&-16
(q/L)^{2 p} p^7 \ln(q/L)^2 -16 (q/L)^{2 p} p^6 \ln(q/L)^3-64
(q/L)^{2 p} p^5 \ln(q/L)\nn &&-24 (q/L)^{2 p} p^5 \ln(q/L)^2-36
(q/L)^{2 p} p^5-4 (q/L)^{2 p} p^3, \nn  n^{(1)Q}_R&=& 28 p^6
(q/L)^{2 p} \ln(q/L)^2 +14 p^4 (q/L)^{2 p} +20 p^5 (q/L)^{2 p}
\ln(q/L)^2 \nn && +50 p^3 (q/L)^p+40 p^3 (q/L)^p \ln(q/L)+146  p^4
\ln(q/L) (q/L)^p \nn && +56 p^4 (q/L)^p \ln(q/L)^2+8 p^6 (q/L)^{2
p} \ln(q/L)^3+4 p^3 (q/L)^{2 p}\nn && +10 p^2 (q/L)^p+48 (q/L)^{2
p} p^5+102 (q/L)^p p^4 +112 p^5 (q/L)^p \ln(q/L)^2 \nn && +32
(q/L)^{2 p} p^6 \ln(q/L)+24 (q/L)^p p^6 \ln(q/L)^2 +32 p^5 (q/L)^p
\ln(q/L)^3 \nn && +104 (q/L)^p p^5 \ln(q/L)+24 (q/L)^p p^6
\ln(q/L)^3 +6 (q/L)^p p^6 \ln(q/L)^4\nn &&+16 p^4 (q/L)^{2 p}
\ln(q/L)+50 p^5 \ln(q/L) (q/L)^{2 p}, \nn n^{(2)Q}_R&=& -6 p^3
(q/L)^{2 p}-8 p^5 (q/L)^{2 p} \ln(q/L)-6 p^4 \ln(q/L) (q/L)^{2 p}
\nn && +6 p^4 (q/L)^{2 p}+10 p^3 (q/L)^p-14 p^2 (q/L)^p -20 p^2-12
p \nn &&-30 p^2 \ln(q/L)+8 (q/L)^p p^6 \ln(q/L)^2 l(p)^2-20 p^3
(q/L)^p \ln(q/L)\nn && -4 p^5 (q/L)^{2 p} \ln(q/L)^2 l(p)^2-90 p^3
\ln(q/L)-36 p^3 \ln(q/L)^2 \nn && -62 (q/L)^p p^4 l(p)^2-16
(q/L)^{2 p} p^5-16 p^4 \ln(q/L)^3 \nn && -8 p^5 \ln(q/L)^2-56 p^4
\ln(q/L)-56 p^4 \ln(q/L)^2 \nn && -8 p^5 \ln(q/L)^3-2 p^5
\ln(q/L)^4-8 (q/L)^p p^5 \ln(q/L)\nn && +8 (q/L)^p p^6 \ln(q/L)^3
+2 (q/L)^p p^6 \ln(q/L)^4-82 p^3 \nn &&-16 p^4 (q/L)^p
\ln(q/L)^2-34 p^4 \ln(q/L) (q/L)^p, \nn n^{(3)Q}_R&=& -15  p^3
\ln(q/L)+2  p^3 (q/L)^{2 p} -4  p^4 (q/L)^{2 p} -8 p^5 \ln(q/L)^2
(q/L)^p \nn &&-9 p^4 \ln(q/L) (q/L)^p-2 p^4 \ln(q/L)^2 (q/L)^p-2
p^5 \ln(q/L)^3 (q/L)^p\nn && + p^3 \ln(q/L) (q/L)^p -8 p^5
\ln(q/L) (q/L)^p-20 p^3 (q/L)^p+9 p^2 (q/L)^p\nn && -5  p^4
\ln(q/L)^3 -4 p^5 \ln(q/L)^2-16 p^4 \ln(q/L) -19 p^4 \ln(q/L)^2
\nn && -4 p^5 \ln(q/L)^3- p^5+16 (q/L)^{-p} p^2 -6 (q/L)^{-p} p +5
(q/L)^{-p} p^2 \ln(q/L)^2 \nn &&+(q/L)^{-p} p^3 \ln(q/L)^3 +10
(q/L)^{-p} p^2 \ln(q/L)+3 (q/L)^{-p} p^3 \ln(q/L)^2 \nn &&+6
(q/L)^{-p} p \ln(q/L) +4 (q/L)^{-p} p^3 \ln(q/L)+6 (q/L)^{-p}-28
p^2 \nn && -24 (q/L)^{2 p} p^4+7 p^3 -3  p^3 \ln(q/L)^2 +3 p^2
\ln(q/L) +12 p. \ea

\begin{figure}
\hspace*{0.5cm}
\includegraphics[width=8.0cm,angle=-90]{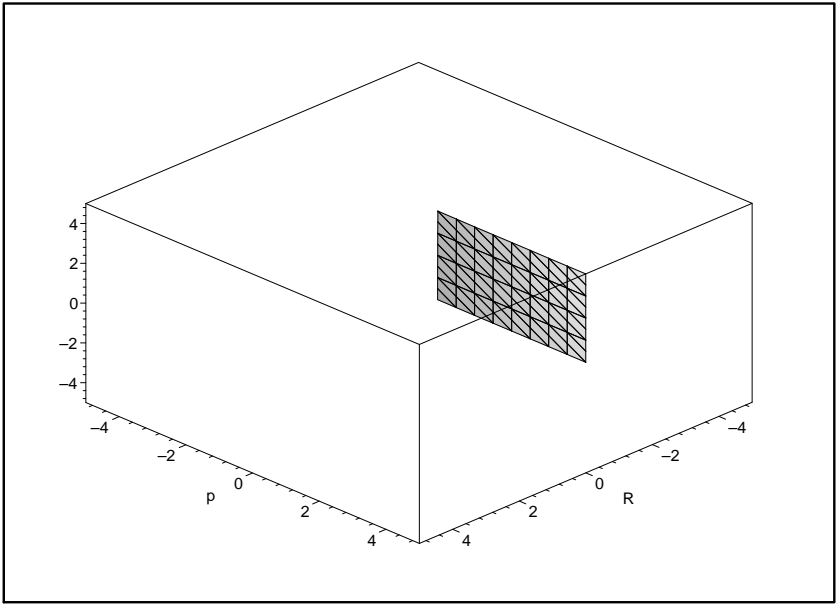}
\caption{The curvature scalar plotted as a function of $q$, and
the index parameter $p$, describing the fluctuations in massless
non-rotating quarkonia.} \label{cur3dPQ} \vspace*{0.5cm}
\end{figure}

\begin{figure}
\hspace*{0.5cm}
\includegraphics[width=8.0cm,angle=180]{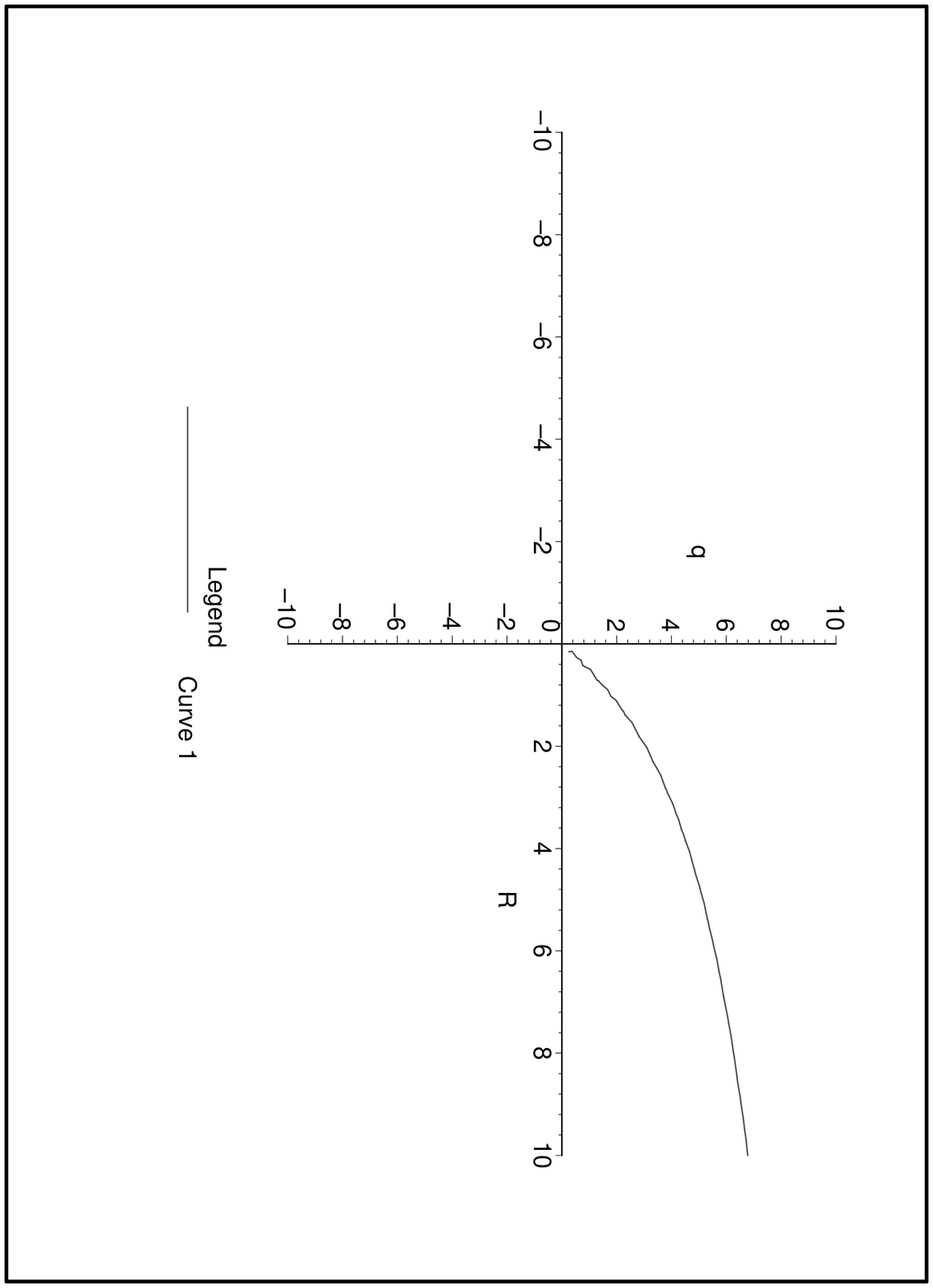}
\caption{The curvature scalar plotted as a function of $q$,
describing the fluctuations in massless non-rotating quarkonia
near the Coulambic regime.} \label{cur2dPQc} \vspace*{0.5cm}
\end{figure}

\begin{figure}
\hspace*{0.5cm}
\includegraphics[width=8.0cm,angle=180]{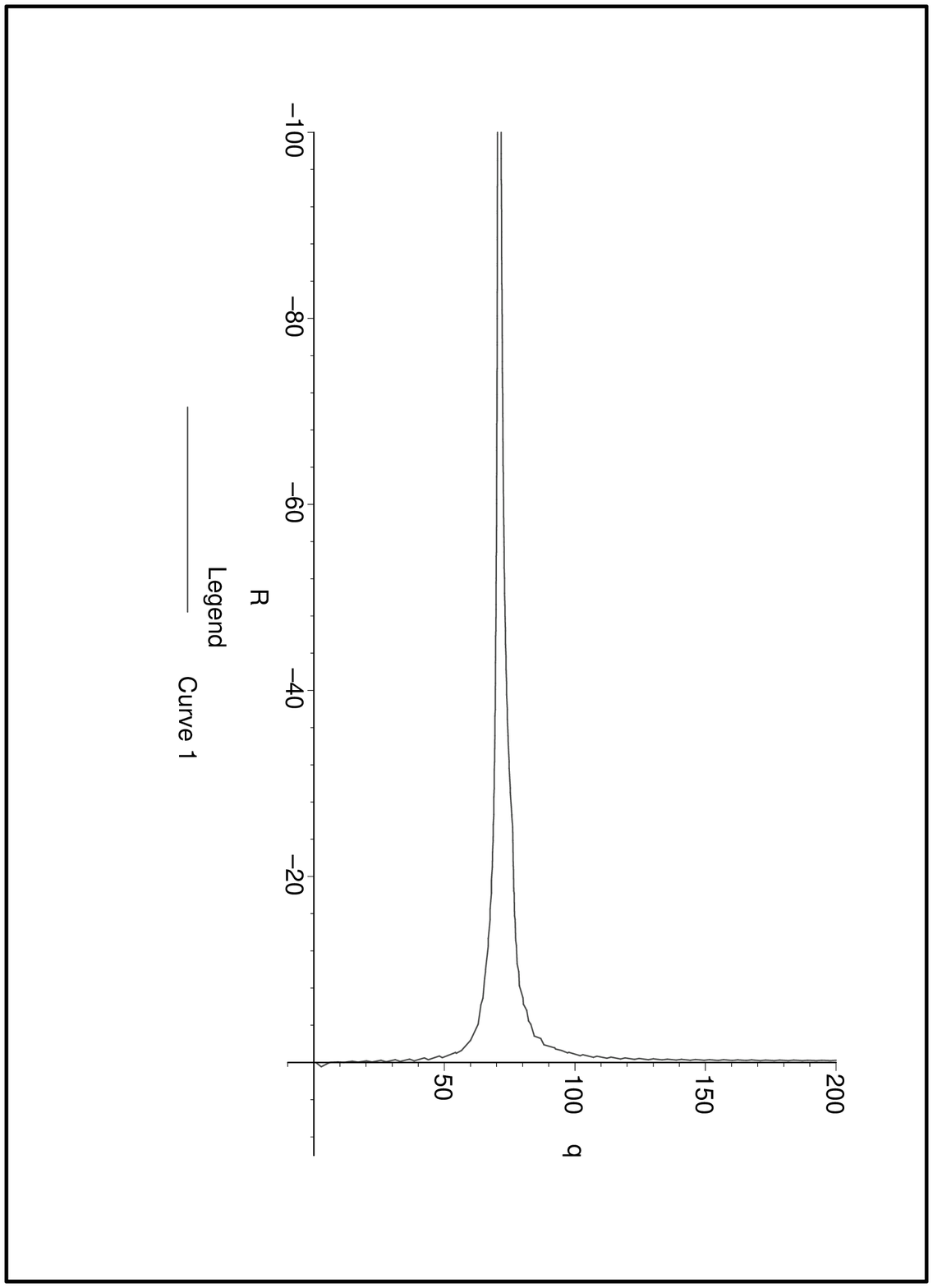}
\caption{The curvature scalar plotted as a function of $q$,
describing the fluctuations in massless non-rotating quarkonia in
the rising regime.} \label{cur2dPQr} \vspace*{0.5cm}
\end{figure}

In the case when the Ricci scalar curvature $R(q,p)$ vanishes, the
underlying quarkonium system is found to be in equilibrium. Such a
state of the configuration can arise with $\{ n^{(i)Q}=0, i=
0,1,2,3 \}$, if the other factors of the scalar curvature remain
non-zero. In the other case, when the $R(q,p)$ diverges, the
configuration goes over a transition. Such an extreme behavior of
the quarkonia is expected to happen, when either the index $p$ or
the the numerator of the determinant of the metric tensor $n^Q_g$
vanish. Geometrically, the intrinsic $qp$-surface becomes the flat
Euclidean plane in the first case, while it gets infinitely curved
in the second one.

For the choice of QCD parameters $L=1000$ and $b=1$, the
Fig.(\ref{det3dPQ}) shows the plot of the determinant of the
metric tensor. This plots explicates the nature of the stability
of non-rotating configurations in the limit of massless quarkonia.
The corresponding plot for the scalar curvature is depicted in the
Fig.(\ref{cur3dPQ}). This plot shows the global nature of the
non-rotating configuration in the limit of massless quarkonia,
under the effects of Gaussian fluctuations of the index $p$ and
the parameter $q$. Notice, for the generic massless non-rotating
quarkonium configurations, that the limiting scalar curvature
interestingly simplifies to the slice shape of
Fig.(\ref{det3dPQ}). Further, it is worth mentioning that the
corresponding peaks of the determinant of the metric tensor, as
depicted in the Fig.(\ref{det3dPQ}) indicate the graphical nature
of the instability present in the general non-rotating massless
quarkonia. Physically, the presence of peaks in the determinant of
the metric tensor shows a non-trivial interaction in the system.

For the regime of the Coulambic potential, the respective surface
plots of the determinant of the metric tensor and scalar curvature
are respectively shown in the Figs.(\ref{det2dPQc}) and
(\ref{cur2dPQc}) for $p=0.1$. We observe that the stability of
massless non-rotating quarkonia exists in certain bands. A closer
view shows that the limiting Coulambic quarkonia are decaying and
interacting particles. For the regime of the rising potential,
e.g., $p=0.51$, the respective surface plots of the determinant of
the metric tensor and scalar curvature are respectively shown in
the Figs.(\ref{det2dPQr}) and (\ref{cur2dPQr}), which shows that
the limiting rising potential quarkonia are stable and
non-interacting particles.  This follows from the fact that the
determinant instability is present only for a specific set $q$.

\subsection{Regge Rotating Quarkonia}
In the present subsection, we analyze the nature of massless
rotating quarkonia generated by the parameters $q, p$ and angular
momentum $J$. Following Eqn.(\ref{Regge}), we find that the
modified strong QCD coupling takes the form

\ba A(q,p,J)= \frac{1}{b} \frac{p}{\ln(1+p (q/L)^p)}+
\frac{1}{b_1} J (J+1). \ea

To focus on the intrinsic geometry of the present case, we chose
the parameters $q, p, J$ as the variables for the QCD coupling. As
in the previous subsection, we may again exploit the definition of
the Hessian function $Hess(A(q,p,J))$ of the QCD coupling.
Considering the analysis of the Regge model, we find that the
components of the metric tensor in the framework has the same
characterizations  for the $g_{qq}$, $g_{qp}$ and $g_{pp}$ as
obtained for the corresponding rotating massless quarkonia. The
component of the metric tensor for the rotation parameter turns
out to be $g_{JJ}= \frac{2}{b_1}$. While, the remaining components
of the metric tensor, involving $J$ and either the $q$ or $p$,
vanish identically.

It follows that the pure pair correlations $\{ g_{qq}, g_{pp},
_{JJ} \}$ between the parameters $\{q, p, J\}$ remain positive
same as in the case of the non-rotating quarkonia. Further, our
computation demonstrates the over-all nature of the parametric
fluctuations. In fact, we find that the determinant of the metric
tensor reduces to the following simple expression

\ba \Vert g \Vert = \frac{2p^3}{b^2b_1q^2 l(p)^5 \exp{(3 l(p))}}
n^Q_g, \ea

where the $n^Q_g(q,p)$ remains the same as for the non-rotating
case. It is worth mentioning that the Regge rotating massless
quarkonia is well-behaved, as long as the corresponding rotating
massless quarkonia remains so. Over the domain of the parameters
$\{ q,p,J \}$, we thus observe that the Gaussian fluctuations have
the same set of thermodynamic metric stability structures, as long
as $b_1>0$. The observation of the metric structure shows that the
Regge rotating massless quarkonia is stable on the $qp$-surface if
the index parameter $p$ and the function $n^Q_g(q,p)$, appearing
in the numerator of the determinant of the metric tensor have the
same sign.

Furthermore, we may easily analyze the underlying important
conclusions for the specific considerations of Regge rotating
massless quarkonia. As in the case of the non-rotating massless
case, the global nature of the scalar curvature and associated
phase transitions may thus be determined over the range of
parameters describing the Regge rotating quarkonia of interest.
Since the previous non-rotating configuration is trivially
embedded in the Regge rotating configuration, we thus find that
the thermodynamic scalar curvature remains exactly the same as for
the non-rotating massless quarkonia. The equality of the two
scalar curvatures shows that the Regge rotation keeps the same
global thermodynamic stability structures as the non-rotating
massless counterpart. In the Regge trajectory model, we thus find
an interesting conclusion that all possible local and global
thermodynamic stability behavior of the massless quarkonia remains
the same up to the sign of $b_1$, as if there were no rotation in
the underlying configuration.

The fact that the efficiency of the rotation induces a mass to the
quarkonium is analyzed by considering the Bloch-Nordsieck
resummation of the angular phases. As per the analysis of the next
subsections, our method offers a non-linear generic
characterization for the quarkonium configurations, which we
consider to be neither purely massless nor purely rotating, but a
strongly coupled QCD.

\section{Massive Quarkonia}

In this section, we extend the intrinsic geometric analysis for
realistic quarkonia and compute the associated local and global
thermodynamic quantities in the subsequent subsections. Following
the notations of the previous section, a given generic quarkonia
can analytically be easily analyzed about some local equilibrium,
if we can fix one of the parameter. The logic simply follows from
the fact that the underlying configurations reduce to an intrinsic
surface, and thus are easy to compute. Such an assumption is
allowed for the specific phases of the QCD, whether one considers
the Coulomb phase or the rising phase. Our analysis certainly does
not stop here, but we indeed extend it for the general quarkonia
with $\{q,p,J\}$ all fluctuating. Let us firstly illustrate the
cases for the two parameter configurations with either $\{q,J\}$
or $\{q,m\}$ fluctuating and then systematically for the generic
quarkonia.

\subsection{Quarkonia in QJ-plane}

In this subsection, we shall use the essential features of
thermodynamic geometry, thus developed, to describe the quarkonia
using modified strong QCD coupling, with an increasing number of
parameters. Let us focus our attention on the geometric nature of
the local and global correlations in the neighbourhood of small
fluctuations, in the chosen quarkonium configurations. As per the
consideration of massive quarkonia, the associated resummed strong
QCD coupling is given by

\ba A(q,J) := \frac{1}{b} \frac{p}{\ln(1+p (q/L)^p)}
\ln(\frac{\sqrt{J}+\sqrt{J-q}}{\sqrt{J}-\sqrt{J-q}}) (1-J(0,a
\sqrt{q})), \ea

where $J(\nu,x)$ is the Bessel function of the first kind of the
order $\nu$. As stated earlier, the thermodynamic metric in the
parameter space is given by the Hessian matrix $Hess(A(q,J))$ of
the strong QCD coupling with respect to the variables defining the
thermodynamic manifolds. In order to simplify the subsequent
expressions, let us define the logarithmic factor of the concerned
Bloch-Nordsieck rotation as

\ba f(q,J):=
\ln(\frac{\sqrt{J}+\sqrt{J-q}}{\sqrt{J}-\sqrt{J-q}}).\ea

In this framework, it turns out that the fluctuation nature of
parametric pair correlations may be easily divulged in terms of
the momentum transfer and rotation parameter of the underlying
quarkonia. Following Eqn.(\ref{eq4}), we find, under the Gaussian
fluctuations of $\{q,J\}$, that the components of the metric
tensor are

\ba g_{qq}&=&-\frac{p}{4 b l(p)^3 \exp{(2l(p))}q^{5/2}(J-q)^{3/2}}
( n^{(0)J}_{11}+ n^{(1)J}_{11} l(p)+ n^{(2)J}_{11} l(p)^2), \nn
g_{qp} &=& \frac{p}{2 b l(p)^2 \exp{(l(p))}q^{3/2}\sqrt{J}
(J-q)^{3/2}} ( n^{(0)J}_{12}+ n^{(1)J}_{12} l(p)), \nn g_{pp}&=&-
\frac{p}{2 b l(p)J^{3/2} (J-q)^{3/2}} (2J-q) (1- J(0,a \sqrt{q})).
\ea

\begin{figure}
\hspace*{0.5cm}
\includegraphics[width=8.0cm,angle=-90]{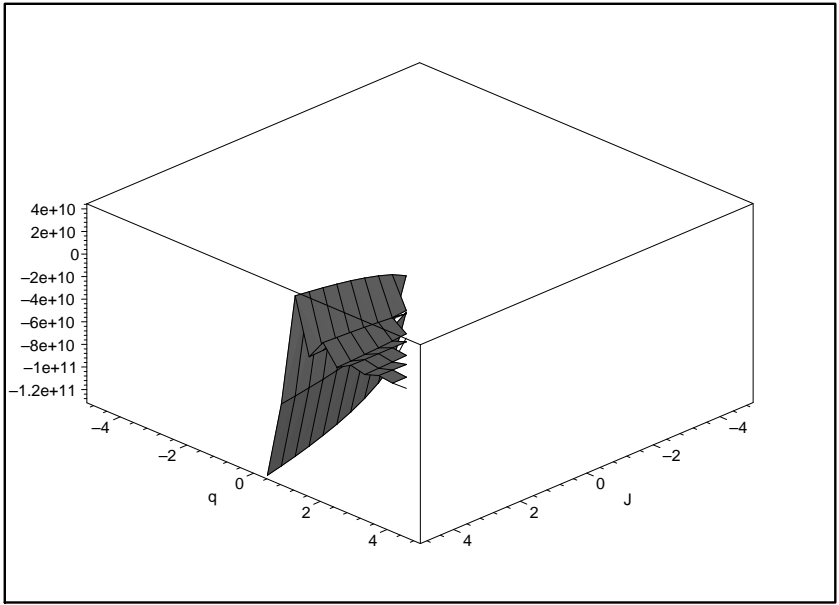}
\caption{The $qq$-component of the metric tensor plotted as a
function of $q$ and the angular momentum $J$, describing the heat
capacity of the charge in massive rotating quarkonia.}
\label{corrQQ3dQJ} \vspace*{0.5cm}
\end{figure}

\begin{figure}
\hspace*{0.5cm}
\includegraphics[width=8.0cm,angle=-90]{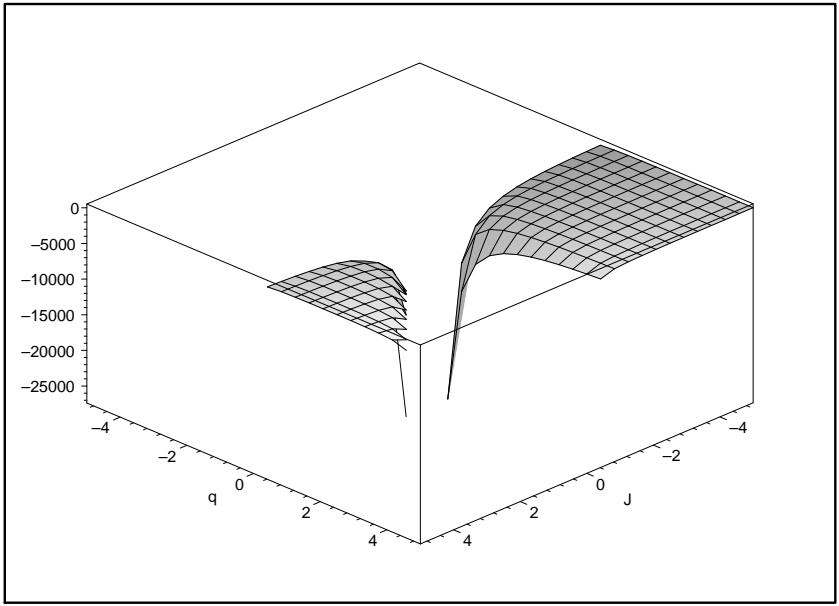}
\caption{The $JJ$-component  of the metric tensor plotted as a
function of $q$ and the angular momentum $J$, describing the heat
capacity of the angular momentum in massive rotating quarkonia.}
\label{corrJJ3dQJ} \vspace*{0.5cm}
\end{figure}

As a result, we find without any approximation that the factors in
the numerator of the local pair correlation associated with the
$qq$-components are expressed as

\ba n^{(0)J}_{11}&=& 8 p^4 f ((q/L)^p)^2 \sqrt{J-q} (q^{3/2} -
q^{3/2} J(0,a \sqrt{q})- \sqrt{q} J + \sqrt{q} J(0,a \sqrt{q})J),
\nn n^{(1)J}_{11}&=& -4 p^2 f \sqrt{J-q} (q/L)^p   (J(1,a
\sqrt{q}) a q^2 +  q^{3/2} - q^{(3/2} J(0,a \sqrt{q}) +  J(1,a
\sqrt{q}) a  q J \nn && - \sqrt{q} J + \sqrt{q} J(0,a \sqrt{q}) J)
+8 p^2 (q/L)^p ( \sqrt{J} q^{3/2} + J^{3/2} \sqrt{q} J(0,a
\sqrt{q}) \nn && - \sqrt{J} q^{3/2} J(0,a \sqrt{q}) - J^{3/2}
\sqrt{q}) +4 p^3 f \sqrt{J-q} (q/L)^p ( \sqrt{q} J- \sqrt{q} J(0,a
\sqrt{q}) J \nn && + q^{3/2} J(0,a \sqrt{q})- q^{3/2}) +4 p^3 f
\sqrt{J-q} ((q/L)^p)^2 ( q^{3/2} + \sqrt{q} J(0,a \sqrt{q}) J  \nn
&& - q^{3/2} J(0,a \sqrt{q}) + J(1,a \sqrt{q}) a q J - \sqrt{q} J
- J(1,a \sqrt{q}) a  q^2)  \nn && +8 p^3 ((q/L)^p)^2 (\sqrt{J}
q^{3/2} - \sqrt{J} q^{3/2} J(0,a \sqrt{q}) + J^{3/2} \sqrt{q}
J(0,a \sqrt{q})- J^{3/2} \sqrt{q}), \nn n^{(2)J}_{11}&=& 2
\sqrt{J} (6 q^{3/2} p (q/L)^p + 3q^{3/2}-4 J^{3/2} \sqrt{q} -3
q^{3/2} J(0,a \sqrt{q})-2 J(1,a \sqrt{q}) a q^2)  \nn && + f
\sqrt{J-q} (J(0,a \sqrt{q}) a^2 q^{5/2} -  J(0,a \sqrt{q}) a^2
q^{3/2} J +2 J(1,a \sqrt{q}) a q J  \nn && -2 J(1,a \sqrt{q}) a
q^2) +4 J^{3/2} (J(1,a \sqrt{q}) a q + \sqrt{q} J(0,a \sqrt{q}))
\nn && +2 p (q/L)^p (f a^2 q^{5/2} \sqrt{J-q} J(0,a \sqrt{q})- f
a^2 q^{3/2} \sqrt{J-q} J(0,a \sqrt{q}) J  \nn && -2  f J(1,a
\sqrt{q}) a q^2 \sqrt{J-q} +2  f J(1,a \sqrt{q}) a q \sqrt{J-q} J
+4 J^{3/2} J(1,a \sqrt{q}) a q \nn &&  -4 J^{3/2} \sqrt{q} +4
J^{3/2} \sqrt{q} J(0,a \sqrt{q}) -4 \sqrt{J} J(1,a \sqrt{q}) a q^2
-6 \sqrt{J} q^{3/2} J(0,a \sqrt{q}))  \nn && +p^2 ((q/L)^p)^2 ( f
a^2 q^{5/2} \sqrt{J-q} J(0,a \sqrt{q}) - f a^2 q^{3/2} \sqrt{J-q}
J(0,a \sqrt{q}) J  \nn && +2  f J(1,a \sqrt{q}) a q \sqrt{J-q} J
-2  f J(1,a \sqrt{q}) a q^2 \sqrt{J-q} -4  \sqrt{J} J(1,a
\sqrt{q}) a q^2  \nn && +4 J^{3/2} J(1,a \sqrt{q}) a q +4  J^{3/2}
\sqrt{q} J(0,a \sqrt{q}) -4  J^{3/2} \sqrt{q} +6 \sqrt{J} q^{3/2}
\nn && -6  \sqrt{J} q^{3/2} J(0,a \sqrt{q})). \ea

While, the factors in the numerator of the associated
$qJ$-component are given by

\ba n^{(0)J}_{12}&=& 2 p^2 (q/L)^p (q^{3/2}- J^{3/2}+J(0,a
\sqrt{J})-J(0,a \sqrt{J})), \nn n^{(1)J}_{12}&=& q^{3/2}+ a q J
J(1,a \sqrt{J})- a q^2 J(1,a \sqrt{J})- q^{3/2} J(0,a \sqrt{J})
\nn && + (q/L)^p p (q^{3/2}+ a q J J(1,a \sqrt{J})- a q^2 J(1,a
\sqrt{J}) - q^{3/2} J(0,a \sqrt{J})).\ea

Herewith, we see that the geometric nature of parametric pair
correlations turns out to be remarkably interesting. The
fluctuating quarkonia may be easily described in terms of the $q$
and $J$. For the configurations with the same sign of the index
$p$ and constant $b$, it is evident that the principle components
of the metric tensor, signifying self pair correlations, are
positive definite functions in a non-trivial range $q$. The local
stability requires that (i) $qq$- fluctuations satisfy the
constraint \ba n^{(0)J}_{11}+ n^{(1)J}_{11}l(p)+
n^{(2)J}_{11}l(p)^2 <0 \ea and (ii) $JJ$- fluctuations be
constrained to the following limiting values of the Bessel
function \ba J(0,a \sqrt{q})&<& \ 1,\ \ q>2J, \nn &>& \ 1, \ \
q<2J. \ea

A straightforward computation demonstrates the over-all nature of
the parametric fluctuations. In this case, we find that the
determinant of the metric tensor reduces to the following
expression

\ba g = -\frac{p^2}{8 b^2 l(p)^4 \exp{(2 l(p))}
q^{5/2}J^{3/2}(J-q)^{3/2}}(n^{(0)J}_{g}+ n^{(1)J}_{g}l(p)+
n^{(2)J}_{g}l(p)^2), \ea

where the coefficients $\lbrace n^{(1)J}_{g}, n^{(2)J}_{g} \rbrace
$ appearing in the determinant of the metric tensor factorize as

\ba n^{(1)J}_{g}&=&4 p^2 (q/L)^p (n^{(12)J}_{g}+
n^{(13)J}_{g}p(q/L)^p), \nn n^{(2)J}_{g}&=& n^{(20)J}_{g}+ 2
n^{(21)J}_{g} p (q/L)^p+ n^{(23)J}_{g}((q/L)^p)^2 p^2. \ea

Explicitly, we find that all factors of the numerator of the
determinant can be presented as

\ba n^{(0)J}_{g}&=& 8 p^4 ((q/L)^p)^2 ( \sqrt{J} \sqrt{q}
\sqrt{J-q} +\sqrt{J} \sqrt{q} \sqrt{J-q} J(0,a \sqrt{q})^2 \nn &&
-f q^{3/2} J(0,a \sqrt{q})^2 -f q^{3/2} +2J f \sqrt{q} +2f q^{3/2}
J(0,a \sqrt{q}) \nn && -2\sqrt{J} \sqrt{q} \sqrt{J-q} J(0,a
\sqrt{q}) +2J f \sqrt{q} J(0,a \sqrt{q})^2 -4J f \sqrt{q} J(0,a
\sqrt{q})), \nn
n^{(12)J}_{g}&=& (f J(1,a \sqrt{q}) a  q^2 -  f q^{3/2} - f J(1,a
\sqrt{q}) a q^2 J(0,a \sqrt{q}) -  f q^{3/2} J(0,a \sqrt{q})^2 \nn
&& +2 f q^{3/2} J(0,a \sqrt{q}) +2 \sqrt{J} q J(1,a \sqrt{q}) a
\sqrt{J-q} J(0,a \sqrt{q})\nn && +2  J f \sqrt{q} +2 J f J(1,a
\sqrt{q}) a q J(0,a \sqrt{q}) +2 J f \sqrt{q} J(0,a \sqrt{q})^2
\nn && -2 J f J(1,a \sqrt{q}) a q -2 \sqrt{J} q J(1,a \sqrt{q}) a
\sqrt{J-q} +4 \sqrt{J} \sqrt{q} \sqrt{J-q} \nn && +4 \sqrt{J}
\sqrt{q} \sqrt{J-q} J(0,a \sqrt{q})^2 -4 J f \sqrt{q} J(0,a
\sqrt{q}) -8 \sqrt{J} \sqrt{q} \sqrt{J-q} J(0,a \sqrt{q})), \nn
n^{(13)J}_{g}&=& f (q^{3/2} + q^{3/2} J(0,a \sqrt{q})^2 -2 J
\sqrt{q} J(0,a \sqrt{q})^2 -2 J \sqrt{q} \nn &&-2 q^{3/2} J(0,a
\sqrt{q}) +4 J \sqrt{q} J(0,a \sqrt{q}))+ ((q/L)^p) (f J(1,a
\sqrt{q}) a q^2 \nn && -  f q^{3/2} J(0,a \sqrt{q})^2- f J(1,a
\sqrt{q}) a q^2 J(0,a \sqrt{q})-  f q^{3/2} \nn && +2 f J \sqrt{q}
-2 \sqrt{J} q J(1,a \sqrt{q}) a \sqrt{J-q} + 2f q^{3/2} J(0,a
\sqrt{q}) \nn && + 2   f J J(1,a \sqrt{q}) a q J(0,a \sqrt{q}) - 2
f J J(1,a \sqrt{q}) a q + 2   f J \sqrt{q} J(0,a \sqrt{q})^2 \nn
&& + 2 \sqrt{J} q J(1,a \sqrt{q}) a \sqrt{J-q} J(0,a \sqrt{q}) +4
\sqrt{J} \sqrt{q} \sqrt{J-q} J(0,a \sqrt{q})^2 \nn && +4 \sqrt{J}
\sqrt{q} \sqrt{J-q} -4 J f \sqrt{q} J(0,a \sqrt{q}) -8 \sqrt{J}
\sqrt{q} \sqrt{J-q} J(0,a \sqrt{q})) , \nn
n^{(20)J}_{g}&=& 2 \sqrt{J} (q^{3/2} J(1,a \sqrt{q})^2 a^2
\sqrt{J-q} +4 \sqrt{q} \sqrt{J-q} +4 \sqrt{q} \sqrt{J-q} J(0,a
\sqrt{q})^2 \nn && +4 J(1,a \sqrt{q}) a q \sqrt{J-q} J(0,a
\sqrt{q}) -4 J(1,a \sqrt{q}) a q \sqrt{J-q} \nn && -8 \sqrt{q}
\sqrt{J-q} J(0,a \sqrt{q}) ) +f (a^2 q^{5/2} J(0,a \sqrt{q})^2 -
a^2 q^{5/2} J(0,a \sqrt{q}) \nn && +2 a q^2 J(1,a \sqrt{q}) -2
J(1,a \sqrt{q}) a q^2 J(0,a \sqrt{q}) +2 J a^2 q^{3/2} J(0,a
\sqrt{q})\nn && -2 J a^2 q^{3/2} J(0,a \sqrt{q})^2 -4 J a q J(1,a
\sqrt{q})+4 J J(1,a \sqrt{q}) a q J(0,a \sqrt{q})), \nn
n^{(21)J}_{g}&=& (f a^2 q^{5/2} J(0,a \sqrt{q})^2 - f a^2 q^{5/2}
J(0,a \sqrt{q}) -2 f J(1,a \sqrt{q}) a q^2 J(0,a \sqrt{q}) \nn &&
+2 f a q^2 J(1,a \sqrt{q}) +2 \sqrt{J} q^{3/2} J(1,a \sqrt{q})^2
a^2 \sqrt{J-q} +2 J f a^2 q^{3/2} J(0,a \sqrt{q})\nn && -2 J f a^2
q^{3/2} J(0,a \sqrt{q})^2 +4 J f J(1,a \sqrt{q}) a q J(0,a
\sqrt{q}) -4 J f a q J(1,a \sqrt{q}) \nn && +8  \sqrt{J} \sqrt{q}
\sqrt{J-q} J(0,a \sqrt{q})^2 -8  \sqrt{J} J(1,a \sqrt{q}) a q
\sqrt{J-q}+8 \sqrt{J} \sqrt{q} \sqrt{J-q} \nn &&  +8 \sqrt{J}
J(1,a \sqrt{q}) a q \sqrt{J-q} J(0,a \sqrt{q}) -16 \sqrt{J}
\sqrt{q} \sqrt{J-q} J(0,a \sqrt{q}) ), \nn
n^{(23)J}_{g}&=& f a^2 q^{5/2} J(0,a \sqrt{q})^2 - f a^2 q^{5/2}
J(0,a \sqrt{q}) +2 f J(1,a \sqrt{q}) a q^2 \nn && -2 f a q^2 J(1,a
\sqrt{q}) J(0,a \sqrt{q}) +2 \sqrt{J} q^{3/2} J(1,a \sqrt{q})^2
a^2 \sqrt{J-q}\nn && +2  J f a^2 q^{3/2} J(0,a \sqrt{q}) -2 J f
a^2 q^{3/2} J(0,a \sqrt{q})^2 \nn && +4  J f a q J(1,a \sqrt{q})
J(0,a \sqrt{q}) -4 J f a q J(1,a \sqrt{q}) \nn && -8 \sqrt{J}
J(1,a \sqrt{q}) a q \sqrt{J-q} +8 \sqrt{J} \sqrt{q} \sqrt{J-q}
J(0,a \sqrt{q})^2 \nn && +8 \sqrt{J} J(1,a \sqrt{q}) a q
\sqrt{J-q} J(0,a \sqrt{q}) +8 \sqrt{J} \sqrt{q} \sqrt{J-q} \nn &&
-16 \sqrt{J} \sqrt{q} \sqrt{J-q} J(0,a \sqrt{q}). \ea

\begin{figure}
\hspace*{0.5cm}
\includegraphics[width=8.0cm,angle=-90]{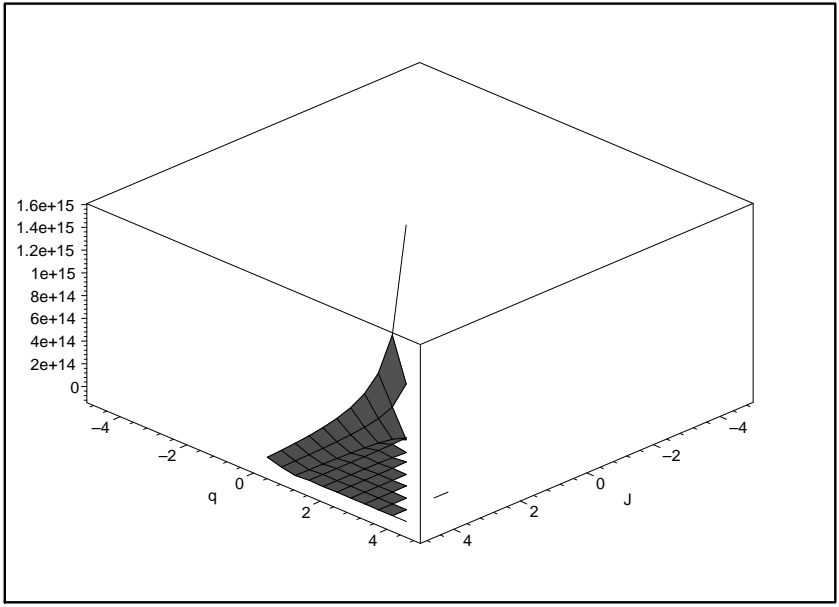}
\caption{The determinant of the metric tensor plotted as a
function of $q$ and the angular momentum $J$, describing the
fluctuations in massive rotating quarkonia.} \label{det3dQJ}
\vspace*{0.5cm}
\end{figure}

From Figs.(\ref{corrQQ3dQJ}) and (\ref{corrJJ3dQJ}), we see how
the pure components of the metric tensor behave, when they are
plotted against $q$ and the angular momentum. For the choice
$L=22500$, $p= 5/6$, $b= 1$ and $a=100000$, the $qq$-heat capacity
as depicted in the Fig.(\ref{corrQQ3dQJ}) shows a large variation
of the minima and maxima, which respectively take in an order of
the amplitude $+10^{10}$ and $-10^{11}$ and occur for $q,J \in
(0,4)$. In this case, we further observe for all $J \in (0,4)$
that the underlying quarkonia become highly unstable as $q$
positively approaches the origin.

On the other hand, the $JJ$-heat capacity in
Fig.(\ref{corrJJ3dQJ}) shows only negative variation of order
$-10^{5}$, which occurs when $q,J$ tend to the origin. In this
case, we notice that all local interactions are present in the
region $q \in (0,4)$ and $J \in (-4,4)$. The strength of these
local interactions depends on the domain chosen in the $(q,J)$
space. These fluctuations signify the thermodynamical interactions
present in the strongly coupled quarkonia.

The corresponding plot for the determinant of the metric tensor is
depicted in Fig.(\ref{det3dQJ}). For the above choice of the
parameters, the Fig.(\ref{det3dQJ}) shows the graphical nature of
the determinant of the metric tensor. These plots explicate the
thermodynamically (un)stability regions for the underlying massive
quarkonia lying in a given QCD phase. Combining the effects of all
fluctuations of the $\{ q,J \}$, we observe that the quarkonia are
stable for $q:= Q^2 \in (1,4)$. In general, the global stability
requires that the determinant of the metric tensor must be
positive definite, which in the present case transform as \ba
n^{J}_g:= n^{(0)J}_{g}+ n^{(1)J}_{g}l(p)+ n^{(2)J}_{g}l(p)^2 &<& \
0. \ea

In this case, it turns out that the thermodynamic curvature may be
written as the series of the charmonium logarithmic factor $l(p)$
and Bloch-Nordsieck logarithmic factor $f(q,J)$ of rotation as the
coefficient of the expansion. Systematically, the exact expression
for the scalar curvature takes the form \ba R(q,J) = \frac{b
l(p)}{2p^2 (n^{J}_g)^2} \sum_n B_n \times (l(p))^n, \ea where the
$B_n$ in the numerator of the scalar curvature are polynomials in
$p$, whose coefficients are the functions of the Bloch-Nordsieck
logarithmic factor $f(q,J)$. While, the denominator of the scalar
curvature precisely takes the numerator of the determinant of the
metric tensor as its square. The quantitative properties of the
scalar curvature and the Riemann curvature tensor remain similar,
as we shall discuss in the next subsection.

\subsection{Quarkonia in QM-plane}
In the present section, we analyze the nature of the fluctuating
quarkonia generated by the momentum $Q$, and mass $m$. To focus on
the general case, we choose the variable $Q$ as the transverse
momentum with the understanding that $k=k_{\bot}$ and the mass
considered as an arbitrary real parameter of the system. Following
the convention of Bloch-Nordsieck resummation, the strong QCD
coupling can be expressed as

\ba A(k,m)=  \frac{1}{b} \frac{p}{\ln(1+p (k^2/L)^p)}
\ln(\frac{m+\sqrt{m^2-k^2}}{m-\sqrt{m^2-k^2}}) (1-J(0,a k)). \ea

When the Bloch-Nordsieck resummed strong QCD coupling $ A(k,m)$ is
allowed to fluctuate as a function of the $\{k,m\}$, we may again
exploit the definition of the Hessian function $Hess( A(k,m))$.
Herewith, we find that the components of the metric tensor are
given by

\ba g_{kk}&=& \frac{p}{b l(p)^3 \exp{(2(l(p))} k^6(m^2-k^2)^{3/2}}
( n^{(0)M}_{11}+ n^{(1)M}_{11} l(p)+ n^{(2)M}_{11} l(p)^2), \nn
g_{km}&=&  \frac{2 p}{b l(p)^2 \exp{(l(p))} k (m^2-k^2)^{3/2}}
(n^{(0)M}_{12}+ n^{(1)M}_{12} l(p)), \nn g_{km}&=&  \frac{2m p}{b
l(p) (m^2-k^2)^{3/2}} (-1+ J(0,ak)),\ea

where the coefficients $\lbrace n^{(1)M}_{g}, n^{(2)M}_{g} \rbrace
$ appearing in the determinant of the metric tensor factorize as

\ba n^{(1)M}_{11}&=& 2(k^2/L)^p p^2 (n^{(12)M}_{11}+
2n^{(13)M}_{11}p(k^2/L)^p), \nn n^{(2)M}_{11}&=& n^{(20)M}_{11}+
2p n^{(21)M}_{11} (k^2/L)^p + p^2 n^{(22)M}_{11} (k^2/L)^{2p}. \ea

\begin{figure}
\hspace*{0.5cm}
\includegraphics[width=8.0cm,angle=-90]{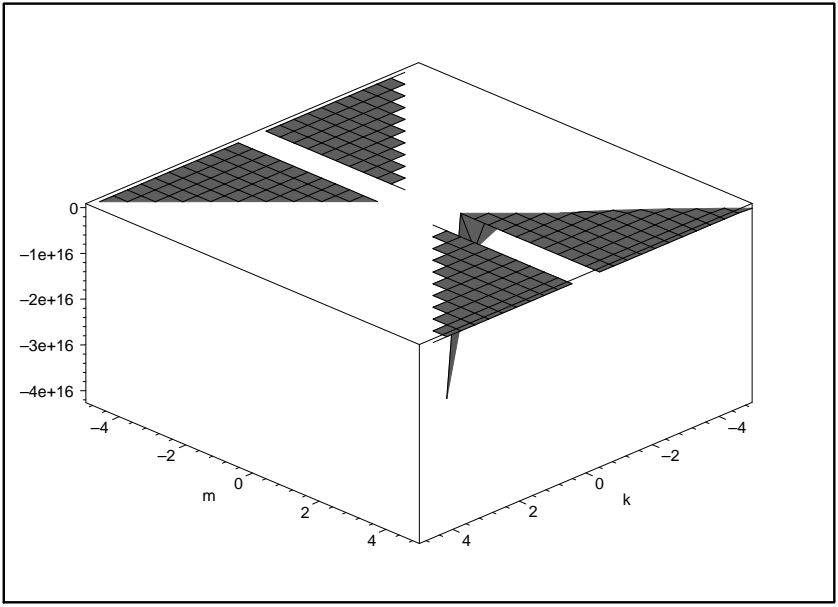}
\caption{The $kk$-component of the metric tensor plotted as a
function of the transverse momentum and mass $k, m$, describing
the heat capacity of transverse momentum for the massive rotating
quarkonia.} \label{corrKK3dkm} \vspace*{0.5cm}
\end{figure}

As it might be expected, we find without any approximation that
the factors in the numerator of the local pair correlation
associated with the $kk$-components can, as before, be easily
expressed as

\ba n^{(0)M}_{11}&=&8 p^4 f ((k^2/L)^p)^2 \sqrt{m^2-k^2} (m^2 -k^2
+ J(0,a k) k^2 -J(0,a k) m^2), \nn
n^{(12)M}_{11}&=& f \sqrt{m^2-k^2} J(0,a k) k^2 - f \sqrt{m^2-k^2}
J(0,a k) m^2 +4 m^3 -4 m^3 J(0,a k) \nn && +4 m k^2 J(0,a k)+ f
\sqrt{m^2-k^2} m^2 - f \sqrt{m^2-k^2} k^2 -4 m k^2  \nn && -2 f
J(1,a k) a k \sqrt{m^2-k^2} m^2 +2 f J(1,a k) a k^3
\sqrt{m^2-k^2}, \nn
n^{(13)M}_{11}&=& f \sqrt{m^2-k^2} k^2 + f ((k^2/L)^p)^2
\sqrt{m^2-k^2} m^2 - f ((k^2/L)^p)^2 \sqrt{m^2-k^2} k^2  \nn && -
f \sqrt{m^2-k^2} J(0,a k) k^2 - f \sqrt{m^2-k^2} m^2  + f
\sqrt{m^2-k^2} J(0,a k) m^2   \nn && - f ((k^2/L)^p)^2
\sqrt{m^2-k^2} J(0,a k) m^2 + f ((k^2/L)^p)^2 \sqrt{m^2-k^2} J(0,a
k) k^2  \nn && -2 f J(1,a k) a ((k^2/L)^p)^2 k \sqrt{m^2-k^2} m^2
 +2 f J(1,a k) a ((k^2/L)^p)^2 k^3 \sqrt{m^2-k^2}  \nn && +4 m k^2
((k^2/L)^p)^2 J(0,a k) -4 m^3 ((k^2/L)^p)^2 J(0,a k) +4 m^3
((k^2/L)^p)^2  \nn && -4 m k^2 ((k^2/L)^p)^2, \nn
n^{(20)M}_{11}&=& f a k^3 \sqrt{m^2-k^2} J(1,a k) -f a k
\sqrt{m^2-k^2} J(1,a k) m^2 +f a^2 k^2 \sqrt{m^2-k^2} J(0,a k) m^2
 \nn && -f a^2 k^4 \sqrt{m^2-k^2} J(0,a k) +2 m^3 +4 m k^2 J(0,a k) -4 m^3
k J(1,a k) a -4 m k^2 \nn && -2 m^3 l(p)^2 J(0,a k) +4 m k^3 J(1,a
k) a, \nn
n^{(21)M}_{11}&=& f a k^3 \sqrt{m^2-k^2} J(1,a k) - f a^2 k^4
\sqrt{m^2-k^2} J(0,a k) - f a k \sqrt{m^2-k^2} J(1,a k) m^2  \nn
&& + f a^2 k^2 \sqrt{m^2-k^2} J(0,a k) m^2 -2 m^3 J(0,a k) +2 m^3
-4 m^3 k J(1,a k) a p -4 m k^2  \nn && +4 m k^3 J(1,a k) a +4 m
k^2 J(0,a k), \nn
n^{(22)M}_{11}&=& f a  k^3 \sqrt{m^2-k^2} J(1,a k) -f a^2 k^4
\sqrt{m^2-k^2} J(0,a k) +f a^2 k^2 \sqrt{m^2-k^2} J(0,a k) m^2 \nn
&& -f a k \sqrt{m^2-k^2} J(1,a k) m^2 +2 m^3 -2 m^3 J(0,a k) -4
m^3 k J(1,a k) a  \nn && -4 m k^2 +4 m k^2 J(0,a k) +4 m k^3 J(1,a
k) a. \ea

\begin{figure}
\hspace*{0.5cm}
\includegraphics[width=8.0cm,angle=-90]{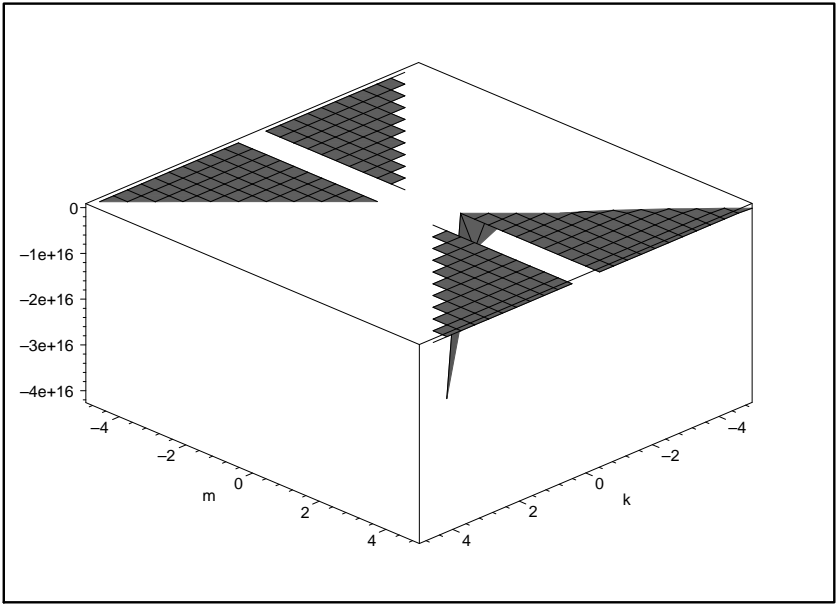}
\caption{The $mm$-component  of the metric tensor plotted as a
function of the transverse momentum and mass $k, m$, describing
the heat capacity of mass for the massive rotating quarkonia.}
\label{corrMM3dkm} \vspace*{0.5cm}
\end{figure}

In the present case, the factors of the numerator of the
$km$-component of the metric tensor are

\ba n^{(0)M}_{12}&=& 2 p^2 (k^2/L)^p (k^2 + J(0,a k) m^2 - J(0,a
k) k^2 -m^2), \nn n^{(1)M}_{12}&=& k^2-k^2 J(0,a k) +J(1,a k) a k
m^2 -J(1,a k) a k^3 \nn && +J(1,a k) a k m^2 p (k^2/L)^p -J(1,a k)
a k^3 p (k^2/L)^p \nn && -k^2 J(0,a k) p (k^2/L)^p +k^2 p
(k^2/L)^p. \ea

\begin{figure}
\hspace*{0.5cm}
\includegraphics[width=8.0cm,angle=-90]{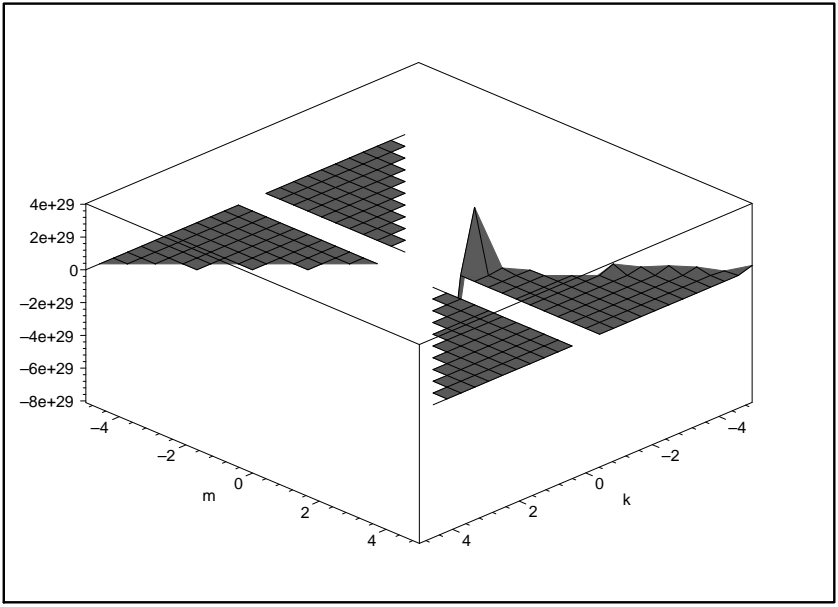}
\caption{The determinant of the metric tensor plotted as a
function of the transverse momentum and mass $k, m$, describing
the fluctuations in massive rotating quarkonia.} \label{det3dkm}
\vspace*{0.5cm}
\end{figure}

When the fluctuations of the quarkonia are described in terms of
the transverse momentum $k$ and mass $m$, we observe that the
principle components of the metric tensor, signifying self pair
correlations, are positive definite functions in a non-trivial
range of $k$. For the configurations with the same sign of the
index $p$ and constant $b$, it turns out that the local stability
requires that (i) $qq$- fluctuations satisfy the constraint \be
n^{(0)M}_{11}+ n^{(1)M}_{11} l(p)+ n^{(2)M}_{11} l(p)^2 > 0, \ee
and (ii) $mm$- fluctuations pose a constraint on the Bessel
function such that the underlying transverse momentum of the
configuration must be limited to the condition of $J(0,a k)> \ 1$.

The determinant of the metric tensor turns out to be in a similar
rational form, as it was obtained in the previous subsection.
However, the simplifications are relatively straightforward, which
in turn implies the following compact expression for the
determinant of the metric tensor

\ba g = -\frac{2p^2}{b^2 l(p)^4 \exp{(l(p)^2)} k^2
(m^2-k^2)^{3/2}} (n^{(0)M}_{g}+ n^{(1)M}_{g}l(p)+
n^{(2)M}_{g}l(p)^2), \ea

with the following factorization \ba n^{(1)M}_{g}&=&  2 p^2
(k^2/L)^p((n^{(12)M}_{g} +p ((k^2/L)^p)n^{(13)M}_{g}), \nn
n^{(2)M}_{g}&=& n^{(20)M}_{g}+ 2 n^{(21)M}_{g} p  (k^2/L)^p+
n^{(22)M}_{g}((k^2/L)^p)^2 p^2. \ea

In this case, it is relatively easy to obtain that the
coefficients are given by

\ba n^{(0)M}_{g}&=& 8p^4( \sqrt{m^2-k^2} ((k^2/L)^p)^2 +
\sqrt{m^2-k^2} ((k^2/L)^p)^2 J(0,a k)^2 \nn && -2 m f
((k^2/L)^p)^2 J(0,a k) + m f ((k^2/L)^p)^2 + m f ((k^2/L)^p)^2
J(0,a k)^2  \nn && -2 \sqrt{m^2-k^2} ((k^2/L)^p)^2 J(0,a k)), \nn
n^{(12)M}_{g}&=& m f-2 m p f +2 m f J(1,a k) a k J(0,a k) + m f
J(0,a k)^2  \nn && +4 m p f J(0,a k) -2 m p f J(0,a k)^2 -2 m f
J(0,a k)  \nn && +4 \sqrt{m^2-k^2} J(0,a k)^2 +4 \sqrt{m^2-k^2}
\nn && +4 \sqrt{m^2-k^2} J(0,a k) J(1,a k) a k -2 m f J(1,a k) a k
\nn && -4 \sqrt{m^2-k^2} J(1,a k) a k -8 \sqrt{m^2-k^2} J(0,a k),
\nn
n^{(13)M}_{g}&=& m f +2 m f J(1,a k) a k J(0,a k) -8
\sqrt{m^2-k^2} J(0,a k)  \nn && +4 \sqrt{m^2-k^2} +4
\sqrt{m^2-k^2} J(0,a k) J(1,a k) a k  \nn && +4 \sqrt{m^2-k^2}
J(0,a k)^2 -4 \sqrt{m^2-k^2} J(1,a k) a k  \nn && -2 m f J(0,a k)
-2 m f J(1,a k) a k + m f J(0,a k)^2, \nn
n^{(20)M}_{g}&=& m (f a^2 k^2 J(0,a k) - f a k J(1,a k) + f a k
J(1,a k) J(0,a k)  \nn && -f a^2 k^2 J(0,a k)^2)+ 2
\sqrt{m^2-k^2}(1+ J(0,a k)^2  \nn && + J(1,a k)^2 a^2 k^2 +2 k
J(0,a k) J(1,a k) a -2 k J(1,a k) a -2 J(0,a k)), \nn
n^{(21)M}_{g}&=& m (f a^2 k^2 J(0,a k) - f a^2 k^2 J(0,a k)^2 + f
a k J(1,a k) J(0,a k)  \nn && - f a k J(1,a k)) +2 \sqrt{m^2-k^2}
(1 + J(1,a k)^2 a^2 k^2  \nn && + J(0,a k)^2 -2 k J(1,a k) a -2
J(0,a k) +2 k J(1,a k) a J(0,a k)), \nn
n^{(22)M}_{g}&=& m f (a k J(1,a k) J(0,a k) + a^2 k^2 J(0,a k) - a
k J(1,a k)  \nn && - a^2 k^2 J(0,a k)^2) +2 \sqrt{m^2-k^2}(1+
J(1,a k)^2 a^2 k^2  \nn && -2 J(0,ak) +2 k J(1,a k) a J(0,a k) +2
J(0,a k)^2 -2 k J(1,a k) a ).\ea

As per the requirement of the positive definiteness of the
determinant of the metric tensor, the global stability of
underlying quarkonia with fluctuating $\{k,m\}$ leads to the
following constraint \ba n^{(0)M}_{g}+ n^{(1)M}_{g}l(p)+
n^{(2)M}_{g}l(p)^2 &<& \ 0. \ea The globally covariant Riemann
curvature tensor $R_{kmkm}$ offers a similar property, as
described in the foregoing subsection. Specifically, we find that
the $R_{kmkm}$ has various factors of $\{ (m^2-k^2)^{n/2} | n \in
Z\}$ and various powers of the logarithmic coupling $l(p)$. Some
of the interesting terms of $R_{kmkm}$ are $-600 m^6
(m^2-k^2)^{11/2} l(p)^4 k^{18}$, $+712 m^{18}(m^2-k^2)^{9/2}
l(p)^4 k^8$, ... and $-1244 m^{28}(m^2-k^2)^{7/2} l(p)^4 k^8$. It
is worth mentioning that $R_{kmkm}$ contains approximately 50,000
terms whose presentation is quite lengthy for the present paper.
Nevertheless, the globally invariant scalar curvature and globally
covariant Riemann curvature tensor can straightforwardly be
obtained. As described in the previous subsection, the globally
covariant physical properties of two parameter quarkonia follow
directly from the product of two geometrically invariant
quantities, viz. $R$ and $\Vert g \Vert$. As per our consideration
of the foregoing subsections for the globally invariant scalar
curvature $R$ and determinant of the metric tensor $\Vert g\Vert
$, the covariant properties, viz., $R_{kmkm}$ follow directly for
the two parameter quarkonia.

From Figs.(\ref{corrKK3dkm}) and (\ref{corrMM3dkm}), we see that
the graphical views of the heat capacities are defined as pure
components of the metric tensor. The plots have been depicted
against the transverse momentum $k$ and mass $m$. In the present
case, the choice of parameters is made as $L=22500$, $p= 5/6$, $b=
1$ and $a=100000$. Notice that the graphical nature of all
thermodynamic interactions remains the same, as long as the
parameter $b$ has a fixed sign.

The $kk$-heat capacity Fig.(\ref{corrKK3dkm}) shows a large
variation of minima when going from a negative $k$ for a given
positive mass $m \in (0,5)$. The order of instability turns out to
be as high as $-10^{16}$. In this case, we further find in the
other quarqants that the underlying quarkonia remains nearly
stable in three distinct pathes.

On the other hand, the $mm$-heat capacity Fig.(\ref{corrMM3dkm})
indicates a similar behavior of the negative variation. In
contrast to the foregoing case, in this case we find that the
local interactions are present only in four disjoint regions, all
of which exclude the origin. As expected, the strength of the
above interactions depends on the domain chosen in the $(k,m)$
plane, and thus the nature of the stability of quarkonia with the
fluctuating $\{k,m\}$.

The corresponding plot for the determinant of the metric tensor
depicted in the Fig.(\ref{det3dkm}), shows the graphical nature of
the determinant of the metric tensor. These plots explicate the of
regions of the thermodynamic (un)stability for the underlying
massive quarkonia lying in a given QCD phase. Specifically, the
significance of the fluctuations of $\{k,m\}$ is that they show
thermodynamical interactions in the strongly coupled massive
quarkonia. The above figures indicate that the quarkonia can be
highly unstable even in the linear Regge regime. As the non-linear
effects become stronger and stronger, it is thus expected that the
thermodynamic instability and correlations would grow further.
This motivates us to extend our analysis to the general quarkonium
configuration.

\subsection{Generic Quarkonia}

In the present subsection, we analyze the properties of general
massive rotating quarkonia, when all parameters of the theory are
allowed to fluctuate. To do so, let us consider the scale $q$,
index of effective potential $p$ and angular momentum $J$ as the
parameters of the present interest. In the framework of the
Bloch-Nordsieck resummation, the strong QCD coupling takes the
following form

\ba
 A(q, p, J)= \frac{1}{b} \frac{p}{\ln(1 + p (q/L)^p)} \ln(\frac{\sqrt{J} +
\sqrt{J - q}}{\sqrt{J} + \sqrt{J - q}})(1 - J(0, a \sqrt{q})). \ea

After some simplification, we find that the components of the
metric tensor are

\ba g_{qq}&=& -\frac{p}{4 b l(p)^3 \exp{(2l(p))}
q^{5/2}(J-q)^{3/2}}( n^{(0)G}_{11}+ n^{(1)G}_{11} l(p)+
n^{(2)G}_{11} l(p)^2), \nn g_{qp}&=&  \frac{1}{2b l(p)^3
\exp{(2l(p))}q^{3/2} (J-q)^{1/2}} (n^{(0)G}_{12}+ n^{(1)G}_{12}
l(p)+ n^{(2)G}_{12} l(p)^2), \nn g_{qJ}&=& \frac{p}{2 b l(p)^2
\exp{(l(p))} q^{3/2} (J-q)^{3/2} J^{1/2}} (n^{(0)G}_{13}+
n^{(1)G}_{13} l(p)), \nn g_{pp}&=& \frac{f}{b l(p)^3
\exp{(2l(p))}} (-1+J(0,a \sqrt{q})) (n^{(0)G}_{22}+ n^{(1)G}_{22}
l(p)), \nn g_{pJ}&=& -\frac{1}{b l(p)^2 \exp{(l(p))} (J-q)^{1/2}
J^{1/2}} (-1+J(0,a \sqrt{q})) (n^{(0)G}_{23}+ (l(p)
n^{(1)G}_{23}), \nn g_{JJ}&=& \frac{p} {2b l(p) (J-q)^{3/2}
J^{3/2}} (-1+J(0,a \sqrt(q))) (2 J-q),\ea

%\ba n^{G}_{11}&=& n^{(0)G}_{11}+ n^{(1)G}_{11} l(p)+ n^{(2)G}_{11}
%l(p)^2,\nn n^{G}_{12}&=& n^{(0)G}_{12}+ n^{(1)G}_{12} l(p)+ n^{(2)G}_{12}
%l(p)^2,\nn n^{G}_{13}&=& n^{(0)G}_{13}+ n^{(1)G}_{13} l(p),\nn
%n^{G}_{22}&=& n^{(0)G}_{22}+ n^{(1)G}_{22} l(p),\nn
%n^{G}_{23}&=& n^{(0)G}_{23}+ (l(p) n^{(1)G}_{23}. \ea

where the coefficients $\lbrace n^{(1)G}_{11}, n^{(2)G}_{11},
n^{(1)G}_{12}, n^{(2)G}_{12} \rbrace $ appearing in the components
of the metric tensor factorize as mentioned before. In the powers
of $p$, the exact factorizations are given as follows: (i) the
$qq$-component

\ba n^{(1)G}_{11}&=& 4 p^2 (q/L)^p n^{(12)G}_{11} +4 p^3
((q/L)^p)^2 n^{(12)G}_{11}, \nn n^{(2)G}_{11}&=& n^{(20)G}_{11}+
2p(q/L)^p n^{(21)G}_{11} +p^2((q/L)^p)^2,\ea

(ii) the $qp$-component

\ba n^{(1)G}_{12}&=& p(q/L)^p n^{(11)G}_{12}+p^2((q/L)^p)^2
n^{(12)G}_{12},\nn
n^{(2)G}_{12}&=& n^{(20)G}_{12}+ 2p(q/L)^p n^{(21)G}_{11}
+p^2((q/L)^p)^2n^{(22)G}_{12}.\ea

Without any approximation, we obtain that the factors in the
numerator of the $qq$-components are given by

\ba n^{(0)G}_{11}&=& 8 p^4 f ((q/L)^p)^2 \sqrt{J-q} (q^{3/2}
+\sqrt{q} J(0,a \sqrt{q}) J -\sqrt{q} J  \nn && -q^{3/2} J(0,a
\sqrt{q})), \nn
n^{(12)G}_{11}&=& f \sqrt{J-q} q^{3/2} - f \sqrt{J-q} q^{3/2}
J(0,a \sqrt{q}) - f \sqrt{J-q} \sqrt{q} J \nn && + f J(1,a
\sqrt{q}) a q \sqrt{J-q} J - f J(1,a \sqrt{q}) a q^2 \sqrt{J-q}
\nn && + f \sqrt{J-q} \sqrt{q} J(0,a \sqrt{q}) J -2 \sqrt{J}
q^{3/2} J(0,a \sqrt{q})  \nn && -2 J^{3/2} \sqrt{q} +2 J^{3/2}
\sqrt{q} J(0,a \sqrt{q}) +2 \sqrt{J} q^{3/2}  \nn && + p f
\sqrt{J-q} \sqrt{q} J - p f  \sqrt{J-q} \sqrt{q} J(0,a \sqrt{q}) J
\nn && - p f \sqrt{J-q} q^{3/2} + p f \sqrt{J-q} q^{3/2} J(0,a
\sqrt{q}), \nn
n^{(13)G}_{11}&=&f J(1,a \sqrt{q}) a q \sqrt{J-q} J - f J(1,a
\sqrt{q}) a q^2 \sqrt{J-q}  \nn && - f \sqrt{J-q} \sqrt{q} J - f
\sqrt{J-q} q^{3/2} J(0,a \sqrt{q})  \nn && + f \sqrt{J-q} \sqrt{q}
J(0,a \sqrt{q}) J + f \sqrt{J-q} q^{3/2}  \nn && -2 \sqrt{J}
q^{3/2} J(0,a \sqrt{q})  +2 J^{3/2} \sqrt{q} J(0,a \sqrt{q})  \nn
&& +2 \sqrt{J} q^{3/2} -2 J^{3/2} \sqrt{q}, \nn
n^{(20)G}_{11}&=& fa( a q^{5/2} \sqrt{J-q} J(0,a \sqrt{q}) - a
q^{3/2} \sqrt{J-q} J(0,a \sqrt{q}) J  \nn && +2 q \sqrt{J-q} J(1,a
\sqrt{q}) J -2 q^2 \sqrt{J-q} J(1,a \sqrt{q}))  \nn &&
+2\sqrt{J}(2 J \sqrt{q} J(0,a \sqrt{q}) -2  J(1,a \sqrt{q}) a q^2
\nn && +2 J J(1,a \sqrt{q}) a q +3  q^{3/2}-4 J^{3/2} \sqrt{q} -3
q^{3/2} J(0,a \sqrt{q})), \nn
n^{(21)G}_{11}&=&  f a^2 q^{5/2} \sqrt{J-q} J(0,a \sqrt{q}) - f
a^2 q^{3/2} \sqrt{J-q} J(0,a \sqrt{q})  J  \nn && +2 f a q
\sqrt{J-q} J(1,a \sqrt{q}) J -2f a q^2 \sqrt{J-q} J(1,a \sqrt{q})
\nn && -4 J^{3/2} \sqrt{q} +4 J^{3/2} J(1,a \sqrt{q}) a q -4
\sqrt{J} J(1,a \sqrt{q}) a q^2  \nn && +4 J^{3/2} \sqrt{q} J(0,a
\sqrt{q}) +6 \sqrt{J} q^{3/2} -6 \sqrt{J} q^{3/2} J(0,a \sqrt{q}),
\nn
n^{(22)G}_{11}&=& f a^2 q^{5/2} \sqrt{J-q} J(0,a \sqrt{q}) -f a^2
q^{3/2} \sqrt{J-q} J(0,a \sqrt{q}) J  \nn && +2 f J(1,a \sqrt{q})
a q \sqrt{J-q} J -2 f J(1,a \sqrt{q}) a q^2 \sqrt{J-q} \nn &&  -4
\sqrt{J} J(1,a \sqrt{q}) a q^2 +4 J^{3/2} \sqrt{q} J(0,a \sqrt{q})
\nn && -4 J^{3/2} \sqrt{q} +4 J^{3/2} J(1,a \sqrt{q}) a q +6
\sqrt{J} q^{3/2} \nn &&  -6 \sqrt{J} q^{3/2} J(0,a \sqrt{q}). \ea

In a parallel way, the factors in the numerator of the
$qp$-components turn out to be

\ba n^{(0)G}_{12}&=& 4 p^3 f ((q/L)^p)^2 (\sqrt{J-q} \sqrt{q}
-\sqrt{J-q} \sqrt{q} J(0,a \sqrt{q})  \nn &&  +p \sqrt{J-q}
\sqrt{q} \ln(q/L) -p \sqrt{J-q} \sqrt{q} J(0,a \sqrt{q})
\ln(q/L)), \nn
n^{(11)G}_{12}&=& 2 \sqrt{J} \sqrt{q} -2 p^2 f \sqrt{J-q} \sqrt{q}
\ln(q/L) -6 p f \sqrt{J-q} \sqrt{q}   \nn && +2 p^2 f \sqrt{J-q}
\sqrt{q} J(0,a \sqrt{q}) \ln(q/L) -2 \sqrt{J} \sqrt{q} J(0,a
\sqrt{q})  \nn &&  - f J(1,a \sqrt{q}) a q \sqrt{J-q} +2 p
\sqrt{J} \sqrt{q} \ln(q/L)   \nn && -2 p \sqrt{J} \sqrt{q} J(0,a
\sqrt{q}) \ln(q/L) +6 p f \sqrt{J-q} \sqrt{q} J(0,a \sqrt{q}) \nn
&& -p f J(1,a \sqrt{q}) a q \sqrt{J-q} \ln(q/L),\nn
n^{(12)G}_{12}&=& 2 \sqrt{J} \sqrt{q} - f J(1,a \sqrt{q}) a  q
\sqrt{J-q} +2 p \sqrt{J} \sqrt{q} \ln(q/L)   \nn && -2 \sqrt{J}
\sqrt{q} J(0,a \sqrt{q}) -2 p \sqrt{J} \sqrt{q} J(0,a \sqrt{q})
\ln(q/L)   \nn && +4 p f \sqrt{J-q} \sqrt{q} J(0,a \sqrt{q})-4 p f
\sqrt{J-q} \sqrt{q}   \nn && -p f J(1,a \sqrt{q}) a  q \sqrt{J-q}
\ln(q/L), \nn
n^{(20)G}_{12}&=& f a q \sqrt{J-q} J(1,a \sqrt{q}) +2 \sqrt{J}
\sqrt{q} J(0,a \sqrt{q}) -2 \sqrt{J} \sqrt{q},\nn
n^{(20)G}_{12}&=& a q \sqrt{J-q} J(1,a \sqrt{q}) +2 \sqrt{J}
\sqrt{q} J(0,a \sqrt{q}) -2 \sqrt{J} \sqrt{q}, \nn
n^{(22)G}_{12}&=& 2 \sqrt{J} \sqrt{q} J(0,a \sqrt{q}) -2 \sqrt{J}
\sqrt{q} +f a q \sqrt{J-q} J(1,a \sqrt{q}).\ea

The factors in the numerator of the $qJ$-components are expressed
as \ba n^{(0)G}_{13}&=& 2 p^2 (q/L)^p (q^{3/2} +\sqrt{q} J(0,a
\sqrt{q}) J -q^{3/2} J(0,a \sqrt{q}) - \sqrt{q} J), \nn
n^{(1)G}_{13}&=& q(q^{1/2} - q^{1/2} J(0,a \sqrt{q}) +J(1,a
\sqrt{q}) a J -J(1,a \sqrt{q}) a q)   \nn && + p (q/L)^p (q^{3/2}-
q^{3/2} J(0,a \sqrt{q}) +J(1,a \sqrt{q}) a q J -J(1,a \sqrt{q}) a
q^2).\ea

The corresponding factors in the numerator of the $pp$-components
are given by \ba n^{(0)G}_{22}&=& -2 p (q/L)^{2 p}(1+2p
\ln(q/L)+p^2 \ln(q/L)^2), \nn n^{(1)G}_{22}&=& (q/L)^p(2+4 p
\ln(q/L) + p^2 \ln(q/L)^2) \nn && +p(q/L)^{2 p}(1 +2 p
\ln(q/L)).\ea

Finally, the factors in the numerator of the $pJ$-components are
\ba n^{(0)G}_{23}&=& -p (q/L)^p -p^2 \ln(q/L) (q/L)^p, \nn
n^{(1)G}_{23}&=& 1 + p (q/L)^p. \ea

Herewith, we observe that the principle components of the metric
tensor, signifying self pair correlations, remain positive
definite functions in non-trivial intervals of the parameters. The
local stability of the configuration requires the following three
simultaneously constraints. For the same sign of $\{p, b$, the
thermodynamic stability enforces: (i) the $qq$- fluctuations
satisfy \ba n^{(0)G}_{11}+ n^{(1)G}_{11} l(p)+ n^{(2)G}_{11}
l(p)^2, \ea

(ii) the $JJ$- fluctuations remain within the limiting values of
the Bessel function \ba J(0,a \sqrt{q})&>& \ 1,\ \ 2J<q, \nn &<& \
1, \ \ 2J>q, \ea

and (iii) the $pp$- fluctuations satisfy \ba n^{(0)G}_{22}+
n^{(1)G}_{22} l(p) &>& \ 0, \ \ J(0,a \sqrt{q})
> 1, \nn &<& \ 0, \ \ J(0,a \sqrt{q}) < 1 \ea

for the same sign of $\{f, b\}$. Correspondingly, the three
dimensional views of the heat capacities are shown in the
following plots for the general rotating massive case. For the
purpose of the present and subsequent diagrammatic views, we shall
focus on the case of $p=5/6$ and do not take any approximation
against the Bessel function. Thereby, we graphically illustrate
the local stability of the rotating quarkonia, which is in
accordance with the above positivity constraints of the specific
heat capacities.

\begin{figure}
\hspace*{0.5cm}
\includegraphics[width=8.0cm,angle=-90]{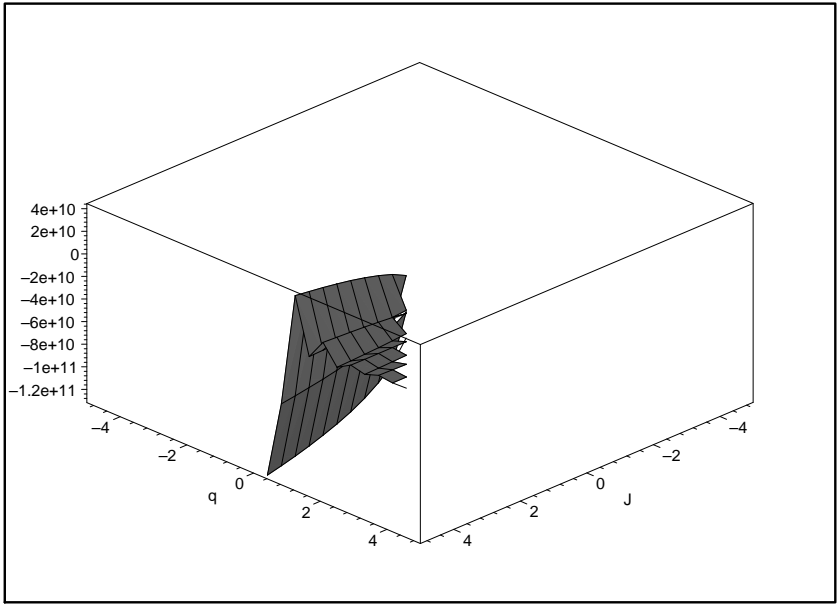}
\caption{The $qq$-component of the metric tensor plotted as a
function of $q$ and the angular momentum, $J$, describing the
nature of the heat capacity of the $JJ$-component for general
massive rotating quarkonia.} \label{corrQQ3dqpJ} \vspace*{0.5cm}
\end{figure}

As a function of scale parameter $q:=Q^2$ and angular momentum
$J$, the $qq$-component is shown in the Fig.(\ref{corrQQ3dqpJ}).
As mentioned before, let us illustrate the present case for the
same choice of the parameters, viz., $L=22500$, $b= 1$ and
$a=100000$. Consequently, we find that the graphical nature of the
underlying thermodynamic nature of the quarkonia remains the same,
as long as the parameter $b$ has a fixed sign.

\begin{figure}
\hspace*{0.5cm}
\includegraphics[width=8.0cm,angle=-90]{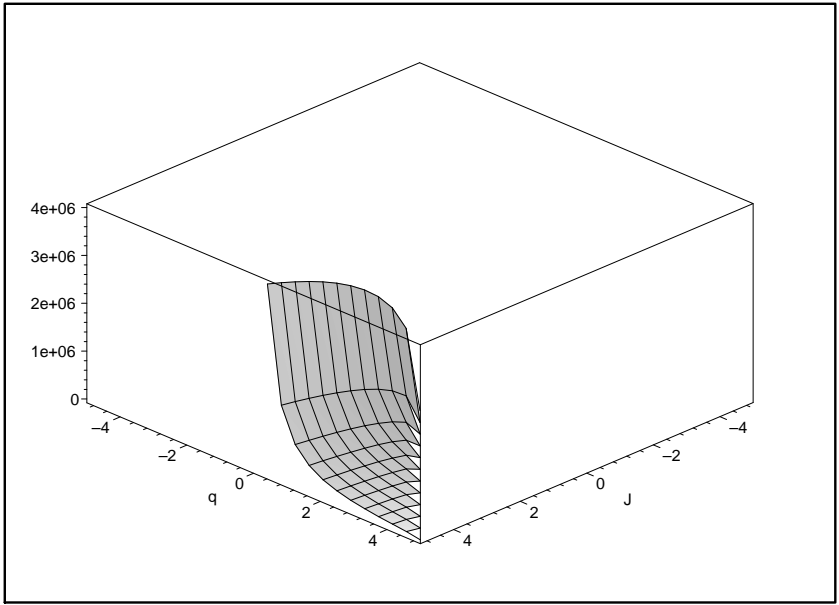}
\caption{The $pp$-component of the metric tensor plotted as a
function of the scale $q$, and the angular momentum, $J$,
describing the nature of the heat capacity of the $JJ$-component
for general massive rotating quarkonia.} \label{corrPP3dqpJ}
\vspace*{0.5cm}
\end{figure}

The $pp$-heat capacity is depicted in the Fig.(\ref{corrPP3dqpJ}).
For a small $q$, this shows a large minimum of depth $10^{6}$, in
the regime of the angular momentum $ J \in (0,5)$. It turns out
that the underlying $pp$-fluctuations of the quarkonia remain
nearly stable in the other quadrants.

\begin{figure}
\hspace*{0.5cm}
\includegraphics[width=8.0cm,angle=-90]{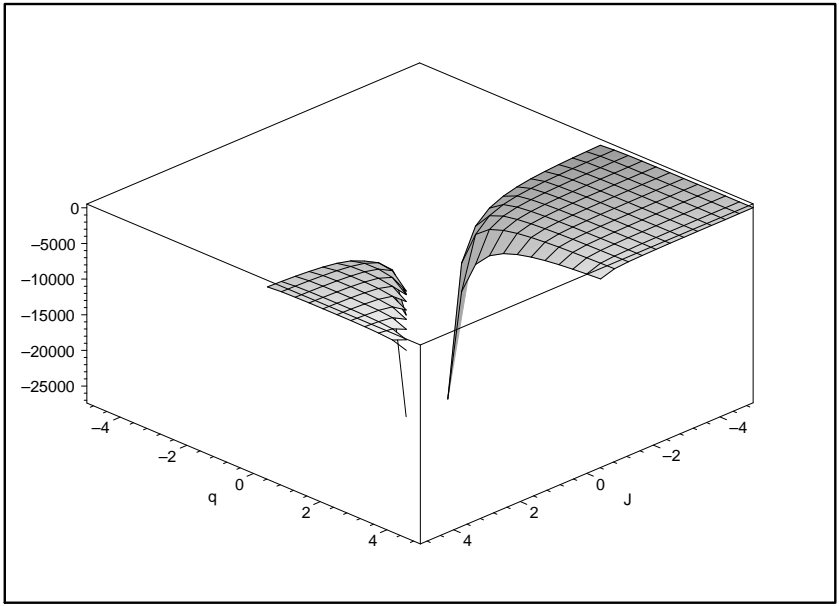}
\caption{The $JJ$-component of the metric tensor plotted as a
function of scale and the angular momentum, viz., $q, J$,
describing the nature of the heat capacity of the $JJ$-component
for general massive rotating quarkonia.} \label{corrJJ3dqpJ}
\vspace*{0.5cm}
\end{figure}

Similarly, the $JJ$-heat capacity is shown in the corresponding
Fig.(\ref{corrJJ3dqpJ}), which indicates a similar behavior of the
negative variation, as mentioned in the case of the $QJ$-plan. The
silent feature of the $JJ$-fluctuations is that we find the two
distinct local behaviors for $J>0$ and $J<0$. In the limit of
vanishing $q$ and $J$, the local interactions reach an order of
$10^{4}$. For fluctuating $\{q,p,J\}$, the strength of the
thermodynamic interactions depends on the domain chosen in the
$(q,p,J)$ plane, and thus so does the thermodynamic stability of
the quarkonia.

We further observe that the Gaussian fluctuations of the most
general quarkonia comply with expected thermodynamic stability.
Physically, the heat capacities under the fluctuations, which are
defined as the self-pair correlations, remain positive quantities
in the domain of the parameters and thus form the well-defined
basis of the manifold $(M_3,g)$. Subsequently, our computation
demonstrates the intrinsic geometric nature of the parametric
fluctuations. The quarkonia remain globally stable under the
fluctuations, if the associated principle minors, viz., $\{p_2,
p_3 \}$ remain positive definite functions. Herewith, the
$qp$-surface is stable if we have a positive surface minor

\ba \label{minor} p_S^G := -\frac{1}{4b^2 l(p)^5 \exp{(3l(p))}
q^{5/2} (J-q)^{3/2}} (n^{(0)G}_{S}+ n^{(1)G}_{g}l(p)+
n^{(2)G}_{S}l(p)^2+ n^{(3)G}_{S}l(p)^3), \ea

where the $n^{(0)G}_{S}$ has the following illuminating
factorization

\ba n^{(0)G}_{S}&=& 8 f^2 ((q/L)^p)^3 (p^4 n^{(01)G}_{S}+ p^5
n^{(02)G}_{S}+ p^6 n^{(03)G}_{S}).\ea

Interestingly, the factors $\{ n^{(01)G}_{S}, n^{(02)G}_{S},
n^{(03)G}_{S} \}$ can be expressed as

\ba n^{(01)G}_{S}&=& q^{3/2} \sqrt{J-q} + q^{3/2} \sqrt{J-q} J(0,a
\sqrt{q})^2 - J \sqrt{q} \sqrt{J-q}  \nn && - J \sqrt{q}
\sqrt{J-q} J(0,a \sqrt{q})^2 -2 q^{3/2} \sqrt{J-q} J(0,a \sqrt{q})
\nn && +2 J \sqrt{q} \sqrt{J-q} J(0,a \sqrt{q}), \nn
n^{(02)G}_{S}&=& 2 q^{3/2} \sqrt{J-q} ln(q/L) -2 J \sqrt{q}
\sqrt{J-q} J(0,a \sqrt{q})^2 \ln(q/L)  \nn && +2 q^{3/2}
\sqrt{J-q} J(0,a \sqrt{q})^2 \ln(q/L) +3 q^{3/2} \sqrt{J-q} \nn &&
-3 \sqrt{q} \sqrt{J-q} J J(0,a \sqrt{q})^2 +3 q^{3/2} \sqrt{J-q}
J(0,a \sqrt{q})^2  \nn && +4 J \sqrt{q} \sqrt{J-q} J(0,a \sqrt{q})
\ln(q/L) -2J \sqrt{q} \sqrt{J-q} ln(q/L) \nn && -6 q^{3/2}
\sqrt{J-q} J(0,a \sqrt{q}) +6 \sqrt{q} \sqrt{J-q} J J(0,a
\sqrt{q})\nn &&  -3 \sqrt{q} \sqrt{J-q} J -4 q^{3/2} \sqrt{J-q}
J(0,a \sqrt{q}) \ln(q/L)), \nn
n^{(03)G}_{S}&=& q^{3/2} \sqrt{J-q} \ln(q/L)^2 + q^{3/2}
\sqrt{J-q} J(0,a \sqrt{q})^2 \ln(q/L)^2 \nn && - J \sqrt{q}
\sqrt{J-q} J(0,a \sqrt{q})^2 \ln(q/L)^2 -2 \sqrt{q} \sqrt{J-q} J
\ln(q/L) \nn &&  +2 q^{3/2} \sqrt{J-q} J(0,a \sqrt{q})^2 \ln(q/L)
+2 J \sqrt{q} \sqrt{J-q} J(0,a \sqrt{q}) \ln(q/L)^2  \nn && +2
q^{3/2} \sqrt{J-q} \ln(q/L) -2 q^{3/2} \sqrt{J-q} J(0,a \sqrt{q})
\ln(q/L)^2  \nn && -2 \sqrt{q} \sqrt{J-q} J J(0,a \sqrt{q})^2
\ln(q/L) +4 \sqrt{q} \sqrt{J-q} J J(0,a \sqrt{q}) \ln(q/L) \nn &&
- J \sqrt{q} \sqrt{J-q} \ln(q/L)^2 -4 q^{3/2} \sqrt{J-q} J(0,a
\sqrt{q}) \ln(q/L). \ea

It is worth mentioning that the factors of the $l(p)$-terms can
further be expressed as

\ba n^{(1)G}_{S}&=&  (q2/L)^{2p} n^{(12)G}_{S} + ((k^2/L)^{3p}
n^{(13)G}_{S})\ea

with the following additional structures

\ba n^{(12)G}_{S}=& n^{(120)G}_{S}+ n^{(121)G}_{S}f+
n^{(122)G}_{S}f^2,\nn
n^{(13)G}_{S}=& n^{(130)G}_{S}+ n^{(131)G}_{S}f+
n^{(132)G}_{S}f^2. \ea

We find in particular that the $120$-factor reduces to the
following explicit expression

\ba n^{(120)G}_{S}&=& 4 ( \sqrt{q} \sqrt{J-q} J p^2 + \sqrt{q}
\sqrt{J-q} p^4 J J(0,a \sqrt{q})^2 \ln(q/L)^2  \nn && -2 q^2
\sqrt{J} p^3 f J(1,a \sqrt{q}) a \ln(q/L) -2 \sqrt{q} \sqrt{J-q} J
J(0,a \sqrt{q}) p^2  \nn && +2 \sqrt{q} \sqrt{J-q} p^3 J J(0,a
\sqrt{q})^2 \ln(q/L) -2 \sqrt{q} \sqrt{J-q} p^4 J J(0,a \sqrt{q})
\ln(q/L)^2 \nn &&  -8 \sqrt{q} \sqrt{J-q} p^3 J J(0,a \sqrt{q})
\ln(q/L)). \ea

The $121$- and $122$-factors factorize as follows

\ba n^{(121)G}_{S}&=&  p^2 n^{(1212)G}_{S}+ 2 p^3 n^{(2113)G}_{S}+
p^4 n^{(1214)G}_{S}, \nn
n^{(122)G}_{S}&=& p^2 n^{(2122)G}_{S}+ 2 p^3 n^{(2123)G}_{S}+ p^4
n^{(2124)G}_{S}+ 4 p^5 n^{(2125)G}_{S}\ea

with the following expressions for the sub-factors

\ba n^{(1212)G}_{S}&=&  \sqrt{J}( J q a J(1,a \sqrt{q}) - q^2
J(1,a \sqrt{q}) a + q^2 J(1,a \sqrt{q}) a J(0,a \sqrt{q})  \nn &&
- J q a J(1,a \sqrt{q}) J(0,a \sqrt{q}) -2 J \sqrt{q} -2 J
\sqrt{q} J(0,a \sqrt{q})^2 +3 q^{3/2}  \nn && +3 q^{3/2} J(0,a
\sqrt{q})^2 +4 J \sqrt{q} J(0,a \sqrt{q}) -6 q^{3/2} J(0,a
\sqrt{q})),\nn
n^{(2113)G}_{S}&=& \sqrt{J}( q^2 J(0,a \sqrt{q}) \ln(q/L) J(1,a
\sqrt{q}) a +J q J(1,a \sqrt{q}) a \ln(q/L)  \nn && - J q J(1,a
\sqrt{q}) a J(0,a \sqrt{q}) \ln(q/L) -2 J \sqrt{q} \ln(q/L)+3
q^{3/2} \nn && -2 J \sqrt{q} J(0,a \sqrt{q})^2 \ln(q/L) -3 J
\sqrt{q} +3 q^{3/2} J(0,a \sqrt{q})^2 \ln(q/L)  \nn && +3 q^{3/2}
J(0,a \sqrt{q})^2 -3 J \sqrt{q} J(0,a \sqrt{q})^2 +3 q^{3/2}
\ln(q/L)  \nn && +4 J \sqrt{q} J(0,a \sqrt{q}) \ln(q/L) +6 J
\sqrt{q} J(0,a \sqrt{q})  \nn && -6 q^{3/2} J(0,a \sqrt{q})
\ln(q/L) -6 q^{3/2} J(0,a \sqrt{q})), \nn
n^{(1214)G}_{S}&=& \sqrt{J} ( q^2 J(1,a \sqrt{q}) a J(0,a
\sqrt{q}) \ln(q/L)^2 + J q \ln(q/L)^2 J(1,a \sqrt{q}) a  \nn && -
q^2 \ln(q/L)^2 J(1,a \sqrt{q}) a - J q J(0,a \sqrt{q}) \ln(q/L)^2
J(1,a \sqrt{q}) a  \nn && -2 J \sqrt{q} J(0,a \sqrt{q})^2
\ln(q/L)^2 +3 q^{3/2}J(0,a \sqrt{q})^2 \ln(q/L)^2  \nn && +3
q^{3/2} \ln(q/L)^2 +4 J \sqrt{q} J(0,a \sqrt{q}) \ln(q/L)^2  \nn
&& +4 q^{3/2} J(0,a \sqrt{q})^2 \ln(q/L) +4 q^{3/2} \ln(q/L) -4 J
\sqrt{q} \ln(q/L)  \nn && -4 J \sqrt{q} J(0,a \sqrt{q})^2 \ln(q/L)
-6 q^{3/2} J(0,a \sqrt{q}) \ln(q/L)^2  \nn && +8 J \sqrt{q} J(0,a
\sqrt{q}) \ln(q/L) -8 q^{3/2} J(0,a \sqrt{q}) \ln(q/L)). \ea

The $122$-factors take five powers of $p$, whose precise
expressions take the form

\ba n^{(2122)G}_{S}&=& a q \sqrt{J-q}( q^{1/2} J a J(1,a
\sqrt{q})^2 -q^{3/2} a J(1,a \sqrt{q})^2 +2 J q^{1/2} a J(0,a
\sqrt{q})^2  \nn && -2 J q^(1/2) a J(0,a \sqrt{q}) +2 q^{3/2} a
J(0,a \sqrt{q}) -2 q^{3/2} a J(0,a \sqrt{q})^2  \nn &&  -4 J J(1,a
\sqrt{q}) J(0,a \sqrt{q}) +4 q J(1,a \sqrt{q}) J(0,a \sqrt{q}) -4
q J(1,a \sqrt{q})  \nn && +4 J J(1,a \sqrt{q}) ), \nn
n^{(2123)G}_{S}&=&  \sqrt{J-q}( q^{3/2} J a^2 J(1,a \sqrt{q})^2
\ln(q/L) - q^{5/2} J(1,a \sqrt{q})^2 a^2 \ln(q/L)  \nn && +2 J
q^{3/2} a^2 J(0,a \sqrt{q})^2 \ln(q/L) +2 q^{5/2} a^2 J(0,a
\sqrt{q}) \ln(q/L)  \nn && -2 q^{5/2} a^2 J(0,a \sqrt{q})^2
\ln(q/L) -2 J q^{3/2} a^2 J(0,a \sqrt{q}) \ln(q/L)  \nn && +4 q^2
a J(1,a \sqrt{q}) J(0,a \sqrt{q}) \ln(q/L) -4 J q a J(1,a
\sqrt{q}) J(0,a \sqrt{q}) \ln(q/L) \nn &&  +4 J q a J(1,a
\sqrt{q}) \ln(q/L) -4 q^{3/2} +4 \sqrt{q} J p^5 J(0,a \sqrt{q})^2
\ln(q/L)  \nn && -4 q^2 a J(1,a \sqrt{q}) \ln(q/L) +6 q J a J(1,a
\sqrt{q}) -6 q J a J(1,a \sqrt{q}) J(0,a \sqrt{q})  \nn && +6 q^2
a J(1,a \sqrt{q}) J(0,a \sqrt{q}) -6 q^2 a J(1,a \sqrt{q}) +8
q^{3/2} J(0,a \sqrt{q}) \nn &&  -8 J \sqrt{q} J(0,a \sqrt{q})),
\nn
n^{(2124)G}_{S}&=& \sqrt{J-q} ( q^{3/2} J J(1,a \sqrt{q})^2 a^2
\ln(q/L)^2 -q^{5/2} J(1,a \sqrt{q})^2 a^2 \ln(q/L)^2  \nn &&  +2
q^{5/2} a^2 J(0,a \sqrt{q}) \ln(q/L)^2 -2 q^{5/2} a^2 J(0,a
\sqrt{q})^2 \ln(q/L)^2  \nn && -2 J q^{3/2} a^2 J(0,a \sqrt{q})
\ln(q/L)^2 +2 J q^{3/2} a^2 J(0,a \sqrt{q})^2 \ln(q/L)^2  \nn &&
-4 J q a J(1,a \sqrt{q}) J(0,a \sqrt{q}) \ln(q/L)^2  +4 J q a
J(1,a \sqrt{q}) \ln(q/L)^2  \nn && +4 q^2 a J(1,a \sqrt{q}) J(0,a
\sqrt{q}) \ln(q/L)^2 -4 q^2 a J(1,a \sqrt{q}) \ln(q/L)^2  \nn &&
+8 q^2 J(0,a \sqrt{q}) \ln(q/L) a J(1,a \sqrt{q}) -8 q J J(0,a
\sqrt{q}) \ln(q/L) a J(1,a \sqrt{q}) \nn &&  +8 q J \ln(q/L) a
J(1,a \sqrt{q}) -8 q^2 J(1,a \sqrt{q}) a \ln(q/L)  \nn && +16 J
\sqrt{q} J(0,a \sqrt{q})^2 \ln(q/L) +16 J \sqrt{q} \ln(q/L) \nn &&
-16 q^{3/2} J(0,a \sqrt{q})^2 \ln(q/L) -16 q^{3/2} \ln(q/L)  \nn
&& +28 \sqrt{q} J J(0,a \sqrt{q})^2 -28 q^{3/2} J(0,a \sqrt{q})^2
\nn && +28 \sqrt{q} J  -28 q^{3/2} -32 J \sqrt{q} J(0,a \sqrt{q})
\ln(q/L)  \nn && +32 q^{3/2} J(0,a \sqrt{q}) \ln(q/L) -56 \sqrt{q}
J J(0,a \sqrt{q}) +56 q^{3/2} J(0,a \sqrt{q})),\nn
n^{(2125)G}_{S}&=& \sqrt{J-q} ( J \sqrt{q} \ln(q/L)^2 + J \sqrt{q}
J(0,a \sqrt{q})^2 \ln(q/L)^2 - q^{3/2} \ln(q/L)^2  \nn && -2
q^{3/2} J(0,a \sqrt{q})^2 \ln(q/L) -2 J \sqrt{q} J(0,a \sqrt{q})
\ln(q/L)^2 +2 \sqrt{q} J \ln(q/L) \nn &&  +2 q^{3/2} J(0,a
\sqrt{q}) \ln(q/L)^2 -4 \sqrt{q} J J(0,a \sqrt{q}) \ln(q/L)  \nn
&& +4 q^{3/2} J(0,a \sqrt{q}) \ln(q/L)). \ea

Further, we take advantage of the following factorizations for the
$131$ and $132$ expressions

\ba n^{(131)G}_{S}&=& 4 p^3 n^{(1313)G}_{S}+ 8 p^4
n^{(1314)G}_{S}+ 4 p^5 n^{(1315)G}_{S}, \nn
n^{(132)G}_{S}&=& p^3 n^{(1323)G}_{S}+ 2 p^4 n^{(1324)G}_{S}+ p^5
n^{(1325)G}_{S}.\ea

Subsequently, the factors of $130$ and $131$ appear in the powers
of the $p$, and are given by

\ba n^{(130)G}_{S}&=& 4 \sqrt{q} \sqrt{J-q}  p^3 J (1+ p^2J(0,a
\sqrt{q})^2 \ln(q/L)^2 +J(0,a \sqrt{q})^2  \nn && +p^2 \ln(q/L)^2
+2p \ln(q/L) +2p J(0,a \sqrt{q})^2 \ln(q/L) -2 J(0,a \sqrt{q}) \nn
&& -2p^2 J(0,a \sqrt{q}) \ln(q/L)^2 -4p J(0,a \sqrt{q}) \ln(q/L)),
\nn
n^{(1313)G}_{S}&=& J^{3/2} q a J(1,a \sqrt{q}) - q^2 \sqrt{J}
J(1,a \sqrt{q}) a + q^2 a J(1,a \sqrt{q}) \sqrt{J} J(0,a \sqrt{q})
\nn && - J^{3/2} q J(1,a \sqrt{q}) a J(0,a \sqrt{q}) -2 J^{3/2}
\sqrt{q}  -2 J^{3/2} \sqrt{q} J(0,a \sqrt{q})^2  \nn && +3 q^{3/2}
\sqrt{J} +3 q^{3/2} \sqrt{J} J(0,a \sqrt{q})^2 +4 J^{3/2} \sqrt{q}
J(0,a \sqrt{q})  \nn && -6  q^{3/2} \sqrt{J} J(0,a \sqrt{q}), \nn
n^{(1314)G}_{S}&=& J^{3/2} q J(1,a \sqrt{q}) a \ln(q/L) + q^2
\sqrt{J} J(1,a \sqrt{q}) a J(0,a \sqrt{q}) \ln(q/L) \nn &&  -
J^{3/2} q J(1,a \sqrt{q}) a J(0,a \sqrt{q}) \ln(q/L) - q^2
\sqrt{J} J(1,a \sqrt{q}) a \ln(q/L)  \nn && -2 J^{3/2} \sqrt{q}
J(0,a \sqrt{q})^2 \ln(q/L) -2 J^{3/2} \sqrt{q} \ln(q/L) +3 q^{3/2}
\sqrt{J} \ln(q/L)  \nn && +3 q^{3/2} \sqrt{J} +3 q^{3/2} \sqrt{J}
J(0,a \sqrt{q})^2 -3 J^{3/2} \sqrt{q}-3 J^{3/2} \sqrt{q} J(0,a
\sqrt{q})^2 \nn && +3 q^{3/2} \sqrt{J} J(0,a \sqrt{q})^2 \ln(q/L)
+4 J^{3/2} \sqrt{q} J(0,a \sqrt{q}) \ln(q/L)  \nn && +6 J^{3/2}
\sqrt{q} J(0,a \sqrt{q}) -6 q^{3/2} \sqrt{J} J(0,a \sqrt{q}) -6
q^{3/2} \sqrt{J} J(0,a \sqrt{q}) \ln(q/L), \nn
n^{(1315)G}_{S}&=& J^{3/2} q \ln(q/L)^2 J(1,a \sqrt{q}) a - q^2
\sqrt{J} J(1,a \sqrt{q}) a \ln(q/L)^2  \nn && + q^2 \sqrt{J} J(0,a
\sqrt{q}) \ln(q/L)^2 J(1,a \sqrt{q}) a - J^{3/2} q J(0,a \sqrt{q})
\ln(q/L)^2 J(1,a \sqrt{q}) a  \nn && -2 J^{3/2} \sqrt{q}
\ln(q/L)^2 -2 J^{3/2} \sqrt{q} J(0,a \sqrt{q})^2 \ln(q/L)^2  +3
q^{3/2} \sqrt{J} \ln(q/L)^2  \nn && +3 q^{3/2} \sqrt{J} J(0,a
\sqrt{q})^2 \ln(q/L)^2 -4 J^{3/2} \sqrt{q} J(0,a \sqrt{q})^2
\ln(q/L) \nn &&  -4 J^{3/2} \sqrt{q} \ln(q/L) +4 q^{3/2} \sqrt{J}
J(0,a \sqrt{q})^2 \ln(q/L) +4 q^{3/2} \sqrt{J} \ln(q/L) \nn &&  +4
J^{3/2} \sqrt{q} J(0,a \sqrt{q}) \ln(q/L)^2 -6 q^{3/2} \sqrt{J}
J(0,a \sqrt{q}) \ln(q/L)^2  \nn && -8 q^{3/2} \sqrt{J} J(0,a
\sqrt{q}) \ln(q/L) +8 J^{3/2} \sqrt{q} J(0,a \sqrt{q}) \ln(q/L).
\ea

The factors of $132$-powers, appearing as the powers of the $p$,
are

\ba n^{(1323)G}_{S}&=& q^{3/2} \sqrt{J-q} J J(1,a \sqrt{q})^2 a^2
-q^{5/2} \sqrt{J-q} a^2 J(1,a \sqrt{q})^2  \nn && -2 J q^{3/2} a^2
\sqrt{J-q} J(0,a \sqrt{q}) -2 q^(5/2) a^2 \sqrt{J-q} J(0,a
\sqrt{q})^2  \nn && +2 q^{5/2} a^2 \sqrt{J-q} J(0,a \sqrt{q}) -4 J
q J(1,a \sqrt{q}) a \sqrt{J-q} J(0,a \sqrt{q})  \nn && +4 J q a
\sqrt{J-q} J(1,a \sqrt{q}) +2 J q^{3/2} a^2 \sqrt{J-q} J(0,a
\sqrt{q})^2  \nn && +4 q^2 a \sqrt{J-q} J(1,a \sqrt{q}) J(0,a
\sqrt{q}) -4 q^2 a \sqrt{J-q} J(1,a \sqrt{q}), \nn
n^{(1324)G}_{S}&=&  q^{3/2} \sqrt{J-q} J a^2 J(1,a \sqrt{q})^2
\ln(q/L) - q^{5/2} \sqrt{J-q} a^2 J(1,a \sqrt{q})^2 \ln(q/L) \nn
&&  +2 J \sqrt{q} \sqrt{J-q} -2 q^{3/2} \sqrt{J-q} +2 J \sqrt{q}
\sqrt{J-q} J(0,a \sqrt{q})^2  \nn && -2 q^{5/2} a^2 \sqrt{J-q}
J(0,a \sqrt{q})^2 \ln(q/L) +2 q^(5/2) a^2 \sqrt{J-q} J(0,a
\sqrt{q}) \ln(q/L)  \nn && -2 J q^{3/2} a^2 \sqrt{J-q} J(0,a
\sqrt{q}) \ln(q/L) +2 J q^{3/2} a^2 \sqrt{J-q} J(0,a \sqrt{q})^2
\ln(q/L)  \nn && -2 q^{3/2} \sqrt{J-q} J(0,a \sqrt{q})^2 -4 q^2
J(1,a \sqrt{q}) a \sqrt{J-q} \ln(q/L)  \nn && +4 J q J(1,a
\sqrt{q}) a \sqrt{J-q} \ln(q/L) -4 J \sqrt{q} \sqrt{J-q} J(0,a
\sqrt{q})  \nn && +4 q^{3/2} \sqrt{J-q} J(0,a \sqrt{q}) -4 J q a
\sqrt{J-q} J(1,a \sqrt{q}) J(0,a \sqrt{q}) \ln(q/L)  \nn && +4 q^2
J(1,a \sqrt{q}) a \sqrt{J-q} J(0,a \sqrt{q}) \ln(q/L)-6 q^2
\sqrt{J-q} J(1,a \sqrt{q}) a \nn && +6 q^2 \sqrt{J-q} J(0,a
\sqrt{q}) J(1,a \sqrt{q}) a +6 q \sqrt{J-q} J J(1,a \sqrt{q}) a
\nn && -6 q \sqrt{J-q} J J(0,a \sqrt{q}) J(1,a \sqrt{q}) a, \nn
n^{(1325)G}_{S}&=& q^{3/2} \sqrt{J-q} J J(1,a \sqrt{q})^2 a^2
\ln(q/L)^2 -q^{5/2} \sqrt{J-q} J(1,a \sqrt{q})^2 a^2 \ln(q/L)^2
\nn && +2 q^{5/2} a^2 \sqrt{J-q} J(0,a \sqrt{q}) \ln(q/L)^2 +2 J
q^{3/2} a^2 \sqrt{J-q} J(0,a \sqrt{q})^2 \ln(q/L)^2  \nn && -2
q^{5/2} a^2 \sqrt{J-q} J(0,a \sqrt{q})^2 \ln(q/L)^2 -2 J q^{3/2}
a^2 \sqrt{J-q} J(0,a \sqrt{q}) \ln(q/L)^2  \nn && +4 q^2 J(1,a
\sqrt{q}) a \sqrt{J-q} J(0,a \sqrt{q}) \ln(q/L)^2 +4 J q a
\sqrt{J-q} J(1,a \sqrt{q}) \ln(q/L)^2 \nn &&  -4 J q J(1,a
\sqrt{q}) a \sqrt{J-q} J(0,a \sqrt{q}) \ln(q/L)^2 -4 q^2 J(1,a
\sqrt{q}) a \sqrt{J-q} \ln(q/L)^2 \nn &&  -8 q^{3/2} \sqrt{J-q}
\ln(q/L) -8 q^{3/2} \sqrt{J-q} J(0,a \sqrt{q})^2 \ln(q/L) +8 J
\sqrt{q} \sqrt{J-q} \ln(q/L)  \nn && +8 J \sqrt{q} \sqrt{J-q}
J(0,a \sqrt{q})^2 \ln(q/L) +8 q \sqrt{J-q} J J(1,a \sqrt{q}) a
\ln(q/L) \nn &&  -8 q \sqrt{J-q} J J(0,a \sqrt{q}) \ln(q/L) J(1,a
\sqrt{q}) a  -8 q^2 \sqrt{J-q} J(1,a \sqrt{q}) a \ln(q/L)  \nn &&
+8 q^2 \sqrt{J-q} J(0,a \sqrt{q}) \ln(q/L) J(1,a \sqrt{q}) a +16
\sqrt{q} \sqrt{J-q} J  \nn && +16 \sqrt{q} \sqrt{J-q} J J(0,a
\sqrt{q})^2 -16 J \sqrt{q} \sqrt{J-q} J(0,a \sqrt{q}) \ln(q/L) \nn
&& -16 q^{3/2} \sqrt{J-q} +16 q^{3/2} \sqrt{J-q} J(0,a \sqrt{q})
\ln(q/L) -16 q^{3/2} \sqrt{J-q} J(0,a \sqrt{q})^2  \nn && -32
\sqrt{q} \sqrt{J-q} J J(0,a \sqrt{q}) +32 q^{3/2} \sqrt{J-q} J(0,a
\sqrt{q}).\ea

It turns out that the factor of the $l(p)^2$ terms can be
expressed as

\ba n^{(2)G}_{S}&=& (q/L)^{p} n^{(21)G}_{S} +(q/L)^{2p}
n^{(22)G}_{S}+ ((q/L)^{3p} n^{(23)G}_{S}) \ea

with the following factorizations

\ba n^{(21)G}_{S}&=& n^{(210)G}_{S}+ n^{(211)G}_{S}f+
n^{(212)G}_{S}f^2,\nn
n^{(22)G}_{S}&=& n^{(220)G}_{S}+ n^{(221)G}_{S}f+
n^{(222)G}_{S}f^2,\nn
n^{(23)G}_{S}&=& n^{(230)G}_{S}+ n^{(231)G}_{S}f+
n^{(232)G}_{S}f^2 \ea

and sub-factorizations of the following specific forms

\ba n^{(211)G}_{S}&=& 4 p n^{(2111)G}_{S}+ 8 p^2 n^{(2112)G}_{S}+
2 p^3 n^{(2113)G}_{S}, \nn
n^{(212)G}_{S}&=& 2 p n^{(2121)G}_{S}+ 2p^2 n^{(2122)G}_{S}+
p^3n^{(2123)G}_{S}.\ea

The precise sub-factors of the $21$-factorizations are

\ba n^{(210)G}_{S}&=& -8 \sqrt{q} \sqrt{J-q} J p( 1+ J(0,a
\sqrt{q})^2+p \ln(q/L) - 2 J(0,a \sqrt{q}) \nn && +p J(0,a
\sqrt{q})^2 \ln(q/L) -2p J(0,a \sqrt{q}) \ln(q/L)), \nn
n^{(2111)G}_{S}&=& 2 J^{3/2} \sqrt{q} J(0,a \sqrt{q})^2 +2 J^{3/2}
\sqrt{q} -2 q^{3/2} \sqrt{J} J(0,a \sqrt{q})^2  \nn && -3 q^{3/2}
\sqrt{J} -4J^{3/2} \sqrt{q} J(0,a \sqrt{q}) +6 q^{3/2} \sqrt{J}
J(0,a \sqrt{q}), \nn
n^{(2112)G}_{S}&=& J^{3/2} q J(1,a \sqrt{q}) a J(0,a \sqrt{q})
\ln(q/L) + q^2 \sqrt{J} J(1,a \sqrt{q}) a \ln(q/L)  \nn && -
J^{3/2} q J(1,a \sqrt{q}) a \ln(q/L) - q^2 \sqrt{J} J(1,a
\sqrt{q}) a J(0,a \sqrt{q}) \ln(q/L)  \nn && +2 J^{3/2} \sqrt{q}
J(0,a \sqrt{q})^2 \ln(q/L) +2 J^{3/2} \sqrt{q} \ln(q/L)  \nn && -3
q^{3/2} \sqrt{J} J(0,a \sqrt{q})^2 \ln(q/L) -3 q^{3/2} \sqrt{J}
J(0,a \sqrt{q})^2  \nn && -3 q^{3/2} \sqrt{J} -3 q^{3/2} \sqrt{J}
\ln(q/L) +3 J^{3/2} \sqrt{q}  \nn && +3 J^{3/2} \sqrt{q} J(0,a
\sqrt{q})^2 -4 J^{3/2} \sqrt{q} J(0,a \sqrt{q}) \ln(q/L)  \nn &&
-6 J^{3/2} \sqrt{q} J(0,a \sqrt{q}) +6 q^{3/2} \sqrt{J} J(0,a
\sqrt{q})  \nn && +6 q^{3/2} \sqrt{J} J(0,a \sqrt{q}) \ln(q/L),
\nn
n^{(2113)G}_{S}&=& 2 J^{3/2} \sqrt{q} \ln(q/L)^2 +2 q^2 \sqrt{J}
J(1,a \sqrt{q}) a \ln(q/L)^2  \nn && -2 J^{3/2} q J(1,a \sqrt{q})
a \ln(q/L)^2 -2 q^2 \sqrt{J} J(1,a \sqrt{q}) a J(0,a \sqrt{q})
\ln(q/L)^2  \nn && +2 J^{3/2} q J(1,a \sqrt{q}) a J(0,a \sqrt{q})
\ln(q/L)^2 -3 q^{3/2} \sqrt{J} \ln(q/L)^2  \nn && -3 q^{3/2}
\sqrt{J} J(0,a \sqrt{q})^2 \ln(q/L)^2 +4 J^{3/2} \sqrt{q} \ln(q/L)
\nn && -4 J^{3/2} \sqrt{q} J(0,a \sqrt{q}) \ln(q/L)^2 -4 q^{3/2}
\sqrt{J} \ln(q/L)  \nn && -4 q^{3/2} \sqrt{J} J(0,a \sqrt{q})^2
\ln(q/L) +4 J^{3/2} \sqrt{q} J(0,a \sqrt{q})^2 \ln(q/L)  \nn && +6
q^{3/2} \sqrt{J} J(0,a \sqrt{q}) \ln(q/L)^2 +6 q^{3/2} \sqrt{J}
J(0,a \sqrt{q}) \ln(q/L)  \nn && +2 J^{3/2} \sqrt{q} J(0,a
\sqrt{q})^2 \ln(q/L)^2 -8 J^{3/2} \sqrt{q} J(0,a \sqrt{q})
\ln(q/L). \ea

For the $212$-components, we have the following factors

\ba n^{(2121)G}_{S}&=&  \sqrt{J-q}( q^{5/2} a^2 J(1,a \sqrt{q})^2
+ J q^{3/2} a^2 J(0,a \sqrt{q}) + q^(5/2) a^2 J(0,a \sqrt{q})^2
\nn && - q^(5/2) a^2 J(0,a \sqrt{q}) - J q^{3/2} a^2 J(0,a
\sqrt{q})^2 - q^{3/2} J a^2 J(1,a \sqrt{q})^2  \nn && +2 J q a
J(1,a \sqrt{q}) J(0,a \sqrt{q}) -2 q^2 a J(1,a \sqrt{q}) J(0,a
\sqrt{q}) +2 q^2 a J(1,a \sqrt{q})  \nn && -2 J q a J(1,a
\sqrt{q})), \nn
n^{(2122)G}_{S}&=& \sqrt{J-q} ( q^{5/2} a^2 J(1,a \sqrt{q})^2
\ln(q/L) - q^{3/2} J a^2 J(1,a \sqrt{q})^2 \ln(q/L)  \nn && -2
q^{5/2} a^2 J(0,a \sqrt{q}) \ln(q/L) -2 J q^{3/2} a^2 J(0,a
\sqrt{q})^2 \ln(q/L)  \nn && +2 q^{5/2} a^2 J(0,a \sqrt{q})^2
\ln(q/L) -4 q^2 a J(1,a \sqrt{q}) J(0,a \sqrt{q}) \ln(q/L)  \nn &&
+4 J q a J(1,a \sqrt{q}) J(0,a \sqrt{q}) \ln(q/L) +4 q J a J(1,a
\sqrt{q}) J(0,a \sqrt{q})  \nn && +2 J q^{3/2} a^2 J(0,a \sqrt{q})
\ln(q/L) -4 J q a J(1,a \sqrt{q}) \ln(q/L)  \nn && +4 q^2 a J(1,a
\sqrt{q}) \ln(q/L) -6 q J a J(1,a \sqrt{q}) \nn &&  -6 q^2 a J(1,a
\sqrt{q}) J(0,a \sqrt{q}) +6 q^2 a J(1,a \sqrt{q})), \nn
n^{(2123)G}_{S}&=& \sqrt{J-q} ( q^{5/2} a^2 J(0,a \sqrt{q})^2
\ln(q/L)^2 -q^{5/2} a^2 J(0,a \sqrt{q}) \ln(q/L)^2  \nn && +J
q^{3/2} a^2 J(0,a \sqrt{q}) \ln(q/L)^2 -J q^{3/2} a^2 J(0,a
\sqrt{q})^2 \ln(q/L)^2  \nn && -2 q^2 a J(1,a \sqrt{q}) J(0,a
\sqrt{q}) \ln(q/L)^2 +2 J q a J(1,a \sqrt{q}) J(0,a \sqrt{q})
\ln(q/L)^2  \nn && +2 q^2 a J(1,a \sqrt{q}) \ln(q/L)^2 -2 J q a
J(1,a \sqrt{q}) \ln(q/L)^2  \nn && +4 q^2 a J(1,a \sqrt{q})
\ln(q/L) +4 q J a J(1,a \sqrt{q}) J(0,a \sqrt{q}) \ln(q/L)  \nn &&
-4 q J a J(1,a \sqrt{q}) \ln(q/L) -4 q^2 a J(1,a \sqrt{q}) J(0,a
\sqrt{q}) \ln(q/L)).\ea

Similarly, it follows that the $22$- factors obey the following
particular relations

\ba n^{(221)G}_{S}&=& 2 p^2 n^{(2121)G}_{S} + 4p^3 n^{(2122)G}_{S}
+2p^4 n^{(2213)G}_{S}, \nn
n^{(222)G}_{S}&=& p^2 n^{(2221)G}_{S} + 2 p^3 n^{(2222)G}_{S} +p^4
n^{(2223)G}_{S}. \ea

In the above relations, the factors of the $220$ and
$221$-components are given by

\ba n^{(220)G}_{S}&=& 16J \sqrt{q} \sqrt{J-q} p^2( 2 J(0,a
\sqrt{q}) +2 p  J(0,a \sqrt{q}) \ln(q/L) \nn && - p  J(0,a
\sqrt{q})^2 \ln(q/L) - p \ln(q/L)- J(0,a \sqrt{q})^2 -1), \nn
n^{(2121)G}_{S}&=& \sqrt{J} (2 J q J(1,a \sqrt{q}) a -2 q^2 J(1,a
\sqrt{q}) a +2 q^2 J(1,a \sqrt{q}) a J(0,a \sqrt{q})  \nn && -2 J
q J(1,a \sqrt{q}) a J(0,a \sqrt{q}) +6 J \sqrt{q} J(0,a
\sqrt{q})^2 -9 q^{3/2} J(0,a \sqrt{q})^2 \nn && +6 J \sqrt{q} -9
q^{3/2} -12 J \sqrt{q} J(0,a \sqrt{q}) +18 q^{3/2} J(0,a
\sqrt{q})), \nn
n^{(2122)G}_{S}&=& \sqrt{J} ( 2 q^2 J(1,a \sqrt{q}) a \ln(q/L) -2
J q J(1,a \sqrt{q}) a \ln(q/L) -40 J \sqrt{q} J(0,a \sqrt{q}) \nn
&& +2 J q J(0,a \sqrt{q}) \ln(q/L) J(1,a \sqrt{q}) a -2 q^2 J(0,a
\sqrt{q}) \ln(q/L) J(1,a \sqrt{q}) a  \nn && +12 J \sqrt{q}
\ln(q/L) +12 J \sqrt{q} J(0,a \sqrt{q})^2 \ln(q/L) -18 q^{3/2}
J(0,a \sqrt{q})^2 \ln(q/L)  \nn && -18 q^{3/2} \ln(q/L) +20 J
\sqrt{q} -20 q^{3/2} +20 J \sqrt{q} J(0,a \sqrt{q})^2 -20 q^{3/2}
J(0,a \sqrt{q})^2  \nn && -24 J \sqrt{q} J(0,a \sqrt{q}) \ln(q/L)
+36 q^{3/2} J(0,a \sqrt{q}) \ln(q/L) +40 q^{3/2} J(0,a \sqrt{q})),
\nn
n^{(2213)G}_{S}&=&\sqrt{J} ( 2 q^2 \ln(q/L)^2 J(1,a \sqrt{q}) a +2
J \sqrt{q} \ln(q/L)^2 -4 q^{3/2} \ln(q/L) \nn && -2 J q \ln(q/L)^2
J(1,a \sqrt{q}) a +2 J q J(0,a \sqrt{q}) \ln(q/L)^2 J(1,a
\sqrt{q}) a \nn &&  -2 q^2 J(0,a \sqrt{q}) \ln(q/L)^2 J(1,a
\sqrt{q}) a +2 J \sqrt{q} J(0,a \sqrt{q})^2 \ln(q/L)^2  \nn && -3
q^{3/2} \ln(q/L)^2 -3 q^{3/2} J(0,a \sqrt{q})^2 \ln(q/L)^2 +4 J
\sqrt{q} \ln(q/L)  \nn && -4 J \sqrt{q} J(0,a \sqrt{q}) \ln(q/L)^2
+4 J \sqrt{q} J(0,a \sqrt{q})^2 \ln(q/L) \nn &&  -4 q^{3/2} J(0,a
\sqrt{q})^2 \ln(q/L) +6 q^{3/2} J(0,a \sqrt{q}) \ln(q/L)^2  \nn &&
-8 J \sqrt{q} J(0,a \sqrt{q}) \ln(q/L) +8 q^{3/2} J(0,a \sqrt{q})
\ln(q/L)).\ea

In this case, we further obtain that the $222$-factors are

\ba n^{(2221)G}_{S}&=& a q( 3 q^{3/2} a \sqrt{J-q} J(0,a
\sqrt{q})^2 +3 J q^{1/2} a \sqrt{J-q} J(0,a \sqrt{q})  \nn && -3
q^{3/2} a \sqrt{J-q} J(0,a \sqrt{q}) -3 J q^{1/2} a \sqrt{J-q}
J(0,a \sqrt{q})^2  \nn && +4 q^{3/2} \sqrt{J-q} J(1,a \sqrt{q})^2
a -4 q^{1/2} \sqrt{J-q} J J(1,a \sqrt{q})^2 a  \nn && -6 q
\sqrt{J-q} J(1,a \sqrt{q}) J(0,a \sqrt{q}) -6 J \sqrt{J-q} J(1,a
\sqrt{q})  \nn && +6 J \sqrt{J-q} J(1,a \sqrt{q}) J(0,a \sqrt{q})
+6 q \sqrt{J-q} J(1,a \sqrt{q})), \nn
n^{(2222)G}_{S}&=& a \sqrt{J-q}( 2 q^{5/2} J(1,a \sqrt{q})^2 a
\ln(q/L) -2 q^{3/2} J J(1,a \sqrt{q})^2 a \ln(q/L)  \nn && +3
q^(5/2) a J(0,a \sqrt{q})^2 \ln(q/L) -3  q^{5/2} a J(0,a \sqrt{q})
\ln(q/L) \nn &&  -3 J q^{3/2} a J(0,a \sqrt{q})^2 \ln(q/L) +3 J
q^{3/2} a J(0,a \sqrt{q}) \ln(q/L)  \nn && +6 J q J(1,a \sqrt{q})
J(0,a \sqrt{q}) \ln(q/L) -6 q^2 J(1,a \sqrt{q}) J(0,a \sqrt{q})
\ln(q/L)  \nn && -6 J q J(1,a \sqrt{q}) \ln(q/L) +6 q^2 J(1,a
\sqrt{q}) \ln(q/L)  -10 q  J J(1,a \sqrt{q}) \nn && +10 q^2 J(1,a
\sqrt{q}) +10 q J J(1,a \sqrt{q}) J(0,a \sqrt{q}) -10 q^2 J(1,a
\sqrt{q}) J(0,a \sqrt{q})), \nn
n^{(2223)G}_{S}&=& a \sqrt{J-q}( q^{5/2} a J(0,a \sqrt{q})^2
\ln(q/L)^2 - q^{5/2} a J(0,a \sqrt{q}) \ln(q/L)^2  \nn && +J
q^{3/2} a J(0,a \sqrt{q}) \ln(q/L)^2 -J q^{3/2} a J(0,a
\sqrt{q})^2 \ln(q/L)^2  \nn && +2 q^2 J(1,a \sqrt{q}) \ln(q/L)^2
-2 q^2 J(1,a \sqrt{q}) J(0,a \sqrt{q}) \ln(q/L)^2  \nn && -2 J q
J(1,a \sqrt{q}) \ln(q/L)^2 +2 J q J(1,a \sqrt{q}) J(0,a \sqrt{q})
\ln(q/L)^2  \nn && -4 q J J(1,a \sqrt{q}) \ln(q/L) +4 q J J(1,a
\sqrt{q}) J(0,a \sqrt{q}) \ln(q/L)  \nn && +4 q^2 J(1,a \sqrt{q})
\ln(q/L) -4 q^2 J(1,a \sqrt{q}) J(0,a \sqrt{q}) \ln(q/L)). \ea

After applying a similar technique, it turns out that the $23$
factors are

\ba n^{(231)G}_{S}&=& 2p^3 n^{(2312)G}_{S} + 4p^4 n^{(2313)G}_{S},
\nn
n^{(232)G}_{S}&=& p^3 n^{(2323)G}_{S} + 2p^4 n^{(2324)G}_{S}, \ea

where the $230$ and $231$ sub-factors are expressed as

\ba n^{(230)G}_{S}&=& 8J p^3 \sqrt{q} \sqrt{J-q}( 2 J(0,a
\sqrt{q}) +2 p J(0,a \sqrt{q}) \ln(q/L) \nn &&  -  p J(0,a
\sqrt{q})^2 \ln(q/L) -  J(0,a \sqrt{q})^2 -  p \ln(q/L) -  1), \nn
n^{(2312)G}_{S}&=& \sqrt{J} ( 2 J q J(1,a \sqrt{q}) a +2 q^2 J(1,a
\sqrt{q}) a J(0,a \sqrt{q}) +2 J \sqrt{q} l(p)^2 \nn &&  -2 J q
J(1,a \sqrt{q}) a J(0,a \sqrt{q}) -2 q^2 J(1,a \sqrt{q}) a +2 J
\sqrt{q} J(0,a \sqrt{q})^2 \nn &&  -3 q^{3/2} ((q/L)^p)^3 -3
q^{3/2} J(0,a \sqrt{q})^2 -4 J \sqrt{q} J(0,a \sqrt{q}) \nn &&  +6
q^{3/2} J(0,a \sqrt{q})), \nn
n^{(2313)G}_{S}&=& \sqrt{q J} ( 2 J \ln(q/L) +2 J J(0,a
\sqrt{q})^2 \ln(q/L) -3 q J(0,a \sqrt{q})^2 \ln(q/L)  \nn && -3 q
\ln(q/L) -4 J J(0,a \sqrt{q}) \ln(q/L) -4 q +4 J +4 J  J(0,a
\sqrt{q})^2  \nn && -4 q J(0,a \sqrt{q})^2 +6 q J(0,a \sqrt{q})
\ln(q/L) +8 q J(0,a \sqrt{q})  \nn && -8 J J(0,a \sqrt{q})).\ea

While, the $232$ factors come as per the following identifications

\ba n^{(2323)G}_{S}&=& a f^2 \sqrt{J-q} ( q^{5/2} a J(0,a
\sqrt{q})^2 +J q^{3/2} a J(0,a \sqrt{q}) -J q^{3/2} a J(0,a
\sqrt{q})^2  \nn && -q^{5/2} a J(0,a \sqrt{q}) +2 q^{5/2} J(1,a
\sqrt{q})^2 a -2 q^{3/2} J J(1,a \sqrt{q})^2 a  \nn && -2 q^2
J(1,a \sqrt{q}) J(0,a \sqrt{q}) -2 J q J(1,a \sqrt{q})  \nn && +2
J q J(1,a \sqrt{q}) J(0,a \sqrt{q})+2 p^3 q^2 J(1,a \sqrt{q})),
\nn
n^{(2324)G}_{S}&=& a f^2 \sqrt{J-q}( q^{5/2} J(1,a \sqrt{q})^2 a
\ln(q/L) + q^{5/2} a J(0,a \sqrt{q})^2 \ln(q/L)  \nn && - q^{5/2}
a J(0,a \sqrt{q}) \ln(q/L) - q^{3/2}  J J(1,a \sqrt{q})^2 a
\ln(q/L)  \nn && + J q^{3/2} a J(0,a \sqrt{q}) \ln(q/L) - J
q^{3/2} a J(0,a \sqrt{q})^2 \ln(q/L)  \nn && +2 J q J(1,a
\sqrt{q}) J(0,a \sqrt{q}) \ln(q/L) -2 J q J(1,a \sqrt{q}) \ln(q/L)
\nn && -2 q^2 J(1,a \sqrt{q}) J(0,a \sqrt{q}) \ln(q/L) +4 q^2
J(1,a \sqrt{q})  \nn && +4 q J J(0,a \sqrt{q}) J(1,a \sqrt{q}) -4
q^2 J(1,a \sqrt{q}) J(0,a \sqrt{q})  \nn &&+2 q^2 J(1,a \sqrt{q})
\ln(q/L) -4 q J J(1,a \sqrt{q}) ).\ea

We notice further that the factors of the $l(p)^3$-terms can be
expressed as

\ba n^{(3)G}_{S}&=& n^{(30)G}_{S} +(q/L)^{p} n^{(31)G}_{S}
+(q/L)^{2p} n^{(32)G}_{S}+ ((q/L)^{3p} n^{(33)G}_{S}). \ea

Following similar factorizations, we have the following terms

\ba n^{(30)G}_{S}&=&  n^{(300)G}_{S}+ n^{(301)G}_{S}f+
n^{(302)G}_{S}f^2,\nn
n^{(31)G}_{S}&=& n^{(310)G}_{S}+ n^{(311)G}_{S}f+
n^{(312)G}_{S}f^2,\nn
n^{(32)G}_{S}&=& n^{(320)G}_{S}+ n^{(321)G}_{S}f+
n^{(322)G}_{S}f^2,\nn
n^{(33)G}_{S}&=& n^{(330)G}_{S}+ n^{(331)G}_{S}f+
n^{(332)G}_{S}f^2, \ea

which factor nicely. As per the expectation, we find that the
$30$-factors are

\ba n^{(300)G}_{S}&=& 4 \sqrt{q} \sqrt{J-q} J(1-2 J(0,a \sqrt{q})+
J(0,a \sqrt{q})^2),\nn
n^{(301)G}_{S}&=& 4 a \sqrt{J} J(1,a \sqrt{q})( q^2 - q^2 J(0,a
\sqrt{q}) - J q + J q J(0,a \sqrt{q})), \nn
n^{(302)G}_{S}&=& f^2 a^2 q^{3/2}  \sqrt{J-q} J(1,a \sqrt{q})^2 (
J -q).\ea

Similarly, the $31$-factors are  obtained as

%n^{(31)G}_{S}&=& n^{(310)G}_{S}+ n^{(311)G}_{S}f+
%n^{(312)G}_{S}f^2,\nn

\ba n^{(310)G}_{S}&=& 12 \sqrt{q} \sqrt{J-q} p J  (1+J(0,a
\sqrt{q})^2-2J(0,a \sqrt{q})), \nn
n^{(311)G}_{S}&=& 12 q p a \sqrt{J}( q  J(1,a \sqrt{q})-q J(1,a
\sqrt{q}) J(0,a \sqrt{q}) \nn && -J J(1,a \sqrt{q}) +J J(1,a
\sqrt{q}) J(0,a \sqrt{q})), \nn
n^{(312)G}_{S}&=&3 q^{3/2} \sqrt{J-q} f^2 a^2 J(1,a \sqrt{q})^2 p
(J- q). \ea

The $32$-factors reduce to

%n^{(32)G}_{S}&=& n^{(320)G}_{S}+ n^{(321)G}_{S}f+
%n^{(322)G}_{S}f^2,\nn

\ba n^{(320)G}_{S}&=& 12 \sqrt{q} \sqrt{J-q} p^2 J(1+  J(0,a
\sqrt{q})^2- 2 J(0,a \sqrt{q})),\nn
n^{(321)G}_{S}&=& 12a q  p^2 \sqrt{J}( q  J(1,a \sqrt{q}) - q
J(1,a \sqrt{q}) J(0,a \sqrt{q})  \nn && + J J(1,a \sqrt{q}) J(0,a
\sqrt{q}) - J J(1,a \sqrt{q})), \nn
n^{(322)G}_{S}&=&3 q^{3/2} \sqrt{J-q} f^2 a^2 J(1,a \sqrt{q})^2
p^2 (J -q).\ea

Finally, it follows that the $33$-factors are given by

%n^{(33)G}_{S}&=& n^{(330)G}_{S}+ n^{(331)G}_{S}f+
%n^{(332)G}_{S}f^2. \ea

\ba n^{(330)G}_{S}&=& 4 \sqrt{q} \sqrt{J-q} p^3 J (J(0,a
\sqrt{q})^2 -2 J J(0,a \sqrt{q}) +1), \nn
n^{(331)G}_{S}&=& 4a q p^3 \sqrt{J} J(1,a \sqrt{q}) ( q- q J(0,a
\sqrt{q}) - J + J  J(0,a \sqrt{q}) ), \nn
n^{(332)G}_{S}&=& +q^{3/2} \sqrt{J-q} f^2 a^2 J(1,a \sqrt{q})^2
p^3 (J-q). \ea

For the general quarkonia with $\{q,p,J\}$ fluctuating, the
stability of the $qp$-surface requires that the principle minor
$p_S^G $ remains positive on $(M_3,g)$. Correspondingly, this
leads to the constraint that the $\{q,p,J\}$ satisfy \ba
n^{(0)G}_{S}+ n^{(1)G}_{g}l(p)+ n^{(2)G}_{S}l(p)^2+
n^{(3)G}_{S}l(p)^3 &<& 0. \ea

\begin{figure}
\hspace*{0.5cm}
\includegraphics[width=8.0cm,angle=-90]{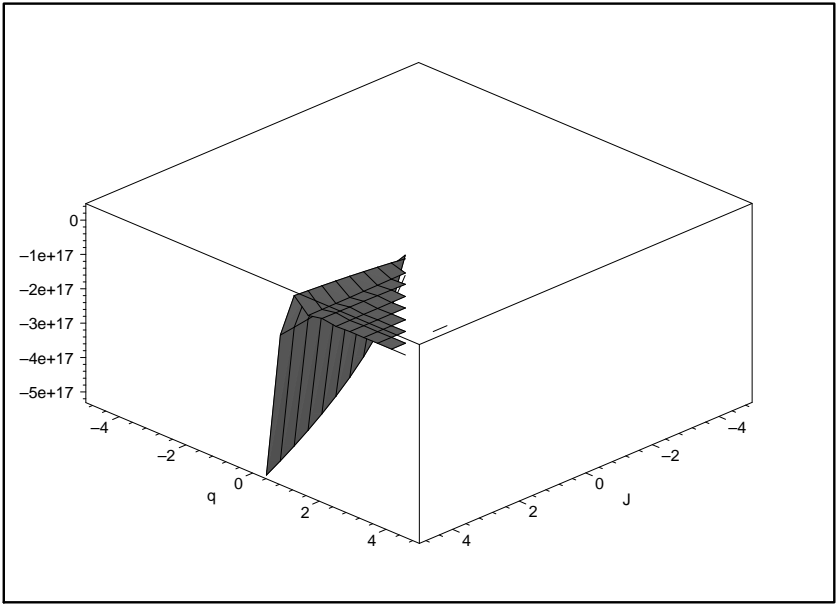}
\caption{The surface minor $p_S^G$ plotted as a function of the
scale and angular momentum, viz., $q, J$, describing the nature of
the stability of the $qp$-surface for general massive rotating
quarkonia.} \label{minor3dqpJ} \vspace*{0.5cm}
\end{figure}

The three dimensional graphical views of the surface minor $p_S^G$
of generic quarkonia is depicted in the Fig.(\ref{minor3dqpJ}).
For the limiting small scale $ q \rightarrow 0^{+} $, the
graphical view of the surface minor $p_S^G$ shows a large negative
minima of the order of the height $-10^{17}$, which occurs in the
regime of a large angular momentum $J \rightarrow 5^{-}$. When all
the parameters are allowed to fluctuate and $b$ is an arbitrary
real number, the exact intrinsic geometric characterization of the
minor is explicitly obtained in the Eqn.(\ref{minor}), which forms
the basis for the stability of the $qp$-surface.

After considering the contributions of the fluctuations to the
strongly coupled massive rotating quarkonia, we obtain the
following general expressions for the determinant of the metric
tensor

\ba \Vert g \Vert = -\frac{p f  (-1+ J(0,a \sqrt{q}))}{8 b^3
l(p)^6 \exp{(3 l(p))} q^{5/2} J^{3/2} (J-q)^{3/2}}
(n^{(0)G}_{g}+ n^{(1)G}_{g}l(p)+ n^{(2)G}_{g}l(p)^2+
n^{(3)G}_{g}l(p)^3), \ea

where the $n^{(0)G}_{g}$-term factorizes as

\ba n^{(0)G}_{g}&=& ((q/L)^p)^3 (8p^4 n^{(04)G}_{g}+ 8p^5
n^{(05)G}_{g}+ 8p^6 n^{(06)G}_{g}). \ea

After simplification, we obtain the sub-factorizations

\ba n^{(04)G}_{g}&=& n^{(040)G}_{g}+ f n^{(041)G}_{g}, \nn
n^{(05)G}_{g}&=& n^{(050)G}_{g}+ f n^{(051)G}_{g},\nn
n^{(06)G}_{g}&=& n^{(060)G}_{g}+ f n^{(061)G}_{g}.\ea

As per our computation, the $\{n^{(040)G}_{g}, n^{(041)G}_{g},
n^{(050)G}_{g}, n^{(051)G}_{g}, n^{(060)G}_{g}, n^{(061)G}_{g}\}$
are expressed as

\ba n^{(040)G}_{g}&=& \sqrt{J} \sqrt{J-q} \sqrt{q} (2J(0,a
\sqrt{q})- J(0,a \sqrt{q})^2-1), \nn
n^{(041)G}_{g}&=& (q^{3/2} + q^{3/2} J(0,a \sqrt{q})^2 -2 q^{3/2}
J(0,a \sqrt{q}) \nn && -2 J \sqrt{q} J(0,a \sqrt{q})^2 -2 J
\sqrt{q} +4 J \sqrt{q} J(0,a \sqrt{q})), \ea

\ba n^{(050)G}_{g}&=& \sqrt{J} \sqrt{J-q} \sqrt{q} (4J(0,a
\sqrt{q}) \ln(q/L) +6  J(0,a \sqrt{q}) \nn &&  -2  \ln(q/L) -2
J(0,a \sqrt{q})^2 \ln(q/L) -3 -3 J(0,a \sqrt{q})^2),\nn
n^{(051)G}_{g}&=& 3  q^{3/2} +3  q^{3/2} J(0,a \sqrt{q})^2 +2
q^{3/2} J(0,a \sqrt{q})^2 \ln(q/L)  \nn && +2 q^{3/2} \ln(q/L) -6
q^{3/2} J(0,a \sqrt{q}) +8 J \sqrt{q} J(0,a \sqrt{q}) \ln(q/L) \nn
&&  +12 J \sqrt{q} J(0,a \sqrt{q}) -4 J \sqrt{q} \ln(q/L) -6 J
\sqrt{q} J(0,a \sqrt{q})^2  \nn && -4 J \sqrt{q} J(0,a \sqrt{q})^2
\ln(q/L)-4 q^{3/2} J(0,a \sqrt{q}) \ln(q/L) -6 J \sqrt{q}, \ea

and

\ba n^{(060)G}_{g}&=& \sqrt{J} \sqrt{J-q} \sqrt{q} (4 J(0,a
\sqrt{q}) \ln(q/L) +2   J(0,a \sqrt{q}) \ln(q/L)^2 \nn && -2
\ln(q/L) - J(0,a \sqrt{q})^2 \ln(q/L)^2 -2 J(0,a \sqrt{q})^2
\ln(q/L)  \nn && - \ln(q/L)^2), \nn
n^{(061)G}_{g}&=& q^{3/2} \ln(q/L)^2 + q^{3/2} J(0,a \sqrt{q})^2
\ln(q/L)^2 +2 q^{3/2} \ln(q/L)  \nn && +2 q^{3/2} J(0,a
\sqrt{q})^2 \ln(q/L) -2 J \sqrt{q} J(0,a \sqrt{q})^2 \ln(q/L)^2
\nn && -2 q^{3/2} J(0,a \sqrt{q}) \ln(q/L)^2 -4 q^{3/2} J(0,a
\sqrt{q}) \ln(q/L)  \nn && -4 J \sqrt{q} J(0,a \sqrt{q})^2
\ln(q/L) +4 J \sqrt{q} J(0,a \sqrt{q}) \ln(q/L)^2 \nn && -2 J
\sqrt{q} \ln(q/L)^2 +8 J \sqrt{q} J(0,a \sqrt{q}) \ln(q/L).\ea

We find that the $l(p)$ terms factorize as follows

\ba n^{(1)G}_{g}&=& ((q/L)^p)^2 n^{(12)G}_{g}+ ((q/L)^p)^3
n^{(13)G}_{g} \ea

with the following sub-factorizations

\ba n^{(12)G}_{g}&=& n^{(120)G}_{g}+ f n^{(121)G}_{g}, \nn
n^{(13)G}_{g}&=& n^{(130)G}_{g}+ f n^{(131)G}_{g}.\ea

The factors of the $12$-components are given by

\ba n^{(120)G}_{g}&=& n^{(1202)G}_{g} p^2+  n^{(1203)G}_{g} p^3+
n^{(1204)G}_{g} p^4, \nn
n^{(121)G}_{g}&=& n^{(1212)G}_{g} p^2 +n^{(1213)G}_{g} p^3
+n^{(1214)G}_{g} p^4 +n^{(1215)G}_{g} p^5 \ea

with the following sub-factorizations

\ba n^{(1202)G}_{g} &=& 2 \sqrt{J} \sqrt{J-q} ( a^2 q^{3/2} J(0,a
\sqrt{q})^2 - a^2 q^{3/2} J(0,a \sqrt{q}) +6 J(1,a \sqrt{q}) a q
\nn && -6 J(1,a \sqrt{q}) a q J(0,a \sqrt{q}) -8 \sqrt{q} J(0,a
\sqrt{q})^2 -8 \sqrt{q} +16 \sqrt{q} J(0,a \sqrt{q})), \nn
n^{(1203)G}_{g}&=& 4 \sqrt{J}  \sqrt{J-q} ( a^2 q^{3/2} J(0,a
\sqrt{q})^2 \ln(q/L) - a^2 q^{3/2} J(0,a \sqrt{q}) \ln(q/L)  \nn
&& +6 J(1,a \sqrt{q}) a q +6 J(1,a \sqrt{q}) a q \ln(q/L) -6 J(1,a
\sqrt{q}) a q J(0,a \sqrt{q}) \ln(q/L)  \nn && -6 q J(1,a
\sqrt{q}) a J(0,a \sqrt{q}) -8 \sqrt{q} \ln(q/L) -8 \sqrt{q} J(0,a
\sqrt{q})^2 \ln(q/L)  \nn && -8 \sqrt{q} J(0,a \sqrt{q})^2 -8
\sqrt{q} +16 \sqrt{q} J(0,a \sqrt{q}) +16 \sqrt{q} J(0,a \sqrt{q})
\ln(q/L)), \nn
n^{(1204)G}_{g}&=& 2 \sqrt{J} \sqrt{J-q} ( a^2 q^{3/2} J(0,a
\sqrt{q})^2 \ln(q/L)^2 - a^2 q^{3/2} J(0,a \sqrt{q}) \ln(q/L)^2
\nn && -6 J(1,a \sqrt{q}) a q J(0,a \sqrt{q}) \ln(q/L)^2 +6 J(1,a
\sqrt{q}) a q \ln(q/L)^2  +8 \sqrt{q} \nn && -8 \sqrt{q} \ln(q/L)
-8 J(1,a \sqrt{q}) a q J(0,a \sqrt{q}) \ln(q/L) +8 \sqrt{q} J(0,a
\sqrt{q})^2  \nn && -8 \sqrt{q} \ln(q/L)^2 -8 \sqrt{q} J(0,a
\sqrt{q})^2 \ln(q/L)^2 -8 \sqrt{q} J(0,a \sqrt{q})^2 \ln(q/L)  \nn
&& +8 q \ln(q/L) J(1,a \sqrt{q}) a -16 \sqrt{q} J(0,a \sqrt{q})
+16 \sqrt{q} J(0,a \sqrt{q}) \ln(q/L)^2  \nn && +16 \sqrt{q} J(0,a
\sqrt{q}) \ln(q/L) ). \ea

The $121$-factors turn out to be

\ba n^{(1212)G}_{g}&=& a q(2 J q^{1/2} J(1,a \sqrt{q})^2 a +2 a
q^{3/2} J(0,a \sqrt{q}) -2 a q^{3/2} J(0,a \sqrt{q})^2  \nn && +4
q J(1,a \sqrt{q}) J(0,a \sqrt{q}) -4 J(1,a \sqrt{q}) q +4 J a
q^{1/2} J(0,a \sqrt{q})^2  \nn && -4 J a q^{1/2} J(0,a \sqrt{q})
-8 J J(1,a \sqrt{q}) J(0,a \sqrt{q}) +8 J J(1,a \sqrt{q}) \nn &&
-q^{3/2} J(1,a \sqrt{q})^2 a ), \nn
n^{(1213)G}_{g}&=& 2( - q^{5/2} J(1,a \sqrt{q})^2 a^2 \ln(q/L) +2
a^2 q^{5/2} J(0,a \sqrt{q}) \ln(q/L)  \nn && -2 a^2 q^{5/2} J(0,a
\sqrt{q})^2 \ln(q/L) +2 J q^{3/2} J(1,a \sqrt{q})^2 a^2 \ln(q/L)
\nn && +4 J(1,a \sqrt{q}) a q^2 J(0,a \sqrt{q}) \ln(q/L) +4 J a^2
q^{3/2} J(0,a \sqrt{q})^2 \ln(q/L)  \nn && -4 q^{3/2} J(0,a
\sqrt{q})^2 -4 J(1,a \sqrt{q}) a q^2 \ln(q/L) -6 J(1,a \sqrt{q}) a
q^2  \nn && -4 J a^2 q^{3/2} J(0,a \sqrt{q}) \ln(q/L) -4 q^{3/2}
+6 J(1,a \sqrt{q}) a q^2 J(0,a \sqrt{q})  \nn && +8 J \sqrt{q} +8
q^{3/2} J(0,a \sqrt{q})-8 J J(1,a \sqrt{q}) a q J(0,a \sqrt{q})
\ln(q/L) \nn && +8 J J(1,a \sqrt{q}) a q \ln(q/L) +8 J \sqrt{q}
J(0,a \sqrt{q})^2 +12 J J(1,a \sqrt{q}) a q \nn && -12 J J(1,a
\sqrt{q}) a q J(0,a \sqrt{q}) -16 J \sqrt{q} J(0,a \sqrt{q}) ),
\nn
n^{(1214)G}_{g}&=& 2 (-q^{5/2} J(1,a \sqrt{q})^2 a^2 \ln(q/L)^2 +2
J q^{3/2} J(1,a \sqrt{q})^2 a^2 \ln(q/L)^2  \nn && +2 a^2 q^{5/2}
J(0,a \sqrt{q}) \ln(q/L)^2 -2 a^2 q^{5/2} J(0,a \sqrt{q})^2
\ln(q/L)^2  \nn && +4 J(1,a \sqrt{q}) a q^2 J(0,a \sqrt{q})
\ln(q/L)^2 -4 a q^2 J(1,a \sqrt{q}) \ln(q/L)^2 \nn && +4 J a^2
q^{3/2} J(0,a \sqrt{q})^2 \ln(q/L)^2 -4 J a^2 q^{3/2} J(0,a
\sqrt{q}) \ln(q/L)^2  \nn && +8 J(1,a \sqrt{q}) a q^2 J(0,a
\sqrt{q}) \ln(q/L) -8 J(1,a \sqrt{q}) a q^2 \ln(q/L)  \nn && +8 J
a q J(1,a \sqrt{q}) \ln(q/L)^2 -8 J J(1,a \sqrt{q}) a q J(0,a
\sqrt{q}) \ln(q/L)^2 \nn &&  -16 q^{3/2} J(0,a \sqrt{q})^2
\ln(q/L) -16 J J(1,a \sqrt{q}) a q J(0,a \sqrt{q}) \ln(q/L)  \nn
&& +16 J J(1,a \sqrt{q}) a q \ln(q/L) -16 q^{3/2} \ln(q/L)-28
q^{3/2} J(0,a \sqrt{q})^2 \nn && -28 q^{3/2} +32 J \sqrt{q}
\ln(q/L) +32 J \sqrt{q} J(0,a \sqrt{q})^2 \ln(q/L) +56 J \sqrt{q}
\nn && +32 q^{3/2} J(0,a \sqrt{q}) \ln(q/L) +56 q^{3/2} J(0,a
\sqrt{q}) +56 J \sqrt{q} J(0,a \sqrt{q})^2  \nn && -64 J \sqrt{q}
J(0,a \sqrt{q}) \ln(q/L) -112 J \sqrt{q} J(0,a \sqrt{q}), \nn
n^{(1215)G}_{g}&=& 4 \sqrt{q} \ln(q/L)( - q \ln(q/L) - q
J(0,a\sqrt{q})^2 \ln(q/L) +2 J \ln(q/L)   \nn && -2 q -2 q J(0,a
\sqrt{q})^2 +2 q J(0,a \sqrt{q}) \ln(q/L) +2 J J(0,a \sqrt{q})^2
\ln(q/L) \nn &&  -4 J J(0,a \sqrt{q}) \ln(q/L) +4 J J(0,a
\sqrt{q})^2 +4 J +4 q J(0,a \sqrt{q})   \nn &&  -8 J J(0,a
\sqrt{q})). \ea

Similarly, the factors of $13$-components are given as

\ba n^{(130)G}_{g}&=& n^{(1303)G}_{g} p^3+  n^{(1304)G}_{g} p^4+
n^{(1305)G}_{g} p^5, \nn
n^{(131)G}_{g}&=& n^{(1313)G}_{g} p^3+ n^{(1314)G}_{g} p^4+
n^{(1315)G}_{g} p^5+ n^{(1316)G}_{g} p^6. \ea

The $130$-sub-factors turn out to be

\ba n^{(1303)G}_{g}&=& 2 \sqrt{J} \sqrt{J-q}( a^2 q^{3/2} J(0,a
\sqrt{q})^2 - a^2 q^{3/2} J(0,a \sqrt{q}) +6 J(1,a \sqrt{q}) a q
\nn && -6 a q J(1,a \sqrt{q}) J(0,a \sqrt{q}) -8 \sqrt{q} -8
\sqrt{q} J(0,a \sqrt{q})^2 +16 \sqrt{q} J(0,a \sqrt{q})),\nn
n^{(1304)G}_{g}&=& 4 \sqrt{J} \sqrt{J-q} ( a^2 q^{3/2} J(0,a
\sqrt{q})^2 \ln(q/L) - a^2 q^{3/2} J(0,a \sqrt{q}) \ln(q/L) \nn &&
+6 q J(1,a \sqrt{q}) a  -6 a q J(1,a \sqrt{q}) J(0,a \sqrt{q})
\ln(q/L) +6 J(1,a \sqrt{q}) a q \ln(q/L)  \nn && -6 q J(1,a
\sqrt{q}) a J(0,a \sqrt{q}) -8 \sqrt{q} -8 \sqrt{q} J(0,a
\sqrt{q})^2 \ln(q/L)-8 \sqrt{q} \ln(q/L) \nn && -8 \sqrt{q} J(0,a
\sqrt{q})^2 +16 \sqrt{q} J(0,a \sqrt{q}) \ln(q/L) +16 \sqrt{q}
J(0,a \sqrt{q})), \nn
n^{(1305)G}_{g}&=& 2 \sqrt{J} \sqrt{J-q} ( a^2 q^{3/2} J(0,a
\sqrt{q})^2 \ln(q/L)^2 - a^2 q^{3/2} J(0,a \sqrt{q}) \ln(q/L)^2
\nn && +6 J(1,a \sqrt{q}) a q \ln(q/L)^2 -6 J(1,a \sqrt{q}) a q
J(0,a \sqrt{q}) \ln(q/L)^2 -8 \sqrt{q} \ln(q/L) \nn && +8 \sqrt{q}
J(0,a \sqrt{q})^2  -8 \sqrt{q} J(0,a \sqrt{q})^2 \ln(q/L)^2 +8
\sqrt{q} -8 \sqrt{q} \ln(q/L)^2  \nn && -8 q J(0,a \sqrt{q})
\ln(q/L) J(1,a \sqrt{q}) a +8 q \ln(q/L) J(1,a \sqrt{q}) a  \nn &&
-8 \sqrt{q} J(0,a \sqrt{q})^2 \ln(q/L) +16 \sqrt{q} J(0,a
\sqrt{q}) \ln(q/L)^2  \nn && +16 \sqrt{q} J(0,a \sqrt{q}) \ln(q/L)
-16 \sqrt{q} J(0,a \sqrt{q}) ).\ea

The $131$-factors take the following explicit expressions

\ba n^{(1313)G}_{g}&=& a q ( -q^{3/2} J(1,a \sqrt{q})^2 a +2 J
q^{1/2} J(1,a \sqrt{q})^2 a +2 a q^{3/2} J(0,a \sqrt{q}) \nn && -2
a q^{3/2} J(0,a \sqrt{q})^2 -4 J(1,a \sqrt{q}) q +4 J(1,a
\sqrt{q}) q J(0,a \sqrt{q})  \nn && -4 J a q^{1/2} J(0,a \sqrt{q})
+4 J a q^{1/2} J(0,a \sqrt{q})^2  +8 J J(1,a \sqrt{q}) \nn && -8 J
J(1,a \sqrt{q}) J(0,a \sqrt{q})), \nn
n^{(1314)G}_{g}&=& 2(-q^{5/2} J(1,a \sqrt{q})^2 a^2 \ln(q/L) -2
q^{3/2} -2 q^{3/2} J(0,a \sqrt{q})^2  \nn && +2 a^2 q^{5/2} J(0,a
\sqrt{q}) \ln(q/L) -2 a^2 q^{5/2} J(0,a \sqrt{q})^2 \ln(q/L)  \nn
&& +2 J q^{3/2} J(1,a \sqrt{q})^2 a^2 \ln(q/L) +4 J \sqrt{q} +4 J
\sqrt{q} J(0,a \sqrt{q})^2  \nn && +4 J a^2 q^{3/2} J(0,a
\sqrt{q})^2 \ln(q/L) +4 a q^2 J(1,a \sqrt{q}) J(0,a \sqrt{q})
\ln(q/L)  \nn && -4 J(1,a \sqrt{q}) a q^2 \ln(q/L) -4 J a^2
q^{3/2} J(0,a \sqrt{q}) \ln(q/L) \nn &&  +4 q^{3/2} J(0,a
\sqrt{q}) -6 J(1,a \sqrt{q}) a q^2 +6 J(1,a \sqrt{q}) a q^2 J(0,a
\sqrt{q})  \nn && -8 J \sqrt{q} J(0,a \sqrt{q}) +8 J J(1,a
\sqrt{q}) a q \ln(q/L)+12 J J(1,a \sqrt{q}) a q \nn && -8 J a q
J(1,a \sqrt{q}) J(0,a \sqrt{q}) \ln(q/L) -12 J J(1,a \sqrt{q}) a q
J(0,a \sqrt{q})), \nn
n^{(1315)G}_{g}&=& -q^{5/2} J(1,a \sqrt{q})^2 a^2 \ln(q/L)^2 +2
a^2 q^{5/2} J(0,a \sqrt{q}) \ln(q/L)^2 \nn &&  -2 a^2 q^{5/2}
J(0,a \sqrt{q})^2 \ln(q/L)^2) +2 J q^{3/2} J(1,a \sqrt{q})^2 a^2
\ln(q/L)^2  \nn && +4 J(1,a \sqrt{q}) a q^2 J(0,a \sqrt{q})
\ln(q/L)^2 -4 J(1,a \sqrt{q}) a q^2 \ln(q/L)^2  \nn && +4 J a^2
q^{3/2} J(0,a \sqrt{q})^2 \ln(q/L)^2 -4 J a^2 q^{3/2} J(0,a
\sqrt{q}) \ln(q/L)^2  \nn && +8 J J(1,a \sqrt{q}) a q \ln(q/L)^2
-8 J J(1,a \sqrt{q}) a q J(0,a \sqrt{q}) \ln(q/L)^2  \nn && +8
J(1,a \sqrt{q}) a q^2 J(0,a \sqrt{q}) \ln(q/L) -8 J(1,a \sqrt{q})
a q^2 \ln(q/L)  \nn && -8 q^{3/2} J(0,a \sqrt{q})^2 \ln(q/L) -8
q^{3/2} \ln(q/L) +16 J \sqrt{q} \ln(q/L) \nn && +16 q^{3/2} J(0,a
\sqrt{q}) \ln(q/L) +16 J J(1,a \sqrt{q}) a q \ln(q/L) -16 q^{3/2}
\nn && -16 J J(1,a \sqrt{q}) a q J(0,a \sqrt{q}) \ln(q/L) +16 J
\sqrt{q} J(0,a \sqrt{q})^2 \ln(q/L) \nn && -16 q^{3/2} J(0,a
\sqrt{q})^2 +32 J \sqrt{q} J(0,a \sqrt{q})^2 +32 q^{3/2} J(0,a
\sqrt{q})  \nn &&  +32 J \sqrt{q} -32 J \sqrt{q} J(0,a \sqrt{q})
\ln(q/L) -64 J \sqrt{q} J(0,a \sqrt{q}),\nn
n^{(1316)G}_{g}&=& -32 J \sqrt{q} \ln(q/L).\ea

%l(p)^2 terms
It follows that the $l(p)^2$ terms factorize as

\ba n^{(2)G}_{g}&=& (q/L)^p n^{(21)G}_{g} +((q/L)^p)^2
n^{(22)G}_{g}+ ((q/L)^p)^3 n^{(23)G}_{g}, \ea

where the sub-factorizations take the following forms

\ba n^{(21)G}_{g}&=& n^{(210)G}_{g}+ f n^{(211)G}_{g}, \nn
n^{(22)G}_{g}&=& n^{(220)G}_{g}+ f n^{(221)G}_{g}, \nn
n^{(23)G}_{g}&=& n^{(230)G}_{g}+ f n^{(231)G}_{g}.\ea

Explicitly, we find that the individual factors appear as

\ba n^{(210)G}_{g}&=& n^{(2101)G}_{g} p+  n^{(2102)G}_{g} p^2+
n^{(2103)G}_{g} p^3, \nn
n^{(211)G}_{g}&=& n^{(2111)G}_{g} p+  n^{(2112)G}_{g} p^2+
n^{(2113)G}_{g} p^3.\ea

The $210$-factors are given by

\ba n^{(2101)G}_{g}&=& 4 \sqrt{J} \sqrt{J-q}( a^2 q^{3/2} J(0,a
\sqrt{q}) - q^{3/2} J(1,a \sqrt{q})^2 a^2 - a^2 q^{3/2} J(0,a
\sqrt{q})^2  \nn && -2 J(1,a \sqrt{q}) a q +2 J(1,a \sqrt{q}) a q
J(0,a \sqrt{q}) +4 \sqrt{q} +4 \sqrt{q} J(0,a \sqrt{q})^2  \nn &&
-8 \sqrt{q} J(0,a \sqrt{q}))),\nn
n^{(2102)G}_{g}&=& 4 \sqrt{J} \sqrt{J-q}( a^2 q^{3/2} J(0,a
\sqrt{q}) \ln(q/L) - a^2 q^{3/2} J(0,a \sqrt{q})^2 \ln(q/L)  \nn
&& +6 J(1,a \sqrt{q}) a q J(0,a \sqrt{q}) \ln(q/L) +6 p^2 J(1,a
\sqrt{q}) a q J(0,a \sqrt{q})  \nn && -6 J(1,a \sqrt{q}) a q -6
J(1,a \sqrt{q}) a q \ln(q/L) +8 \sqrt{q} \ln(q/L)  \nn && +8
\sqrt{q} J(0,a \sqrt{q})^2 \ln(q/L) +10 \sqrt{q} +10 \sqrt{q}
J(0,a \sqrt{q})^2  \nn && -16 \sqrt{q} J(0,a \sqrt{q}) \ln(q/L)
-20 \sqrt{q} J(0,a \sqrt{q})), \nn
n^{(2103)G}_{g}&=& 2 \sqrt{J} \sqrt{J-q}(  q^{3/2} J(1,a
\sqrt{q})^2 a^2 \ln(q/L)^2 +4 J(1,a \sqrt{q}) a q J(0,a \sqrt{q})
\ln(q/L)^2  \nn && +4 q J(0,a \sqrt{q}) \ln(q/L) J(1,a \sqrt{q}) a
+4 \sqrt{q} J(0,a \sqrt{q})^2 +4 \sqrt{q} \ln(q/L)^2  \nn && -4 q
\ln(q/L) J(1,a \sqrt{q}) a -4 J(1,a \sqrt{q}) a q \ln(q/L)^2 -8
\sqrt{q} J(0,a \sqrt{q}) \nn && +4 \sqrt{q} +4 \sqrt{q} J(0,a
\sqrt{q})^2 \ln(q/L)^2 +8 \sqrt{q} \ln(q/L) +8 \sqrt{q} J(0,a
\sqrt{q})^2 \ln(q/L) \nn && -8 \sqrt{q} J(0,a \sqrt{q}) \ln(q/L)^2
-16 \sqrt{q} J(0,a \sqrt{q}) \ln(q/L)).\ea

While, the $211$-factors are given as

\ba n^{(2111)G}_{g}&=& 2 a ( a q^{5/2} J(0,a \sqrt{q})^2 - a
q^{5/2} J(0,a \sqrt{q}) + q^{5/2} J(1,a \sqrt{q})^2 a  \nn && +2
J(1,a \sqrt{q}) q^2 -2 J(1,a \sqrt{q}) q^2 J(0,a \sqrt{q}) -2 J
q^{3/2} J(1,a \sqrt{q})^2 a  \nn && -2 J a q^{3/2} J(0,a
\sqrt{q})^2 +2 J p a q^{3/2} J(0,a \sqrt{q}) -4 J J(1,a \sqrt{q})
q  \nn && +4 J J(1,a \sqrt{q}) q J(0,a \sqrt{q})),\nn
n^{(2112)G}_{g}&=& 2 a( q^{5/2} J(1,a \sqrt{q})^2 a \ln(q/L) +2 a
q^{5/2} J(0,a \sqrt{q})^2 \ln(q/L)  \nn && -2 J q^{3/2} J(1,a
\sqrt{q})^2 a \ln(q/L) -2 a q^{5/2} J(0,a \sqrt{q}) \ln(q/L)  \nn
&& +4 J a q^{3/2} J(0,a \sqrt{q}) \ln(q/L) +4 J(1,a \sqrt{q}) q^2
\ln(q/L)  \nn && -4 J(1,a \sqrt{q}) q^2 J(0,a \sqrt{q}) \ln(q/L)
-4 J a q^{3/2} J(0,a \sqrt{q})^2 \ln(q/L)  \nn && +6 q^2 J(1,a
\sqrt{q}) -6 q^2 J(0,a \sqrt{q}) J(1,a \sqrt{q}) -12 J q J(1,a
\sqrt{q}) \nn && +8 J J(1,a \sqrt{q}) q J(0,a \sqrt{q}) \ln(q/L)
-8 J J(1,a \sqrt{q}) q \ln(q/L)  \nn && +12 J q J(0,a \sqrt{q})
J(1,a \sqrt{q})), \nn
n^{(2113)G}_{g}&=& a q( a q^{3/2} J(0,a \sqrt{q})^2 \ln(q/L)^2 +2
q J(1,a \sqrt{q}) \ln(q/L)^2  \nn && -2 J(1,a \sqrt{q}) q J(0,a
\sqrt{q}) \ln(q/L)^2 - a q^{3/2} J(0,a \sqrt{q}) \ln(q/L)^2  \nn
&& -4 q J(0,a \sqrt{q}) \ln(q/L) J(1,a \sqrt{q}) +4 q \ln(q/L)
J(1,a \sqrt{q})  \nn && +8 J J(0,a \sqrt{q}) \ln(q/L) J(1,a
\sqrt{q}) -8 J \ln(q/L) J(1,a \sqrt{q})  \nn && -4 J J(1,a
\sqrt{q}) \ln(q/L)^2 +4 J J(1,a \sqrt{q}) J(0,a \sqrt{q})
\ln(q/L)^2  \nn && +2 J a q^{1/2} J(0,a \sqrt{q}) \ln(q/L)^2 -2 J
a q^{1/2} J(0,a \sqrt{q})^2 \ln(q/L)^2). \ea

The $22$-factors can be cascaded into the following form

%n^{(22)G}_{g}&=& n^{(220)G}_{g}+ f n^{(221)G}_{g}, \nn

\ba n^{(220)G}_{g}&=& n^{(2202)G}_{g} p^2+  n^{(2203)G}_{g} p^3+
n^{(2204)G}_{g} p^4, \nn
n^{(221)G}_{g}&=& n^{(2212)G}_{g} p^2+  n^{(2213)G}_{g} p^3+
n^{(2214)G}_{g} p^4 \ea

with the individual sub-factors

\ba n^{(2202)G}_{g}&=& 2  \sqrt{J} \sqrt{J-q} ( 4 J(1,a \sqrt{q})
a q J(0,a \sqrt{q}) -4 a^2 q^{3/2} J(0,a \sqrt{q})^2  \nn && -4
J(1,a \sqrt{q}) a q +4 a^2 q^{3/2} J(0,a \sqrt{q}) -5 q^{3/2}
J(1,a \sqrt{q})^2 a^2  \nn && +12 \sqrt{q} +12 \sqrt{q} J(0,a
\sqrt{q})^2 -24 \sqrt{q} J(0,a \sqrt{q})), \nn
n^{(2203)G}_{g}&=& 4 \sqrt{J} \sqrt{J-q} ( - q^{3/2} J(1,a
\sqrt{q})^2 a^2 \ln(q/L) +2 a^2 q^{3/2} J(0,a \sqrt{q}) \ln(q/L)
\nn && -2 a^2   q^{3/2} J(0,a \sqrt{q})^2 \ln(q/L) +8 J(1,a
\sqrt{q}) a q J(0,a \sqrt{q}) \ln(q/L)  \nn && -8 J(1,a \sqrt{q})
a q \ln(q/L) +10 J(1,a \sqrt{q}) a q J(0,a \sqrt{q})  \nn && -10 q
J(1,a \sqrt{q}) a +12 \sqrt{q} J(0,a \sqrt{q})^2 \ln(q/L) +12
\sqrt{q} \ln(q/L)  \nn && +16 \sqrt{q} +16 \sqrt{q} J(0,a
\sqrt{q})^2 -24 \sqrt{q} J(0,a \sqrt{q}) \ln(q/L) -32 \sqrt{q}
J(0,a \sqrt{q})), \nn
n^{(2204)G}_{g}&=& 2 \sqrt{J} \sqrt{J-q} ( q^{3/2} J(1,a
\sqrt{q})^2 a^2 \ln(q/L)^2 +4 q J(0,a \sqrt{q}) \ln(q/L) J(1,a
\sqrt{q}) a  \nn && +4 \sqrt{q} -4 J(1,a \sqrt{q}) a q \ln(q/L)^2
+4 J(1,a \sqrt{q}) a q J(0,a \sqrt{q}) \ln(q/L)^2  \nn && -4 q
\ln(q/L) J(1,a \sqrt{q}) a +4 \sqrt{q} J(0,a \sqrt{q})^2
\ln(q/L)^2 +4 \sqrt{q} \ln(q/L)^2  \nn && +4 \sqrt{J-q} J(0,a
\sqrt{q})^2 -8 \sqrt{q} J(0,a \sqrt{q}) -8 \sqrt{q} J(0,a
\sqrt{q}) \ln(q/L)^2  \nn && +8 \sqrt{q} \ln(q/L) +8 \sqrt{q}
J(0,a \sqrt{q})^2 \ln(q/L) -16 \sqrt{q} J(0,a \sqrt{q}) \ln(q/L)).
\ea

The $221$-factors are given by the following expressions

\ba n^{(2212)G}_{g}&=& a( 3 a   q^{5/2} J(0,a \sqrt{q})^2 -3 a
q^{5/2} J(0,a \sqrt{q}) +4 q^{5/2}J(1,a \sqrt{q})^2 a  \nn && +6 J
a q^{3/2} J(0,a \sqrt{q}) -6 J(1,a \sqrt{q}) q^2 J(0,a \sqrt{q})
+6 J(1,a \sqrt{q}) q^2  \nn && -6 J a q^{3/2} J(0,a \sqrt{q})^2 -8
J q^{3/2} J(1,a \sqrt{q})^2 a -12 J q J(1,a \sqrt{q}) \nn && +12 J
J(1,a \sqrt{q}) q J(0,a \sqrt{q}) ), \nn
n^{(2213)G}_{g}&=& 2 a q( 2 q^{3/2} J(1,a \sqrt{q})^2 a \ln(q/L)
+3 a q^{3/2} J(0,a \sqrt{q})^2 \ln(q/L)  \nn && -3 a q^{3/2} J(0,a
\sqrt{q}) \ln(q/L) -4 J q^{1/2} J(1,a \sqrt{q})^2 a \ln(q/L) \nn
&& +6 J(1,a \sqrt{q}) q \ln(q/L) -6 J(1,a \sqrt{q}) q J(0,a
\sqrt{q}) \ln(q/L)  \nn && +6 J a q^{1/2} J(0,a \sqrt{q}) \ln(q/L)
-6 J a q^{1/2} J(0,a \sqrt{q})^2 \ln(q/L) \nn && -10 q J(0,a
\sqrt{q}) J(1,a \sqrt{q}) +12 J J(1,a \sqrt{q}) J(0,a \sqrt{q})
\ln(q/L) \nn && +10 q J(1,a \sqrt{q}) -12 J J(1,a \sqrt{q})
\ln(q/L) -20 J J(1,a \sqrt{q}) \nn && +20 J J(0,a \sqrt{q}) J(1,a
\sqrt{q})), \nn
n^{(2214)G}_{g}&=& a q( a q^{3/2} J(0,a \sqrt{q})^2 \ln(q/L)^2 -a
q^{3/2} J(0,a \sqrt{q}) \ln(q/L)^2  \nn && -2 J(1,a \sqrt{q}) q
J(0,a \sqrt{q}) \ln(q/L)^2 +2 J a q^{1/2} J(0,a \sqrt{q})
\ln(q/L)^2  \nn && -2 J a q^{1/2} J(0,a \sqrt{q})^2 \ln(q/L)^2 +2
q J(1,a \sqrt{q}) \ln(q/L)^2  \nn && +4 q \ln(q/L) J(1,a \sqrt{q})
-4 q J(0,a \sqrt{q}) \ln(q/L) J(1,a \sqrt{q})  \nn && -4 J J(1,a
\sqrt{q}) \ln(q/L)^2 +4 J J(1,a \sqrt{q}) J(0,a \sqrt{q})
\ln(q/L)^2 \nn && +8 J J(0,a \sqrt{q}) \ln(q/L) J(1,a \sqrt{q}) -8
J \ln(q/L) J(1,a \sqrt{q})).\ea

Similarly, the $23$-factors turn out to be

%n^{(23)G}_{g}&=& n^{(230)G}_{g}+ f n^{(231)G}_{g}

\ba n^{(230)G}_{g}&=& n^{(2303)G}_{g} p^3+ n^{(2304)G}_{g} p^4,
\nn
n^{(231)G}_{g}&=& n^{(2313)G}_{g} p^3+ n^{(2314)G}_{g} p^4.\ea

\ba n^{(2303)G}_{g}&=& 2 \sqrt{J} \sqrt{q} \sqrt{J-q} ( 2 a^2   q
J(0,a \sqrt{q}) -2 a^2   q J(0,a \sqrt{q})^2 -3 q J(1,a
\sqrt{q})^2 a^2 \nn && +4 J(0,a \sqrt{q})^2 +4 -8 J(0,a
\sqrt{q})), \nn
n^{(2304)G}_{g}&=& 4\sqrt{J} \sqrt{J-q} ( a^2 q^{3/2} J(0,a
\sqrt{q}) \ln(q/L) -a^2 q^{3/2} J(0,a \sqrt{q})^2 \ln(q/L) \nn &&
-q^{3/2} J(1,a \sqrt{q})^2 a^2 \ln(q/L) +2 a q J(1,a \sqrt{q})
J(0,a \sqrt{q}) \ln(q/L) \nn && -2 a q J(1,a \sqrt{q}) \ln(q/L) +4
\sqrt{q} \ln(q/L) +4 J(1,a \sqrt{q}) a q J(0,a \sqrt{q}) \nn && +4
\sqrt{q} J(0,a \sqrt{q})^2 \ln(q/L) -4 J(1,a \sqrt{q}) a q
+6\sqrt{q} +6 \sqrt{q} J(0,a \sqrt{q})^2 \nn && -8 \sqrt{q} J(0,a
\sqrt{q}) \ln(q/L) -12 \sqrt{q} J(0,a \sqrt{q})),\ea

and

\ba n^{(2313)G}_{g}&=& a q ( a q^{3/2} J(0,a \sqrt{q})^2 -a
q^{3/2} J(0,a \sqrt{q}) +2 q^{3/2} J(1,a \sqrt{q})^2 a \nn && +2 q
J(1,a \sqrt{q}) -2 J(1,a \sqrt{q}) q J(0,a \sqrt{q}) +2 J a
q^{1/2} J(0,a \sqrt{q}) \nn && -2 J a q^{1/2} J(0,a \sqrt{q})^2 -4
J q^{1/2} J(1,a \sqrt{q})^2 a -4 J J(1,a \sqrt{q}) \nn && +4 J
J(1,a \sqrt{q})  J(0,a \sqrt{q})), \nn
n^{(2314)G}_{g}&=& 2 a q ( a q^{3/2} J(0,a \sqrt{q})^2 \ln(q/L)
+q^{3/2} J(1,a \sqrt{q})^2 a \ln(q/L) \nn && -a q^{3/2} J(0,a
\sqrt{q}) \ln(q/L) -2J q^{1/2} J(1,a \sqrt{q})^2 a \ln(q/L)  \nn
&& -2J a q^{1/2} J(0,a \sqrt{q})^2 \ln(q/L) -2J(1,a \sqrt{q}) q
J(0,a \sqrt{q}) \ln(q/L)  \nn && +2q J(1,a \sqrt{q}) \ln(q/L) +2J
a q^{1/2} J(0,a \sqrt{q}) \ln(q/L) +4q J(1,a \sqrt{q})  \nn && -4q
J(0,a \sqrt{q}) J(1,a \sqrt{q}) +4J J(1,a \sqrt{q}) J(0,a
\sqrt{q}) \ln(q/L)  \nn && -4J J(1,a \sqrt{q}) \ln(q/L) +8J J(0,a
\sqrt{q}) J(1,a \sqrt{q}) -8J J(1,a \sqrt{q})).\ea

%l(p)^3 terms

Finally, the $l(p)^3$ terms are given by

\ba n^{(3)G}_{g}&=&  n^{(30)G}_{g} +(q/L)^p n^{(31)G}_{g}
+((q/L)^p)^2 n^{(32)G}_{g}+ ((q/L)^p)^3 n^{(33)G}_{g}, \ea

where the factors are expressed as

\ba n^{(30)G}_{g}&=& n^{(300)G}_{g}+ f n^{(301)G}_{g}, \nn
n^{(31)G}_{g}&=& n^{(310)G}_{g}+ f n^{(311)G}_{g}, \nn
n^{(32)G}_{g}&=& n^{(320)G}_{g}+ f n^{(321)G}_{g}, \nn
n^{(33)G}_{g}&=& n^{(330)G}_{g}+ f n^{(331)G}_{g}. \ea

The $30$-factors are given by the following expressions

\ba n^{(300)G}_{g}&=& 2 a q \sqrt{J} \sqrt{J-q} ( a q^{1/2} J(0,a
\sqrt{q})^2 - a q^{1/2} J(0,a \sqrt{q})  \nn && +2 J(1,a \sqrt{q})
J(0,a \sqrt{q}) +2 q^{1/2} J(1,a \sqrt{q})^2 a -2 J(1,a
\sqrt{q})), \nn
n^{(301)G}_{g}&=& a^2 q^{3/2}  J(1,a \sqrt{q})^2 (2 J -q). \ea

The $31$-factors take the following forms

\ba n^{(310)G}_{g}&=& 6 a p q \sqrt{J} \sqrt{J-q} ( a q^{1/2}
J(0,a \sqrt{q})^2 - a q^{1/2} J(0,a \sqrt{q})  \nn && -2 J(1,a
\sqrt{q}) +2 J(0,a \sqrt{q}) J(1,a \sqrt{q}) +2 q^{1/2} J(1,a
\sqrt{q})^2 a), \nn
n^{(311)G}_{g}&=& 3 a^2 q^{3/2} J(1,a \sqrt{q})^2 p (2 J -q). \ea

Similarly, the $32$-factors turn out to be

\ba n^{(320)G}_{g}&=& 6 a q  p^2 \sqrt{J} \sqrt{J-q} ( a q^{1/2}
J(0,a \sqrt{q})^2 - a q^{1/2} J(0,a \sqrt{q})  \nn && -2 J(1,a
\sqrt{q}) +2 q^{1/2} J(1,a \sqrt{q})^2 a +2 J(0,a \sqrt{q}) J(1,a
\sqrt{q})), \nn
n^{(321)G}_{g}&=& 3 a^2 p^2 q^{3/2}  J(1,a \sqrt{q})^2 (2 J
-q).\ea

Finally, the $33$-factors can be obtained as follows

\ba n^{(330)G}_{g}&=& 2a q p^3 \sqrt{J} \sqrt{J-q}( a q^{1/2}
J(0,a \sqrt{q})^2 - a q^{1/2} J(0,a \sqrt{q})  \nn && +2 J(1,a
\sqrt{q}) J(0,a \sqrt{q}) -2 J(1,a \sqrt{q}) +2 q^{1/2} J(1,a
\sqrt{q})^2 a), \nn
n^{(331)G}_{g}&=&  a^2 p^3  J(1,a \sqrt{q})^2 J(1,a \sqrt{q})^2 (2
J -q). \ea

\begin{figure}
\hspace*{0.5cm}
\includegraphics[width=8.0cm,angle=-90]{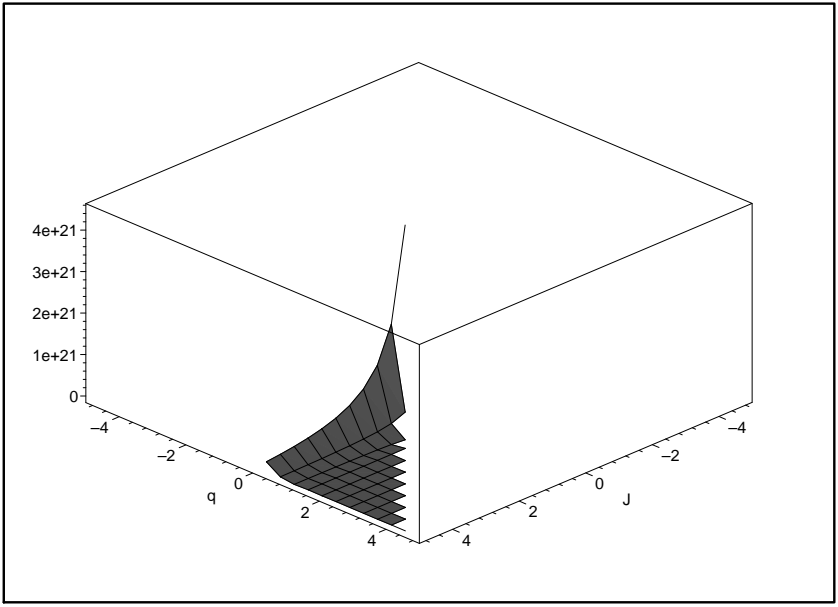}
\caption{The dimensional behavior of the determinant of the metric
tensor $\Vert g \Vert $ plotted as a function of the scale
 and angular momentum, viz., $q, J$, describing the nature
of the stability of the general massive rotating quarkonia.}
\label{det3dqpJ} \vspace*{0.5cm}
\end{figure}

From the three parameter quarkonia, the Fig.(\ref{det3dqpJ})
offers the general three dimensional behavior of the determinant
of the metric tensor. From the viewpoint of the fixed $p$
quarkonia in the Regge regime as previously described, we notice
in the present case that the graph of the determinant of the
metric tensor shows the similar nature of the Gaussian
fluctuations of the strong QCD coupling. Notice that the
underlying quarkonium configuration may become unstable in
specific regions of the parameters. For instance, for small values
of the momentum scale and angular momentum, viz $q, J \rightarrow
0^{+}$, we observe that the determinant of the metric tensor picks
up a large amplitude of order $10^{21}$. However, for intermediate
values of $q$ and angular momentum $q,J$, the Fig.(\ref{det3dqpJ})
shows that the underlying configuration is well-stable  and so is
the basis for the three dimension intrinsic manifold $(M_3,g)$.
From the observation of the Fig.(\ref{det3dqpJ}), we predict that
the regions of the thermodynamic stability are present in the
interval $q,J \in (1,4)$. Globally, the stability of $(M_3,g)$
constraint the principle minors $\{g_{ii}, p_S^G, \Vert g \Vert
\}$ to remain positive. Specifically, for the same sign of $\{ b,
p, f\}$, the volume stability of the $(M_3,g)$ imposes the
following constraint \ba n^{(0)G}_{g}+ n^{(1)G}_{g}l(p)+
n^{(2)G}_{g}l(p)^2+ n^{(3)G}_{g}l(p)^3 &>& \ 0, \ J(0,a \sqrt{q})
< 1, \nn &<& \ 0, \ J(0,a \sqrt{q}) < 1. \ea

Importantly,  it is worth mentioning that both the limiting
configurations with $J=q$ and $J(0,a \sqrt{q})= 1$ are abided from
the thermodynamic stability constraints. To summarize the phases
of generic quarkonia, the exact formula for the scalar curvature
may analogously be deduced as the one we have offered for the
$qJ$-plane. With some modification of the pre-factors, we find
that the summation over $l(p)$ naturally arises with the $B_n$ as
the polynomials in $p$, whose coefficients can be expressed as the
functions of the Bloch-Nordsieck logarithmic factor $f(q,J)$. In
the present case, the denominator of the scalar curvature involves
the third and second power of the factors in the numerator of the
determinant of the metric tensor. Interestingly, the quantitative
property of the scalar curvature follows, except the fact that its
explicit presentation is a bit prolonged. Up to a phase of QCD, we
observe that the global properties of the three parameter
quarkonia remain the same as we have exactly indicated for the
$QJ$- and $QM$-planes.

\section{Discussion and Conclusion}
We have studied the thermodynamic geometric properties of
non-abelian quarkonium bound states arising from the consideration
of the one-loop gluon confining QCD potentials. Following the
Polyakov argument, the thermodynamic intrinsic geometric nature of
the QCD coupling is analyzed for the case of the Coulambic and
rising confining regimes. Without any approximation, the effect of
Regge and Bloch-Nordsieck rotations is explored form the stability
perspective of the strongly coupled quarkonia with and without the
inclusion
of the effects coming from the mass.%the effects of the mass.
The fact that once the quark matter being formed should be stable,
motivated us to determine the values of the parameters of strongly
coupled quarkonia. We have obtained the allowed range of
parameters from the stability conditions of the thermodynamic
geometry. We have initiated our analysis with the consideration of
massless non-rotating quarkonia. In this case, we have derived
that the thermodynamic stability constraints impose the
requirements $g_{ii}>0$ and $\Vert g \Vert >0$. Given the momentum
scale $q$ and index $p$ of the configuration, our analysis
determines an intrinsically stable inter dependence of the
parameter from one and the other. The Regge rotating quarkonia is
geometrically trivial because it only requires the positive sign
of the rotation term. With and without the Regge rotation, the
domain of the index $p(q)$ can be offered as \ba \mathcal
P_{m=0}:= \{p \ | \ g_{ii}> 0, \ \Vert g \Vert >0 \}. \ea

In the case of massive rotating quarkonia, there is a set of
thermodynamic stability constraints on the parameters which arises
from the fluctuations of the q-angular momentum plane, q-mass
plane and the entire q-angular momentum-index manifold, when all
the parameters of the configuration are allowed to fluctuate. It
turns out that one can fix all the parameters of the configuration
following from one constraint to the other. One such procedure is
to respectively consider the fluctuations on the $QJ$-plane, then
on the $QM$-plane and finally on the $QpJ$-manifold. In the case
of the $QJ$-plane, it follows that we determine the
thermodynamically stable $Q^2(J)$ directly from the fluctuations
of the Bloch-Nordsieck rotation. By taking an account of the
thermodynamic stability constraint of the $Q^2(J)$, we can
subsequently determine the thermodynamically stable domain of the
mass $M(Q)$ on the $QM$-plane. Finally, the thermodynamically
stable domain of the index $p$ can be determined as the function
of $Q$ and $J$ on the $QpJ$-manifold. Thus, the consideration of
the thermodynamic fluctuation theory offers the determination of
the stable domains of the QCD strong coupling. This follows from
the fact that the stable domain of $q$ and the Bloch-Nordsieck
rotation can be determined from one another, and thus the index
can solely be expressed as a function of one the either two. This
shows the power of the thermodynamic stability condition that
determines all possible stable phases of the strongly coupled
quarkonia. Based on the analysis of the present paper, the
thermodynamically stable index of the Bloch-Nordsieck rotating
massive quarkonia is constrained by the following set \ba \mathcal
P_{m \ne 0}:= \{p \ | \ n^{G}_{ii}> 0,\ n^{G}_{S}> 0,\ n^{G}_{g}>
0 \}. \ea

To be specific, we have introduced the intrinsic geometric notion
to QCD thermodynamics, and thereby it has been applied to study
the behavior of constant and variable index strong QCD coupling in
several simple thermodynamic considerations. Our explicit
computations of the local and global thermodynamic corrections
provide a prominent realization of the limiting equilibrium
thermodynamic configuration of the strongly coupled QCD. This
offers an intrinsic geometric exercise of the quarkonia, which has
been described by the parameters of the strong coupling potential.
The scalar curvature of the underlying thermodynamic metric tensor
takes an exact form, which signifies the correlation volume of the
underlying quarkonia. With and without the variable index strong
QCD coupling, our consideration of the underlying thermodynamic
fluctuations indicates interesting relationships between the
geometrical concepts of quarkonium thermodynamics and phases of
the massless, Regge rotating and Bloch-Nordsieck rotating
quarkonia. For the case of the Bloch-Nordsieck rotating quarkonia,
we have implemented the above constraints one by one, and thereby
we conclude that the thermodynamically stable domain of the
parameters is highly constrained on the $(M_3,g)$. For the value
of $p=5/6$ which generically corresponds to the Regge rotating
quarkonia, we observe that all the stability constraints are
satisfied, whenever $q,J \in (1,4)$. Thermodynamically, this
determines all possible intermediate stable phases of the strongly
coupled quarkonia. As the gluonic interactions become softer and
softer, we find in the limit of Bloch-Nordsieck resummation that
the underlying Sudhakov form factor offers the thermodynamic
stability properties of the $\pi$, $K$ and $D_s$ particles.

Mathematically, we have computed the intrinsic geometric
properties of the Bessel function of first kind convoluted with
two logarithmic functions of the respective weights $(0,2p)$.
Explicitly, we determine the domain of the convolution of the
zeroth order Bessel function $J(0,a\sqrt{q})$, where the
configuration finds globally stable fixed points and remains
globally regular on the intrinsic three manifold $(M_3,g)$. Notice
that the exact determination of the global stability set $\mathcal
P_{m \ne 0}$ is consistent with the local stability constraints on
$(M_3,g)$. Geometrically, we find that the global stability
requirement of $(M_3,g)$ avoids the limit
$J(0,a\sqrt{q})\rightarrow 1$.  From the perspective of the
intrinsic Riemannian geometry, it would be worth to find the
globally stable and globally regular domains of the convoluted
Bessel functions with finitely many elementary functions of
distinct weights. This study herewith anticipates further
examination of the globally stable and regular mapping class
convolutions from the perspective of intrinsic Riemannian
geometry.

Physically, it would be interesting to analyze whether this kind
of approach can be pushed further to explore intrinsic geometric
properties of strong coupling QCD thermodynamic correlations.
Specifically, we would like to understand the modifications
towards the elements of the set $\mathcal P_{m \ne 0}$ and the
associated Ricci scalar curvature, which we have explicitly
offered for the $QJ$-plane. This geometric initiation would take a
better shape once we take finite size effects of higher order
perturbative QCD into account, and would reach its final goal with
the complete account of non-perturbative QCD. The geometric
feature of these explorations is left for the future.

\section*{Acknowledgment}
The work of S. B. has been supported in part by the European
Research Council grant n. 226455, \textit{``SUPERSYMMETRY, QUANTUM
GRAVITY AND GAUGE FIELDS (SUPERFIELDS)''}. BNT thanks Prof. V.
Ravishankar for support and encouragement. The work of BNT has
been supported by a postdoctoral research fellowship of the
\textit{``INFN-Laboratori Nazionali di Frascati, Roma, Italy''}.

\end{document}